\documentclass[aps,prb,twocolumn,groupedaddress,floatfix,longbibliography]{revtex4-2}

\usepackage{graphicx}
\usepackage{dcolumn}
\usepackage{bm}
\usepackage{amsmath}
\usepackage{xcolor}
\usepackage{braket}
\usepackage[utf8]{inputenc}
\usepackage[T1]{fontenc}
\usepackage{hyperref}
\hypersetup{
breaklinks=true,
    colorlinks=true,
    linkcolor=blue,
    citecolor=blue,
    filecolor=magenta,
    urlcolor=cyan,
}

\usepackage{blkarray}
\usepackage{multirow}

\usepackage{natbib}      

\newcommand{\circlez}{\raisebox{.5pt}{\textcircled{\raisebox{-.9pt} {\bf 0}}}}
\newcommand{\circlem}{{\large ${\bm \ominus}$}}
\newcommand{\circlep}{{\large ${\bm \oplus}$}}
\newcommand{\circleps}{{\scriptsize ${\bm \oplus}$}}

\newcommand{\hbv}{\widehat{\bf V}}
\newcommand{\hbs}{\widehat{\bf S}}
\newcommand{\igs}{{\it g}}
\newcommand{\ies}{{\it e}}

\newcommand{\ve}{\varepsilon}
\newcommand{\cs}{{\cal S}}
\newcommand{\cv}{{\cal V}}

\newcommand{\br}{{\bf r}}

\begin{document}

\title{Valleytronic full configuration-interaction approach: An application to the excitation 
spectra of Si double-dot qubits}

\author{Constantine Yannouleas}
\email{Constantine.Yannouleas@physics.gatech.edu}
\author{Uzi Landman}
\email{Uzi.Landman@physics.gatech.edu}

\affiliation{School of Physics, Georgia Institute of Technology,
             Atlanta, Georgia 30332-0430}

\date{submitted: 04 July 2022}

\begin{abstract}
A study of the influence of strong electron-electron interactions and Wigner-molecule (WM) formation 
on the spectra of $2e$ singlet-triplet double-dot Si qubits is presented based on a full 
configuration interaction (FCI) approach that incorporates the valley degree of freedom (VDOF) in the 
context of the continuous (effective mass) description of semiconductor materials. Our FCI solutions 
correspond to treating the VDOF as an isospin in addition to the regular spin. A major advantage of our 
treatment is its capability to assign to each energy curve in the qubit's spectrum a complete set of 
good quantum numbers for both the spin and the valley isospin. This allows for the interpretation of the 
Si double-dot spectra according to an underlying SU(4) $\supset$ SU(2) $\times$ SU(2)
group-chain organization. Considering parameters in the range of actual experimental situations, we 
demonstrate for the first time in a double-dot qubit that, in the (2,0) charge configuration and compared 
to the expected large, and dot-size determined, single-particle (orbital) energy gap, the strong $e-e$ 
interactions drastically quench the spin-singlet$-$spin-triplet energy gap, $E_{\rm ST}^{\mbox \circleps}$, 
within the same valley, making it competitive to the small energy gap, $E_V$, between the two valleys. We 
present results for both the $E_{\rm ST}^{\mbox \circleps} < E_V$ and $E_{\rm ST}^{\mbox \circleps} > E_V$ 
cases, which have been reported to occur in different experimental qubit devices. In particular, we investigate 
the spectra as a function of detuning and demonstrate the strengthening of the all-important avoided 
crossings due to a lowering of the interdot barrier and/or the influence of valley-orbit coupling. We 
further  demonstrate, as a function of an applied magnetic field, the emergence of avoided crossings in the 
(1,1)  charge configuration due to the more general spin-valley coupling, in agreement with experiments.
The valleytronic FCI method formulated and implemented in this paper, and demonstrated for the case of two 
electrons confined in a tunable double quantum dot, offers also a most effective tool for analyzing the spectra
of Si qubits with more than two wells and/or more than two electrons, in field-free conditions, as well as 
under the influence of an applied magnetic field. Furthermore, it can also be straightforwardly extended to 
the case of bilayer graphene quantum dots.
\end{abstract}

\maketitle

\section{Introduction}
\label{intr}

Studies of silicon qubits are developing into a major research focus \cite{copp13,vand19,burk21,kuem21},
due to their inherent long electron decoherence times. However, the presence of the additional valley degree 
of freedom (VDOF) introduces a higher degree of complexity (compared to cases where the VDOF is absent, 
e.g., GaAs nanodevices \cite{pett05,yann06,kim21}) which 
has not as yet been satisfactorily deciphered, experimentally or theoretically. Indeed, a number of 
publications consider it as a challenge to be overcome or to do away with
(see, e.g., Refs.\ \cite{copp13,maun12,kawa14,coppwigg21}), whereas a second group of publications 
(see, e.g., Refs.\ \cite{pett16,pett18,nich21,erca22,pett22})
considers it as a potential resource to be further explored. Experimental efforts in both directions are 
intensely pursued, but a definitive resolution has not been reached as yet.

The current state of understanding of the VDOF complexity in solid-state qubits is further compounded by 
the recent realization that the energy gaps in the relevant excitation spectra (with the potential to be
involved in the operation of the qubit) depend crucially on Wigner-molecule (WM) 
\cite{yann99,grab99,yann00,yann00.2,fili01,yann02.1,yann02.2,szaf03,yann06,ront06,yann07.2,yann07,yann09} 
formation and the strong electron-electron interactions
that naturally arise in these silicon nanodevices \cite{corr21,erca22,erca21.2,urie21,corr21.2} (also found 
in GaAs 3e hybrid qubits \cite{kim21,yann22,yann22.2}), due to an interplay between materials 
parameters and size that yields large values for the Wigner parameter $R_W > 1$; see Sec.\ \ref{sec:rw}.

Among the various many-body approaches employed in this area of research, 
it is becoming increasingly transparent that the strong $e-e$ 
interaction regime in few-electron nanosystems can be best described theoretically through the use of the 
microscopic full configuration interaction (FCI) methodology \cite{shav98}, 
which treats in a most efficient way the two-body part of the Hamiltonian governing the system. Indeed, the 
FCI (referred to also as exact diagonalization) has been successfully applied 
\cite{yann03,szaf03,yann06,ront06,yann07.2,yann09,yann22,yann22.2}
in the last two decades to two-dimensional (2D) quantum dots (QDs) with single-band semiconductor materials 
substrates, like GaAs. FCI calculations for condensed-matter nanostructures 
that incorporate aspects of the VDOF have been also considered recently \cite{erca21.2,more21}. 
However, a CI approach that unequivocally 
relates to the field of valleytronics \cite{been07,been08,nori18,scur20,mrud21} by properly incorporating
the valley degree of freedom as an isospin \cite{hein70}, in complete analogy with the regular spin, is 
still missing. 

Here, we fill this gap by introducing a valleytronic FCI (VFCI) approach that employs single-particle 
(one-body) bases associated with the continuum-model (effective mass approximation \cite{kohn55}
of the two-band structure in Si QDs 
\cite{[{For the application of the effective mass theory to the band structure of single carbon-nanotube QDs, 
see Ref.\ \cite{maks12} and (a)}][{}]sech10,
*[{for the application of the effective mass theory to the band structure of single graphene QDs, see 
Ref.\ \cite{maks12} }][{}]bece14,
*[{for the application of the effective mass theory to the band structure of Si QDs,
see (c) }][{; (d) }]eto03,*frie07,*[{and (e) }][{}]frie10}). 
Our VFCI enables the acquisition of numerical results complete with full 
spin-isospin assignments that reveal an underlying SU(4) $\supset$ SU(2) $\times$ SU(2)
\cite{[{For the earliest application of the SU(4) group theory to physics (specifically nuclear 
physics, based on the approximate proton-neutron equivalence), and for further original references related 
to the concept of the isospin (referred to also as isotopic spin), see~}] [{}] wign37, 
[{For an earlier application of the SU(4) group theory to condensed-matter spin-lattice models with  
{Hamiltonians} of a generalized Heisenberg type that take into account the orbital degeneracy present in the
metal ions of many transition metal oxides, see (a) }] [{, and references therein. }] shi98,
*[{ See also (b) }][]auer95, 
[{For Hubbard-type applications of the SU($N$) group theory to ultracold-atom optical lattices, 
see (a) }] [{, and references therein. }] capp16,*[{ See also (b) }][{}]hone04} 
group-chain organization of the spectra of Si double-quantum-dot (DQD) qubits. 
Specifically, the valley isospin assignments in our VFCI consist of a pair of indices $({\cal V},V_z)$ in 
analogy with the two indices $({\cal S},S_z)$ of the regular spin. 
(This is in contrast with earlier CI implementations in single QDs constructed from other materials 
exhibiting a VDOF, which have been restricted 
\cite{[{For bilayer graphene QDs, see (a) }] [{}]zare13,
*[{for carbon nanotube QDs, see, e.g., (b) }][{}]maks12}
to characterizing the CI states using one valley index only, 
namely the valley projection {$V_z$} \footnote{
Also CI implementations (see, e.g., Ref.\ \cite{erca21.2}) for Si/SiGe single QDs that employ lattice 
tight-binding single-particle bases, as well as CI simulations of exchange-coupled donors in silicon using
multi-valley effective mass theory \cite{more21},  are using the regular spin indices $({\cal S}, S_z)$ 
only.}.)

In this paper, we show that such an organizing principle is essential in deciphering the complexity of the 
two-electron-DQD (2e-DQD) spectra arising from the interplay of both the VDOF and the selective suppression 
of spectral energy gaps associated with the emergence of Wigner molecules in the strong-correlation regime.
This endeavor is most desirable, given that the recent CI investigations of the WM effects on the spectra 
and behavior of Si qubits have been restricted \cite{corr21,erca22,erca21.2,urie21} to the case of a single
QD, in addition to overlooking the systematic patterns arising from the underlying group theoretical 
properties of the valley isospin.   

\begin{figure*}[t]
\centering\includegraphics[width=17.5cm]{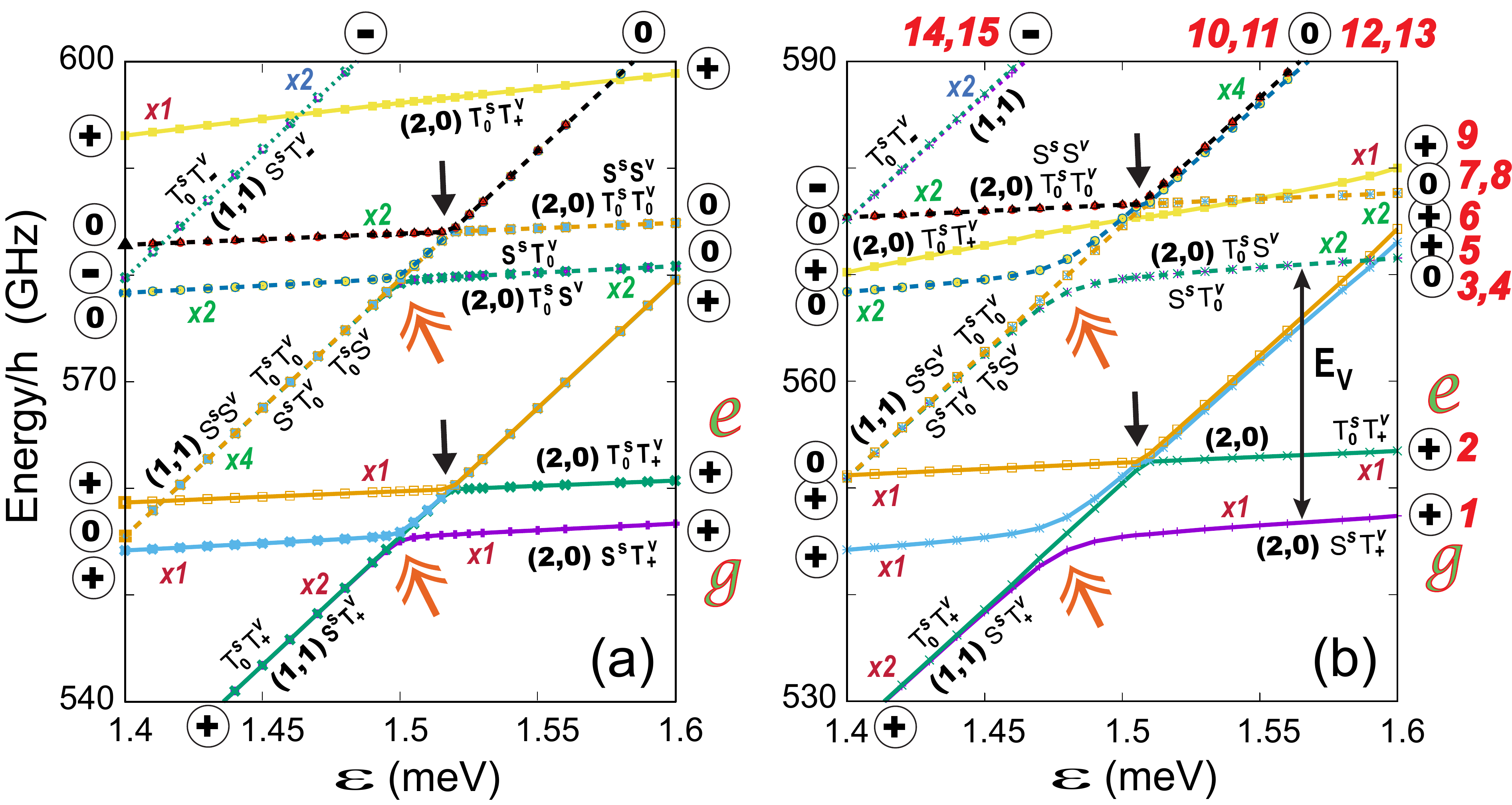}
\caption{
The case with $E_{\rm ST}^{\mbox \circleps} < E_V$.
VFCI lowest energy spectrum, associated with the Hamiltonian $H_{\rm MB}+H_{\rm VS}$ (see text), for the case 
of a 2e Si DQD with a weaker individual-dot confinement, $\hbar \omega_0 = 0.40$ meV $=96.719h$ GHz, for both 
dots, at a vanishing magnetic field, $B=0$. The spectrum is plotted in the transition region between the 
(1,1) and the (2,0) charge configurations, as a function of the detuning, $\ve$, and for a total-spin 
projection $S_z=0$. The interdot separation was taken as $d=150$ nm, and the valley gap was set as 
$E_V=100$ $\mu$eV $=24.180h$ GHz. 
{\it The first excited state in the (2,0) configuration has both electrons in the same lowest-energy 
valley [g and e denote the ground and first-excited states in the (2,0) configuration, respectively].}
The energy difference between the $e$ and $g$ states at $\ve=1.60$ meV equals 
$E_{\rm ST}^{\mbox \circleps} \approx 4.05h$ GHz in panel (a) and 
$E_{\rm ST}^{\mbox \circleps} \approx 6.09h$ GHz in panel (b). 
(a) A high interdot barrier was implemented by setting $\epsilon_1^{b,{\rm inp}}=0.65$ (see Sec.\ 
\ref{sec:mbh} for the meaning of the input barrier-controlling parameter). (b) A low interdot barrier was 
implemented by setting $\epsilon_1^{b,{\rm inp}}=0.50$.
The in-plane Si effective mass was taken as 0.191$m_e$ and the dielectric constant as $\kappa=11.4$. 
The symbols \circlep, \circlez, and \circlem ~indicate states with both electrons in the low-energy valley, 
with the electrons in different valleys, and with both electrons in the high-energy valley, respectively. 
The 5 near-horizontal lines correspond to the (2,0) charge configuration (as indicated). The three
quasi-parallel lines with a larger slope correspond to the (1,1) charge configuration (as indicated). 
The valley-orbit and spin-isospin couplings were neglected, and as a result all states are associated with 
good total spin [$\cs(\cs+1)$,$S_z$] and valley isospin [$\cv(\cv+1)$,$V_z$] 
quantum numbers (as indicated). The symbols made from
a combination of the capital letters S and T have the following meaning: A capital S denotes a singlet state,
whereas a capital T denotes a triplet state. The subscripts $\pm$ denote an $S_z=\pm 1$ spin projection or a 
$V_z=\pm 1$ valley projection, respectively, whereas a subscript $0$ denotes an $S_z=0$ or $V_z=0$ 
projection. A superscript ``s'' denotes the regular spin, 
whereas a superscipt ``v'' denotes the valley isospin. 
``$xn$'' (with $n=1$, 2, or 4) denotes the degeneracy associated with a given energy curve; red color is used 
for the \circlep ~states (with $V_z=+1$), green color for the \circlez ~states (with $V_z=0$), and blue 
color for the \circlem ~ones (with $V_z=-1$). 
The double-headed red arrows indicate avoided crossings between two spin singlet states in the transition 
from the (1,1) to the (2,0) configuration. These crossings are underdeveloped in the case of a high interdot
barrier [see panel (a)], but become pronounced for low interdot barriers [see panel (b)]. 
The single-head black arrows indicate avoided crossings between two spin triplet states in the transition 
from the (1,1) to the (2,0) configuration. These crossings are underdeveloped in both the cases of a high 
interdot barrier [see panel (a)], as well as a low interdot barrier [see panel (b)]. The red numbers at the 
border of panel (b) will assist with the correspondence between charge densities (see Fig.\ \ref{dens0450}) 
and the states whose energies are plotted here. In all figures and the values mentioned in the text, the 
energies are referenced to $2 \hbar \sqrt{\omega_0^2+\omega_c^2/4}$, where $\omega_c=eB/(m^*c)$ is the 
cyclotron frequency. The dots in this figure and in all subsequent figures are equidistant from the origin.
}
\label{sp046550}
\end{figure*}

We analyze characteristic cases of Si 2e-DQD theoretical spectra that can be associated with experimentally 
measured ones. In particular, we address the following cases: 
\begin{enumerate}
\item
The first-excited state in the (2,0) configuration \footnote{
The notation $(n_L,n_R)$ indicates charge configurations with $n_L$ electrons in the left dot, and $n_R$ 
electrons in the right dot.}
is a spin triplet with both electrons in the lower-energy valley. This case is the result of strong $e-e$ 
interactions which drastically quench the spin-singlet/spin-triplet energy gap, 
$E_{\rm ST}^{\mbox \circleps}$, within the same valley, making it smaller than the small
energy gap, $E_{V}$, between the two Si valleys, i.e., $E_{V} > E_{\rm ST}^{\mbox \circleps}$. 
Related experimental situations have been reported in Refs.\ \cite{corr21,dods22}. 
The VFCI results concerning this case are presented in Sec.\ \ref{sec:same}.
\item
The first-excited state in the (2,0) charge configuration involves the excitation of one electron to the
higher-energy valley, namely, one has $E_{V} < E_{\rm ST}^{\mbox \circleps}$. 
Related experimental situations have been reported in Ref.\ \cite{pett22,nich21}.
The VFCI results concerning this case are presented in Sec.\ \ref{sec:diff}.
\item
In the (1,1) charge configuration, the first-excited state involves the promotion of one electron to the
higher-energy valley and a complete set of three multiplets [containing an SU(4)-characteristic total of 16
states] are resolved by lifting their degeneracies through the application of a magnetic field. The related
experimental investigation was reported in Ref.\ \cite{taha14}.
This investigation parallels the recent investigations of the 2e spectra in single bilayer graphene quantum 
dots \cite{enss19,stam21}.
The VFCI results concerning this case are presented in Sec.\ \ref{sec:mfs}.
\end{enumerate}

Before proceeding with the description of the VFCI results, we mention here that in the context of
valleytronics, the valley isospin is defined by a three-dimensional vector $\hbv$ (in analogy with the 
regular spin vector $\hbs$) which has three projections $(V_x,V_y,V_z)$ [in analogy with the regular-spin 
projections  $(S_x,S_y,S_z)$]. The three valley projections $V_q$, $q=x,y,z$, obey the same Lie algebra 
(commutation relations) as the three spin projections $S_q$, $q=x,y,z$. Likewise, the valley Casimir 
operator is given by $\hbv^2=V_x^2+V_y^2+V_z^2$ [with eigenvalues $\cv(\cv+1)$], in analogy with the 
regular-spin Casimir operator, $\hbs^2=S_x^2+S_y^2+S_z^2$ [with eigenvalues $\cs(\cs+1)$]. An electron 
having $V_z=\pm 1/2$ means that it lies in the low- or high-energy valley, respectively.
An expanded presentation of the mathematics of the VFCI approach is given in Sec.\ \ref{sec:meth}.

\section{Results}
\label{sec:res}

In this section, we present VFCI results for the low energy spectra of Si 2e-DQD devices with parameters
similar to those of actual quantum qubits investigated experimentally and reported in recent and current 
literature. Prior to presentation and discussion of the results of our calculations, it is pertinent to comment
here in some detail about certain aspects of our calculations, originating from the intrinsic properties of the
material (silicon) used in making the DQD qubits addressed by our study. To this end we focus specifically on 
the valletronic nature of the electronic structure of the Si quantum dots studied here, and the terminology  
used in characterizing and discussing their properties.

The band structure of crystalline silicon (having a covalently-bonded, cubic, diamond lattice structure) is 
known to exhibit in the conduction band, electron states that show six equivalent (degenerate) minimum 
energies, associated  with  crystal momenta (${\bf k}$) that are 0.85 of the way to the Brillouin-zone 
boundary; these six states are termed ``valleys'' \cite{cohe88,card01,phil62,schf97,ando82}. 
In nanoscale devices the degeneracy of the valleys is 
broken by various effects, including strain, confinement effects (such as lattice mismatch and/or abruptness 
of the interface between the nano-feature and the confining material) and electric-field effects. Due to 
strain in Si/SiGe quantum wells and (interfacial) quantum dots (in particular, in heterostructure 
semiconductors) \cite{schf97}, and higher subband quantization energy in MOS devices \cite{ando82}, the 
energies of the (four) in-plane valleys are raised, resulting in a remaining double (two-fold) degeneracy (in 
the direction normal to the dot’s plane), which itself has been shown \cite{copp13} to be lifted by electronic 
confinement due to electric field and the effects of the QD boundary structure (including interfacial disorder
and/or steps for Si/SiGe quantum dots). 

As remarked at the end of the introductory section, to characterize and classify the VDOF of the two remaining 
valleys, we introduce in this paper an isospin designation that is constructed in analogy [including the SU(2) 
algebra generated by the $i$-multiplied Pauli matrices] with that of the regular spin of the electrons 
(replacing $\hbs$ with $\hbv$ when referring to the VDOF). Obviously, in the absence of two-body interactions, 
the occupation of the single-particle energy states \footnote{
These confinement-induced single-particle energy states are also referred to as ``orbitals'', see, e.g.,
the expressions ``atomic orbitals'', ``space orbitals'', and ``spin-orbitals'' in chemistry \cite{szabo}}
in the dots would depend on an interplay between the confinement (including an applied magnetic field) and the 
valley effects. For the case of valley degeneracy or near-degeneracy (determined by the intervalley-splitting, 
$E_V$, also referred to as valley gap), and with a confinement gap 
(that is the energy spacing between successive confinement–induced 
single-particle states) which is much larger than $E_V$, this interplay results in ``doubling'' of the spectrum
(two near-degenerate states for each confinement state, each corresponding to a different valley). 

Moving next to the studying of the many-body states in Si QDs, we start by considering a many-body reference
Hamiltonian, $H_{\rm MB}$, which includes the confinement potential defining the quantum dots, applied magnetic 
fields, and the interelectron Coulomb  potential, assuming the case of a full valley degeneracy; see Sec.\ 
\ref{sec:mbh}. The valley splitting is then included by adding a one-body Hamiltonian term $H_{\rm VS}$; see 
Eq.\ (\ref{hvg}) in Sec.\ \ref{sec:sic}. Furthermore, from among others, we consider in this paper two other 
one-body interaction terms that are of particular interest (for details, see Sec.\ \ref{sec:sic}). 
Namely, we consider a spin- and isospin-dependent coupling, $H_{\rm SIC}$, that consists of 
two contributions: (i) A contribution that acts only within the isospin (valley) degree of freedom and mixes 
the valleys, but not the real spins; this term is referred to as the valley-orbit coupling, 
$H_{\rm VOC}$, in analogy with the real spin-orbit interaction and (ii) A contribution, $H_{\rm SVOC}$, that 
couples simultaneously both the (real) spin and isospin degrees of freedom, termed as the spin-valley-orbit 
coupling, or simply the spin-valley coupling. 

The VFCI calculations for the total energies discussed in Secs.\ \ref{sec:same} and 
\ref{sec:diff} were carried out in the Hilbert-space sector specified by the total-spin projection $S_z=0$.  
This is sufficient for the purpose of these sections, because the total energies of the reference many-body 
Hamiltonian, $H_{\rm MB}$, as well as its extensions $H_{\rm MB} + H_{\rm VS}$ and 
$H_{\rm MB} + H_{\rm VS} + H_{\rm VOC}$ that include the valley splitting (VS) and/or the pure valley-orbit 
coupling (VOC), do not depend on the value $\pm 1$ or 0 of the total-spin projection. In cases when VFCI 
eigenstates with $S_z=\pm 1$ values need to be considered, e.g., for counting the degeneracy of the states 
participating in a given multiplet (see Sec.\ \ref{sec:count} below), an explicit mention of the VFCI results 
will be made without showing the corresponding energy spectra. On the other hand, no restriction on the 
total-spin projection $S_z$ (and on the total-isospin projection $V_z$ as well) is placed in Sec.\ 
\ref{sec:mfs} where the full spin-isospin coupling, $H_{\rm SIC}$, which flips both the valley-isospin and 
regular-spin indices, is taken into consideration.   

\subsection{First-excited state with both electrons in the same valley}
\label{sec:same}

\subsubsection{Low-energy spectra}
\label{sec:same_spec}

The VFCI lowest energy spectrum in the transition region between the (1,1) and the (2,0) charge 
configurations, as a function of the detuning, $\ve$, is displayed in Fig.\ \ref{sp046550}. A weaker 
individual-dot confinement of $\hbar \omega_0 = 0.40$ meV was employed for both dots, along with an interdot 
separation of $d=150$ nm \footnote{
The dots in all instances in this paper are  placed at equal distances from the origin.}.
The valley gap was assumed to be $E_V=100$ $\mu$eV $=24.180 h$ GHz, whereas the 
in-plane Si effective mass was taken as 0.191$m_e$ and 
the dielectric constant of Si as $\kappa=11.4$. A high interdot 
barrier was implemented in Fig.\ \ref{sp046550}(a) by setting the input interdot-barrier parameter to 
$\epsilon_1^{b,{\rm inp}}=0.65$ (see Sec.\ \ref{sec:mbh} for the meaning of this input barrier-controlling 
parameter). A low interdot barrier was implemented in Fig.\ \ref{sp046550}(b) by setting 
$\epsilon_1^{b,{\rm inp}}=0.50$. [For the definition and an illustration of the two-center-oscillator 
(TCO) two-well confinement employed in this paper, see Eq.\ (\ref{vtco}) and Fig.\ \ref{tco3d} 
in Sec.\ \ref{sec:mbh}, respectively.]

The symbols \circlep, \circlez, and \circlem ~indicate states with both electrons in the low-energy valley, 
with the electrons in different valleys, and with both electrons in the high-energy valley, respectively. 
The spin-isospin coupling was neglected, and as a result all states are associated with good total 
spin ($\cs$, $S_z$) and valley isospin ($\cv$, $V_z$) quantum numbers, 
as indicated via the symbols made from a combination of the capital letters S and T. 
These symbols have the following meaning: A capital S denotes a 
singlet state, whereas a capital T denotes a triplet state \footnote{
A spin singlet has eigenvalues $\cs(\cs+1)=0$ and $S_z=0$, whereas a spin triplet has eigenvalues 
$\cs(\cs+1)=2$ and $S_z=0$ or $S_z=\pm 1$. Correspondingly, an isospin singlet has eigenvalues $\cv(\cv+1)=0$
and $V_z=0$, whereas an isospin triplet has eigenvalues $\cv(\cv+1)=2$ and $V_z=0$ or $V_z=\pm 1$.}.
The subscripts $\pm$ denotes an $S_z=\pm1 $ or $V_z=\pm 1$ projection, respectively, whereas a subscript $0$ 
denotes an $S_z=0$ or $V_z=0$ projection. A superscript ``s'' denotes the regular spin, whereas a superscipt 
``v'' denotes the valley isospin. The symbol ``$xn$'' (with $n=1$, 2, or 4) denotes the degeneracy associated 
with a given energy curve; red color is used for the \circlep ~states (with $V_z=+1$), green color for the 
\circlez ~states (with $V_z=0$), and blue color for the \circlem ~ones (with $V_z=-1$). 

The energy difference at $\ve=1.6$ meV between the (2,0) states indicated as $e$ [first-excited state, no. 2 
in Fig.\ \ref{sp046550}(b)] and $g$ [ground state, no. 1 in Fig.\ \ref{sp046550}(b)] equals 
$E_{\rm ST}^{\mbox \circleps}$. The energy difference at $\ve=1.6$ between the (2,0) lines indicated in Fig.\
\ref{sp046550}(b) as nos. 3,4 (a doubly degenerate pair) and the line no. 1 (ground state) equals $E_V$. 
$E_V$ equals also the energy difference between the middle and any outer of the three (1,1) parallel lines. 
It is clear that $E_{\rm ST}^{\mbox \circleps} < E_V$ for the case illustrated in this section.  

As is apparent from Fig.\ \ref{sp046550}, the VFCI solutions are able to capture both the (1,1) and (2,0)
charge configurations, whether ground or excited states, and their interconversion as a function of
the detuning. Specifically, the 5 quasi-parallel and quasi-horizontal lines correspond to the (2,0) charge 
configuration (as indicated), whereas the 3 quasi-parallel lines with a large slope correspond to the (1,1) 
charge configuration (again, as indicated). Of particular interest are the avoided crossings (marked by 
double-head red arrows) between two spin singlet states in the transition from the (1,1) to the (2,0) 
configuration. These crossings are underdeveloped in the case of a high interdot barrier [see Fig.\ 
\ref{sp046550}(a)], but become pronounced for low interdot barriers [see Fig.\ \ref{sp046550}(b)]. The 
single-head black arrows indicate avoided crossings between two spin triplet states in the transition from 
the (1,1) to the (2,0) configuration, which however remain underdeveloped in both the cases of a high 
interdot barrier [see Fig.\ \ref{sp046550}(a)], as well as a low interdot barrier [see Fig.\ 
\ref{sp046550}(b)]. 

A main conclusion form the VFCI results in Fig.\ \ref{sp046550} is that, for a given $V_z$, the energy gaps 
between the spin singlet and and spin triplet states in the (2,0) configuration are drastically suppressed 
compared to the orbital (single-particle) gap of $\hbar \omega_0 = 0.40$ meV $=96.719h$ GHz associated with the 
non-interacting limit; see, e.g., the spin-singlet$-$spin-triplet gap, $E_{\rm ST}^{\mbox \circleps}$, at 
$\ve=1.60$ meV between the ground and first-excited states (denoted as $\igs$ and $\ies$, 
respectively), which is $E_{\rm ST}^{\mbox \circleps} \approx 4.05h$ GHz in Fig.\ \ref{sp046550}(a) and 
$E_{\rm ST}^{\mbox \circleps} \approx 6.09h$ GHz in Fig.\ \ref{sp046550}(b). This quenching of the gaps, which 
recently was observed experimentally in Si \cite{corr21,dods22} (but also in GaAs \cite{kim21}) DQD qubits,
is the result of strong-electron correlations and of the formation of Wigner molecules. Namely, the ensuing 
spatial localization of the electrons within the left or right QD reduces the Coulomb repulsion between them, 
a process that leads to the convergence of the energies between the states with symmetric and antisymmetric 
space parts. For two electrons, this process mimicks the dissociation of the natural H$_2$ molecule, and it 
was discovered earlier in the case of GaAs quantum dots 
\cite{yann01,yann02.1,yann02.2,yann06,yann06.2,yann07}.    


\begin{figure*}[t]
\centering\includegraphics[width=15.0cm]{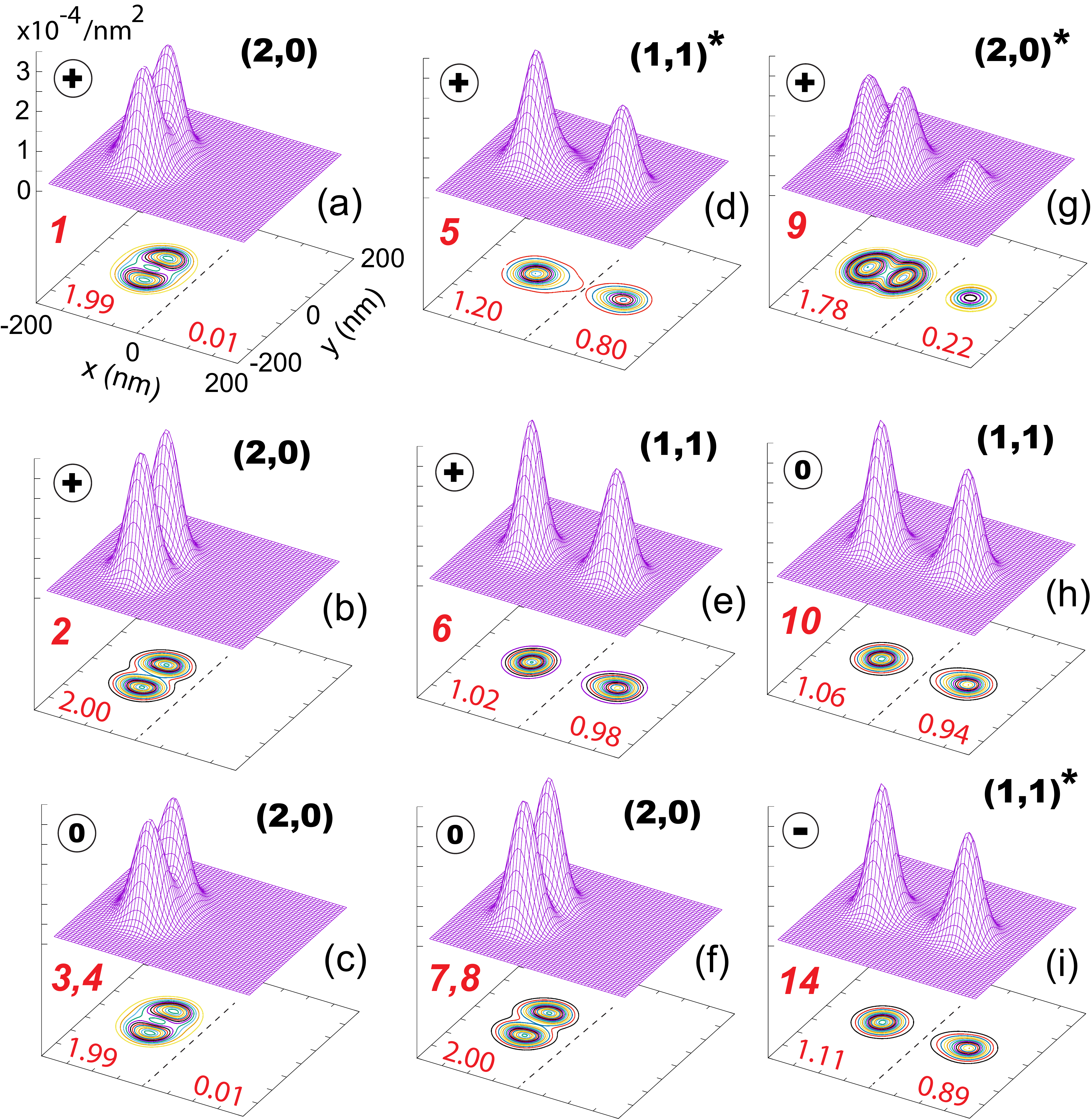}
\caption{
Charge densities associated with the VFCI states whose energies are plotted in Fig.\ \ref{sp046550}(b),
at $\ve=1.60$ meV for curves nos. 1-9, at $\ve=1.56$ meV for curve no. 10, and
at $\ve=1.46$ meV for curve no. 14. This is the case with a low interdot barrier that was implemented 
by setting $\epsilon_1^{b,{\rm inp}}=0.50$ (see Sec.\ \ref{sec:mbh} for the meaning of the input
barrier-controlling parameter).  The displayed red decimal numbers in each panel are the VFCI calculated 
electron occupancies in the left and right wells of the DQD. A {\Large ``*''} denotes a state participating
in an avoided crossing, to a lesser or greater extent.
}
\label{dens0450}
\end{figure*}

\subsubsection{Charge densities}
\label{sec:samedens}

The formation of WMs (in the case of asymmetric confinements) is graphically illustrated through the charge 
densities, which are displayed in Fig.\ \ref{dens0450}. For the reader's convenience and for helping with 
establishing the correspondence between the charge densities in Fig.\ \ref{dens0450} and the states whose 
energies are plotted in Fig.\ \ref{sp046550}, we assigned to the VFCI energy curves the numbers displayed at 
the border of Fig.\ \ref{sp046550}(b). These numbers are used in this section below. 

To facilitate the identification and elucidation of the main trends, we display, along with the charge 
densities, the VFCI calculated electron occupancies (red lettering) in the left and right wells of the DQD 
(rounded to the second decimal point). [These VFCI occupations are further rounded to the closest integer
in order to obtain the $n_L$'s and $n_R$'s $(n_L + n_R = 2)$ used in the notation $(n_L, n_R)$ \cite{Note2}
for the charge configurations.] Naturally, the charge densities are normalized to the total number of 
electrons, $N = 2$.

The ground-state at $\ve=1.60$ meV (curve no. 1) with a symmetric space part and both electrons in the 
low-energy valley (note the symbol \circlep ~in Fig.\ \ref{dens0450}) exhibits a rather well developed WM 
inside the left well, which is aligned parallel to the $y$-axis. Curve no. 2 with an antisymmetric space 
part and both electrons in the low-energy valley exhibits an even better developed WM due to its nodal 
structure; again this WM resides within the left QD and is aligned parallel to the $y$-axis.

Promoting one electron to the high-energy valley [states marked as \circlez] leaves the charge densities
unaltered. Indeed, the density in Fig.\ \ref{dens0450}(a) coincides with the densities in Fig.\ 
\ref{dens0450}(c) (curves nos. 3 and 4), and the density in Fig.\ \ref{dens0450}(b) coincide with those
in Fig.\ \ref{dens0450}(f) (curves nos. 7 and 8).    

State no. 9 at $\ve=1.60$ meV with both electrons in the low-energy valley deviates from a pure (2,0) 
configuration, as is apparent from the density in Fig.\ \ref{dens0450}(g). Indeed, at this point, state
no. 9 starts forming an avoided crossing with state no. 5 [an (1,1) state with the same spin-isospin quantum 
numbers (T$_0^s$T$_+^v$)], and thus it becomes a superposition of both a (2,0) and an (1,1) configuration;
this behavior is denoted with a {\Large ``*''} as a superscript. Note that the (2,0) component of state no. 9 
is associated with a 2e-WM aligned along the $x$-axis. Naturally, state no. 5 is also a superposition,
although weaker, of both a (2,0) and an (1,1) configuration and is denoted with a {\Large ``*''} as a 
superscript. Another state that belongs in this category is state no. 14. 

The two remaining panels [i.e., Figs.\ \ref{dens0450}(e,h)], corresponding to curves no. 6 at $\ve=1.60$ meV, 
and to the curve no. 10 at $\ve=1.56$ meV, respectively, display densities associated with an (1,1) 
configuration. Similar (1,1) densities (not shown) are also associated with curves nos. 11, 12, 13.

\subsubsection{Counting of states and the degeneracy of multiplets}
\label{sec:count}

\begin{figure*}[t]
\centering\includegraphics[width=17.5cm]{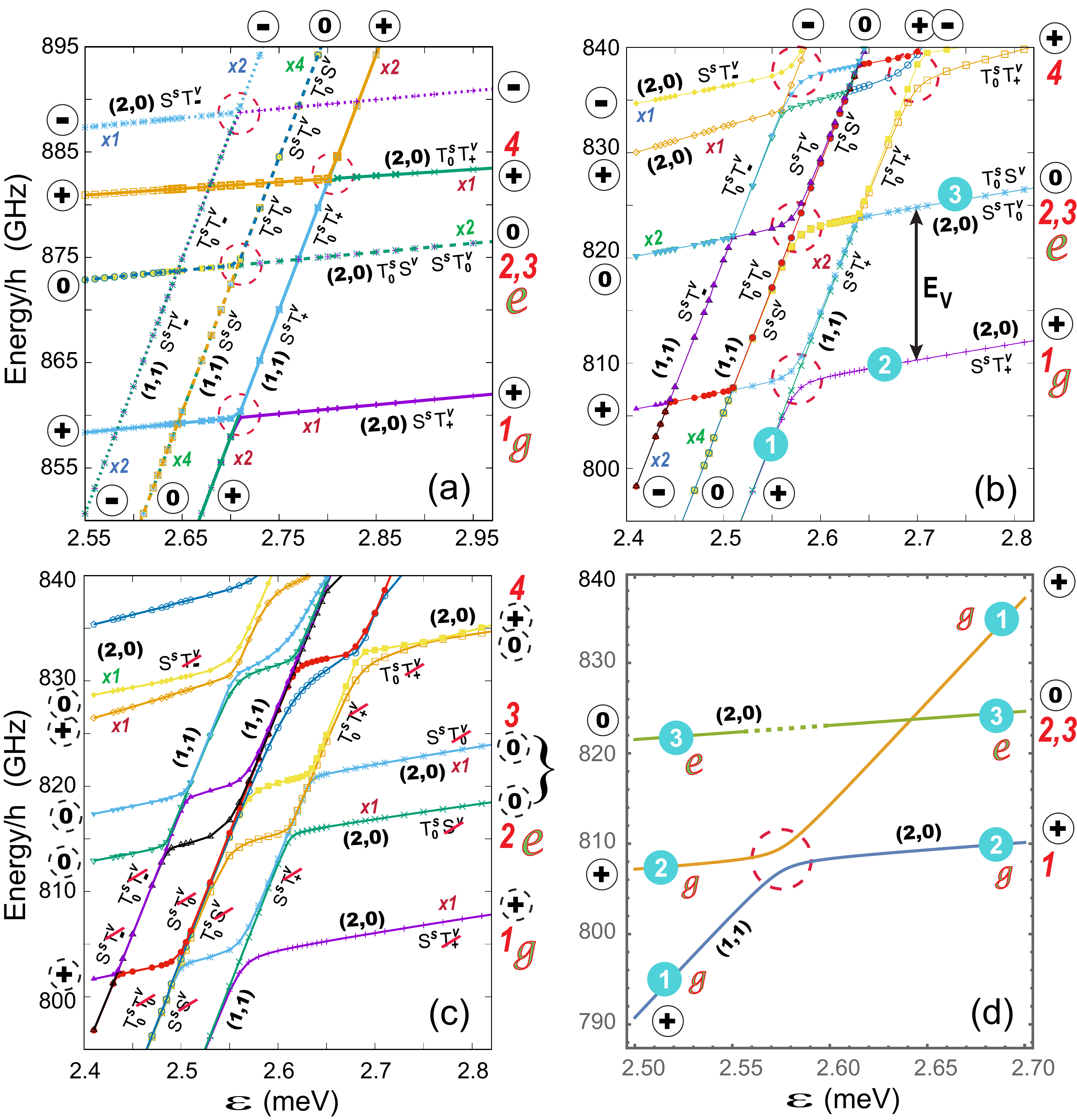}
\caption{
The case with $E_{\rm ST}^{\mbox \circleps} > E_V$.
VFCI lowest energy spectrum for the case of a Si 2e-DQD with a stronger individual-dot confinement,
$\hbar \omega_0 = 0.80$ meV, for both dots, at a vanishing magnetic field, $B=0$.
The spectrum is plotted in the transition region between the (1,1) and the
(2,0) charge configurations, as a function of the detuning, $\ve$, and for a total-spin projection
$S_z=0$. The interdot separation was taken as $d=150$ nm and the valley gap as $E_V=60$ $\mu$eV $=14.508h$
GHz. {\it The first excited state in the (2,0) configuration (state no. 2) is characterized by the symbol
{\rm \circlez}, that is, it has one electron in the low-energy valley and the second electron in the 
higher-energy valley [g and e denote the ground and first-excited states in the (2,0) configuration, 
respectively]}.
(a) With a Hamiltonian $H_{\rm MB}+H_{\rm VS}$, a high interdot barrier was implemented by setting 
$\epsilon_1^{b,{\rm inp}}=0.65$ (see the text for the meaning of this input barrier-controlling parameter). 
(b) With a Hamiltonian $H_{\rm MB}+H_{\rm VS}$, a low interdot barrier was implemented by setting
$\epsilon_1^{b,{\rm inp}}=0.35$. 
(c) Same as in (b), but with a $H_{\rm VOC}$ term (pure valley-orbit coupling, see text) with $\Delta=0.05$ 
meV and $\phi_0=0$ added in the Hamiltonian. The red strikethrough bars and the dashed circles indicate that 
the valley isospin in this panel does not possess good quantum numbers due to the intervalley mixing.
(d) The three lines labeled 1, 2, and 3 on a blue disk in panel (b) reproduced according to the toy 
effective Hamiltonian (\ref{heff}). For an analysis of the trends in the spectra of this figure, see the 
text. The symbols \circlep, {\rm \circlez}, and \circlem ~have the same meaning as in Fig.\ \ref{sp046550}.
The in-plane Si effective mass in panels (a), (b), and (c) was taken as 0.191$m_e$ and the dielectric 
constant as $\kappa=11.4$. 
}
\label{sp080065}
\end{figure*}

{\it The case of the (1,1) charge configuration.\/}
For $S_z=0$, the valleytronic FCI produces a group of 8 states with an (1,1) configuration (shown in Figs.\ 
\ref{sp046550} and \ref{sp080065}) that are grouped as 2 \circlep$-$4 \circlez$-$2 \circlem. 
Likewise, for $S_z=+1$, the valleytronic FCI produces a group of 4 states (not shown) with an (1,1) 
configuration grouped as 1 \circlep$-$2 \circlez$-$1 \circlem. Finally, for $S_z=-1$, the valleytronic FCI 
produces another group of 4 states (not shown) with an (1,1) configuration grouped again as 
1 \circlep$-$2 \circlez$-$1 \circlem. In total, one obtains 16 states that are grouped in multiplets as 4 
\circlep$-$8 \circlez$-$4 \circlem. 

The number of 16 states is the hallmark of a fully developed SU(4) symmetry that would be achieved in the 
2e-VFCI in the absence of any spin-isospin coupling (i.e., neglecting the $H_{\rm VOC}$ and/or $H_{\rm SIC}$
terms) and for the case of a vanishing valley gap ($E_V=0$). 
These sixteen states are the product of the four spin states (one singlet, S$^s$, and three triplets, 
T$^s_\pm$ and T$^s_0$) and the four valley isospin states (one singlet, S$^v$, and three triplets, 
T$^v_\pm$ and T$^v_0$). Their quantum numbers are explicitly given as follows: there are 6 antisymmetric 
combinations S$^s$T$^v_+$, S$^s$T$^v_0$, S$^s$T$^v_-$, T$^s_+$S$^v$, T$^s_0$S$^v$, T$^s_-$S$^v$ and 10 
symmetric combinations S$^s$S$^v$, T$^s_+$T$^v_+$, T$^s_+$T$^v_0$, T$^s_+$T$^v_-$, T$^s_0$T$^v_+$, 
T$^s_0$T$^v_0$, T$^s_0$T$^v_-$, T$^s_-$T$^v_+$, T$^s_-$T$^v_0$, T$^s_-$T$^v_-$.

In the case of a single elliptic dot, these 16 SU(4)-states, that form the lowest-energy part of the
spectrum, organize in two multiplets \footnote{
We have indeed checked the accuracy of this statement for a single elliptic Si QD using the VFCI code.} 
in analogy with the case of an SU(4) Heisenberg lattice dimer (see (a) in Ref.\ \cite{shi98} and 
Ref.\ \cite{sark17}), six of them in one multiplet with a symmetric space part and the remaining 
ten in a second higher-energy multiplet with an antisymmetric space part \footnote{
An additional multiplet of 10 states with an antisymmetric space part appears also in a single elliptic 
Si QD at even higher energy. Indeed, for a circular dot, there are two degenerate antisymmetric space 
wave functions (associated with the two angular momenta $L=\pm 1$) and thus the second higher-energy 
multiplet consists of 20 states. Including the 6-member lower-energy multiplet, this results in 26 SU(4) 
states. For an elliptic dot, the circular symmetry of the space wave functions is lifted and the multiplet 
of 20 states splits in two 10-state multiplets. Although overlooked, and even misinterpreted in earlier CI 
calculations \cite{corr21}, this underlying SU(4) $\supset$ SU(2) $\times$ SU(2) organization
of the spectrum of a single elliptic Si quantum dot is operational in other variants of CI calculations
as well, as one can attest by a careful examination of the associated CI results. Indeed in Fig.\ 3         
of Ref.\ \cite{corr21.2}, panel (b) contains 16 states and panel (c) contains 10 states, for a total of
26 states.}.  

In the case of two well separated QDs in the strict SU(4) limit, the energy gap between space-symmetric and 
space-antisymmetric multiplets vanishes due to a vanishing left-right spatial overlap, and this yields a total 
of 16 degenerate SU(4)-states in the (1,1) configuration. However, in Si DQDs, this 
degeneracy is lifted due to the independent emergence of a valley gap that results from finite-size effects, 
and the SU(4) symmetry is lowered to an SU(2) $\times$ SU(2) one, characterized by the 
4 \circlep-8 \circlez-4 \circlem ~multiplet organization discussed above. 
       
{\it The case of the (2,0) and (0,2) charge configurations.\/}
From Fig.\ \ref{sp046550}, it can easily be seen that, according to the VFCI results, the hallmark number of 
the 16 states of the SU(4) $\supset$  SU(2) $\times$ SU(2) chain is preserved in the (2,0) configuration; 
simply the exhange gap $J$ between the space-symmetric and space-antisymmetric states acquires a non-vanishing 
finite value. For example, as seen in Fig.\ \ref{sp046550}(a), the two degenerate \circlep ~states, 
S$^s$T$^v_+$ and T$^s_0$T$^v_+$, in the (1,1) configuration transition into the $g$ and $e$ states in the (2,0)
configuration, exhibiting a gap of $4.05h$ GHz at $\ve=1.60$ meV. A similar transition applies also in the case
of the four degenerate (1,1) states in the \circlez ~manifold, which splits into two doubly degenerate
manifolds in the (2,0) configuration. Finally, a transition of the doubly degenerate \circlem ~manifold of
the (1,1) configuration to two non-degenerate states in the (2,0) configuration is not shown in
Fig.\ \ref{sp046550}(a), but it was observed in our extensive VFCI computational results.

Our verification in the VFCI spectra of the presence of the 16 hallmark states associated 
with the SU(4) $\supset$  SU(2) $\times$ SU(2) chain contrasts with the counting from a Hubbard two-site
modeling \cite{burk12} of a Si 2e-DQD that incorporates the VDOF. Indeed, instead of the expected number
of 16 states, the model in Ref.\ \cite{burk12} allows only for 6 states in the (2,0) configuration. This
incomplete conclusion follows directly from the assumption that each Hubbard site has one level only,
an assumption that does not allow the construction of 2e antisymmetric space wave functions in the case
of the (2,0) configuration.

We note that adding a spin-isospin coupling term (including one or both contributions, see Sec.\ \ref{sec:sic}) 
in the many-body Hamiltonian will lift the degeneracies illustrated in the spectra of Fig.\ \ref{sp046550}, 
however, the overall organization of such Si-DQD spectra will be traceable back to that in Fig.\ 
\ref{sp046550}, as long as the strength of the spin-isospin coupling is not extreme.  
          
\subsection{First-excited state with electrons in different valleys}
\label{sec:diff}

We turn now to the case when the first-excited state in the (2,0) configuration has one electron in the 
low-energy valley and the second electron in the higher-energy valley, i.e., it exhibits a valley
isospin projection $V_z=0$ (denoted as \circlez).
The evolution of the VFCI low-energy spectra in this case is investigated in 
Fig.\ \ref{sp080065}, as a function of the height of the interdot barrier [Fig.\ \ref{sp080065}(a) and
Fig.\ \ref{sp080065}(b)] and in response to the inclusion of the pure intervalley-coupling Hamiltonian term,
$H_{\rm VOC}$ (see Sec.\ \ref{sec:sic}), in the Hamiltonian $H_{\rm MB}+H_{VS}$ [Fig.\ \ref{sp080065}(c)]. 

Specifically, for a total-spin projection $S_z=0$, Fig.\ \ref{sp080065} displays, as a function of the 
detuning, $\ve$, the VFCI low-energy spectrum for a Si 2e-DQD with a stronger individual-dot 
confinement of $\hbar \omega_0 = 0.80$ meV in the transition region between the (1,1) and the (2,0) charge 
configurations. The valley gap was taken as $E_V=60$ $\mu$eV $=14.508h$ GHz. As in the weaker 
individual-dot-confinement case of Fig.\ \ref{sp046550}, the interdot separation was set to $d=150$ nm,
the in-plane Si effective mass was taken as 0.191$m_e$, and the dielectric constant as $\kappa=11.4$.  
A high interdot barrier ($\epsilon_1^{b,{\rm inp}}=0.65$) was chosen for Fig.\ \ref{sp080065}(a), whereas
a low interdot barrier ($\epsilon_1^{b,{\rm inp}}=0.35$) was used for Fig.\ \ref{sp080065}(b). In Fig.\ 
\ref{sp080065}(c), the low interdot barrier with $\epsilon_1^{b,{\rm inp}}=0.35$ was maintained, but as 
aforementioned, a pure intervalley-coupling term (VOC, see Sec.\ \ref{sec:sic}) 
with $\Delta=0.05$ meV and $\phi_0=0$ was added to the many-body Hamiltonian $H_{\rm MB}+H_{VS}$.
The red strikethrough bars over the capital letters with a ``v'' subscript, as well as the dashed circles
over ``$+$'', $0$, and ``$-$'', indicate that the valley isospin in this panel does not possess good quantum 
numbers due to the intervalley mixing. Nevertheless, the number of states remains unaltered and the 
associated topology of the spectrum in Fig.\ \ref{sp080065}(c) can be traced back to that in 
Fig.\ \ref{sp080065}(b).

For $\ve>2.8$ meV, the energy difference between the (2,0) lines indicated as $e$ [doubly degenerate 
first-excited state, nos. 2,3 lines in Fig. \ref{sp080065}(b)] and the line $g$ [ground state, no. 1 in Fig.\
\ref{sp080065}(b)] equals $E_V$. $E_V$ equals also the energy difference between the middle and any outer 
of the three (1,1) parallel lines. The energy difference at $\ve>2.8$ between the (2,0) line indicated in 
Fig. \ref{sp080065}(b) as no. 4 and the line no. 1 (ground state) equals $E_{\rm ST}^{\mbox \circleps}$. 
It is clear that $E_{\rm ST}^{\mbox \circleps} > E_V$ for the case illustrated in this section.

In all three panels, Figs.\ \ref{sp080065}(a), \ref{sp080065}(b), and \ref{sp080065}(c), 
the three quasi-parallel lines with a large slope correspond to the (1,1) charge 
configuration (as indicated), whereas the quasi-horizontal lines  correspond to the (2,0) charge
configuration. In Fig.\ \ref{sp080065}(a), where the intervalley coupling was neglected and a high barrier
was applied, the avoided crossings expected between curves with the same quantum numbers 
(see dashed circles) are very weak. Lowering the interdot barrier, however, while still neglecting 
intervalley coupling, yields the pronounced avoided crossings enclosed in the dashed circles of Fig.\ 
\ref{sp080065}(b). 

Introducing a non-negligible intervalley coupling in Fig.\ \ref{sp080065}(c) has three effects: 
1) the valley isospin is not conserved, 2) the degeneracy of states with the same valley isospin quantum 
numbers is lifted; see the two separate \circlez ~lines (nos. 2 and 3) in  Fig.\ \ref{sp080065}(c) that 
developed out from the doubly degenerate \circlez ~line (marked as ``2,3'') in Fig.\ \ref{sp080065}(b), 
and 3) the pure crossings between curves with different valley quantum numbers in Fig.\ \ref{sp080065}(b) 
transform to avoided crossings; contrast Fig.\ \ref{sp080065}(b) with Fig.\ \ref{sp080065}(c).

As mentioned earlier, experimental reports \cite{corr21,dods22} from the Wisconsin-Madison group indicated
that the first excited state in the (2,0) configuration in a Si/SiGe DQD is the spin-triplet state with both 
electrons in the same lower-energy valley, whereas experimental measurements \cite{pett22,nich21} from other 
groups on Si-DQD devices (with apparently different parameters) indicated that the first excited
energy level in this configuration is associated with a state having each electron in a different valley [see 
states nos. 2 or 3 in Fig.\ \ref{sp080065}(b)]. In particular, using microwave-frequency scanning gate 
microscopy, Ref.\ \cite{pett22} reported a measured spectrum of three lowest-energy states in the detuning 
window covering the transition from the (1,1) to the (2,0) configuration \footnote{
Ref.\ \cite{pett22} proposes a more general framework. The case of the transition from the (1,1) to the (2,0)
configuration happens by setting $N_1=1$ and $N_2=0$ in Fig.\  4(b) of Ref.\ \cite{pett22}.}.
In Fig.\ \ref{sp080065}(b), one can identify a triad of lowest-energy VFCI levels (denoted by nos. 1, 2, and 3
inside a blue disk) that have the same topology as the group of the (1,1) and the two (2,0) states in 
Fig.\  4(b) of Ref.\ \cite{pett22}. To further demonstrate the analogies with the experimental trends, we
note that this triad of energy levels can be isolated from the full VFCI spectrum and that it can be
reproduced [see Fig.\ \ref{sp080065}(d)] by an effective three-level Hamiltonian as follows:
\begin{align}
H_{\rm eff}= \left( 
\begin{array}{ccc}
\alpha_1 \widetilde{\ve}+C & \delta & 0 \\
\delta & \alpha_2 \widetilde{\ve}+C & 0 \\
0           & 0 & \alpha_3 \widetilde{\ve}+C+E_V 
\end{array}
\right),
\label{heff}
\end{align}
where $\widetilde{\ve}=\ve-\ve_0$ with $\ve_0=621.63h$ GHz ($=2.575$ meV), $\alpha_1=0.957$, 
$\alpha_2=\alpha_3=0.065$, $C=808.25h$ GHz, $\delta=1.3h$ GHz, and $E_V=14.51h$ GHz ($=60$ $\mu$eV), i.e., 
the valley splitting used in the VFCI calculation.

The interaction between the (2,0) \circlep ~ground state and the (1,1) \circlep ~ground state
generates a visible avoided crossing, in agreement with the experiment. In contrast,
using the parameters above, the (2,0) \circlez ~first-excited state in Fig.\ \ref{sp080065}(d) does not 
develop any avoided crossing with the (1,1) \circlep ~ground-state curve. This is in remarkable agreement 
with the behavior of the experimental curves, suggesting that the valley-orbit coupling in the experimental
device is either absent or rather weak.

We note that, although the value of the Wigner parameter $R_W=7.07$ in this section is not strong enough 
(compared to $R_W=10.0$ in Sec.\ \ref{sec:same}) to suppress the $E_{\rm ST}^{\mbox \circleps}$ energy below
$E_V$, the value of $E_{\rm ST}^{\mbox \circleps}$ at $\ve=2.8$ is still drastically lower than the orbital 
gap $\hbar \omega_0=0.80$ meV $=193.44h$ GHz, i.e., at $\ve=2.95$ meV, one has 
$E_{\rm ST}^{\mbox \circleps} \approx 21.52h$ GHz in Fig.\ \ref{sp080065}(a) and, at $\ve=2.81$ meV, 
$E_{\rm ST}^{\mbox \circleps} \approx 27.84h$ GHz in Fig.\ \ref{sp080065}(b); for the definition of
the Wigner parameter $R_W$, see Sec.\ \ref{sec:rw}.

\begin{figure*}[t]
\centering\includegraphics[width=17cm]{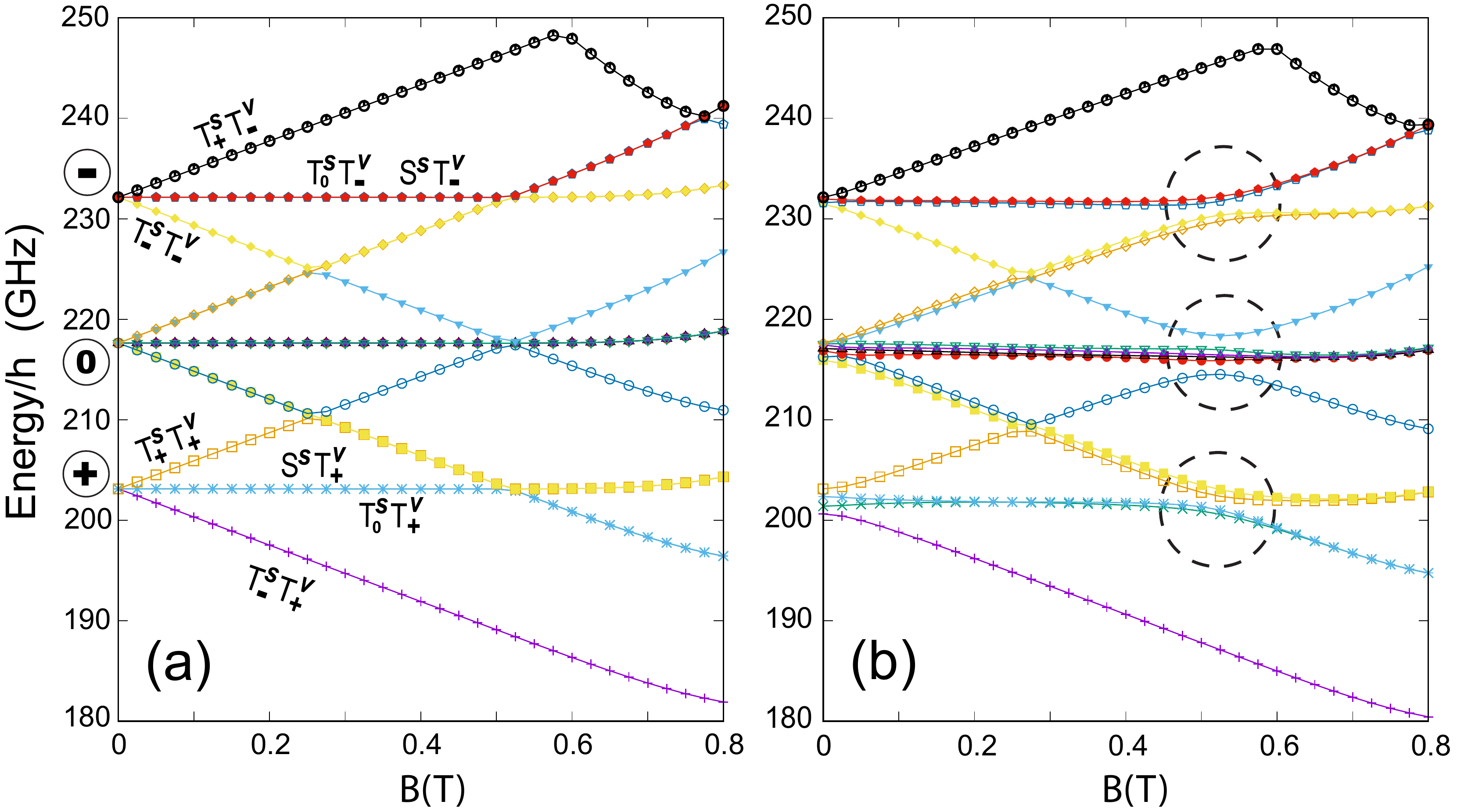}
\caption{
VFCI spectra of the extended Hamiltonian $H_{\rm MB} + H_{\rm VS} + H_{\rm SIC}$ for the 16 low-energy 
states associated with the (1,1) configurations of a singlet-triplet Si 2e-DQD qubit at small detuning as a 
function of a perpendicular magnetic field $B$. $H_{\rm SIC}$ includes both contributions according
to Eq.\ (\ref{hsic}). (a) Case of a very small 
SIC coupling parameter $\Delta=0.00001$ meV. This practically vanishing value does not generate visible 
avoided crossings, but it helps to enforce good spin and isospin quantum numbers by lifting the degeneracies
in the spectrum by imperceptible amounts. For the \circlep ~(both electrons in the lower-energy valley)
and \circlem ~(both electrons in the higher-energy valley) states, these quantum numbers
are indicated in the figure. For the \circlez ~states (one electron in each valley), these quantum numbers
are as follows:  T$^s_+$S$^v$ and T$^s_+$T$^v_0$ for the upper branch; 
S$^s$S$^v$, S$^s$T$^v_0$, T$^s_0$S$^v$, and T$^s_0$T$^v_0$ for the middle branch; 
T$^s_-$S$^v$ and T$^s_-$T$^v_0$ for the lower branch. These quantum numbers correspond to the VFCI results 
for the spin and isospin quantum numbers; for a detailed example, see Table \ref{vfci_si} in Appendix 
\ref{sec:vfci_si}.
(b) Case of a larger SIC parameter $\Delta=0.03$ meV. The generation of avoided crossings is visible in 
the three encircled areas. There is good overall agreement with the experimental results and 
phenomenological analysis in Ref.\ \cite{taha14}. The remaining parameters for the DQD are:
individual-dot confinement $\hbar \omega_0=0.40$ meV (for both dots), $E_V=0.06$ meV, $\ve=0.05$ meV, 
$m^*=0.191\;m_e$, $\epsilon_1^{b,{\rm inp}}=0.50$, and Land\'{e} factor $g^*=2$; see Sec.\
\ref{sec:mbh} for the meaning of the input barrier-controlling parameter $\epsilon_1^{b,{\rm inp}}$.
}
\label{figmf}
\end{figure*}

\subsection{Magnetic-field spectra}
\label{sec:mfs}

To further illustrate the capabilities of the present VFCI, we investigate in this section the dependence on 
the magnetic field, $B$, of the spectra of a Si 2e-DQD in the (1,1) charge configuration. For the example 
case here, we use an individual-dot confinement $\hbar \omega_0=0.40$ meV, an interdot separation $d=150$ nm, 
and a valley splitting $E_V =0.06$ meV = $14.508h$ GHz, as was the case in Sec.\ \ref{sec:same}.
For the detuning, a small value of $\ve=0.05$ meV was used to guarantee that the DQD remains in the (1,1)
charge configuration. The Land\'{e} factor was taken as $g^*=2$, appropriate for silicon.

Fig.\ \ref{figmf} displays, as a function of $B$, the spectra associated with the 16 low-energy states 
of the 2e-DQD specified in the previous paragraph. We stress again that the number 16 is a hallmark of the 
underlying SU(4) $\supset$ SU(2) $\times$ SU(2) symmetry-group chain, as discussed in Sec.\ \ref{sec:count}.
In particular, for comparison, Fig.\ \ref{figmf}(a) displays the 16-state spectrum in the absence of any 
spin-isospin coupling. Actually, considering both terms in Eq.\ (\ref{hsic}), we implemented such a SIC 
coupling with a very small strength $\Delta=0.00001$ meV. This small value does not generate visible avoided 
crossings, but it helps to enforce good spin and isospin quantum numbers by lifting the degeneracies in the 
spectrum by an imperceptible amount.

From Fig.\ \ref{figmf}(a),
it is seen that the three original multiplets at $B=0$ (grouped as 4 \circlep$-$8 \circlez$-$
4 \circlem) break down and fan out with increasing $B$. Indeed, the states with $S_z=0$ run parallel to 
the $B$-axis, whereas states with $S_z=1$ exhibit an ascending sloping and states with $S_z=-1$ exhibit a 
descending sloping. No avoided crossings are visible in Fig.\  \ref{figmf}(a), and the energy lines can be
characterized by good spin and valley-isospin quantum numbers [as indicated in Fig.\ \ref{figmf}(a)].

We note that consideration of the full spin-isospin coupling requires the enlargement of the Hilbert space 
defined by the basis Slater determinants (see Sec.\ \ref{sec:meth}), i.e., Slater determinants preserving 
individually all three values (0, $\pm 1$) of the total spin projection, $S_z$, and isospin projection,
$V_z$, must be included in the basis, and this was done for the calculations in both Figs.\ \ref{figmf}(a) and
\ref{figmf}(b). 

In contrast to Fig.\ \ref{figmf}(a), the results displayed in Fig.\ \ref{figmf}(b) correspond to an 
$H_{\rm SIC}$ term with a rather large strength, $\Delta = 0.03$ meV. 
The spin and valley isospin do not have good quantum numbers anymore, but the number
of states remains unaltered and the associated topology can be traced back to that in Fig.\ \ref{figmf}(a).
On the other hand, well visible avoided crossings develop in three spots (highlighted within circles)
whenever the valley gap $E_V$ equals the Zeeman energy $E_Z = g^* \mu_B B$ ($\mu_B = 5.788383\;10^{-5}$ 
eV/Tesla is the Bohr magneton). 

Our VFCI results for the case of a Si 2e-DQD portrayed in Fig.\ \ref{figmf}(b) are in 
agreement with the experimental results and the phenomenological analysis of Ref.\ \cite{taha14};
see, e.g., Fig.\ 5(b) therein. We note further that such an avoided crossing associated with the 
condition $E_V=E_Z$ has been observed experimentally for other Si nanostructures, e.g., in the case of a
single QD \cite{dzur13}. In Ref.\ \cite{dzur17}, the avoided crossing at $E_V=E_Z$ was produced in a Si 
2e-DQD by keeping the magnetic field constant while varying the valley splitting as a result of the
application of a changing gate voltage.   

\section{Methods}
\label{sec:meth}

In this section, we present the mathematics for the VFCI formalism. In addition, we give the 
definition for the Wigner parameter and the variants of the many-body Hamiltonian used.  

\subsection{Wigner parameter}
\label{sec:rw}

At zero magnetic field and in the case of a single circular harmonic QD, the degree of electron 
localization and Wigner-molecule pattern formation can be associated with the so-called Wigner parameter 
\cite{yann99,yann07,erca21.2,yann22.2},
\begin{align}
R_W = Q/(\hbar\omega_0),
\label{rw}
\end{align}
where $Q$ is the Coulomb interaction strength and $\hbar \omega_0$ is the energy quantum of the harmonic 
potential confinement (being proportional to the one-particle kinetic energy); $Q = e^2/(\kappa l_0 )$,
with $l_0 = ( \hbar/(m^* \omega_0) )^{1/2}$ the spatial extension of the lowest state’s wave function in 
the harmonic (parabolic) confinement. Naturally, experimental signatures for the formation of Wigner 
molecules are expected for values $R_W > 1$, with the WM pattern being more robust the larger the 
value of $R_W$.

As mentioned in the last paragraph of Sec.\ \ref{sec:diff}, the values of $R_W$ corresponding to the Si DQDs 
studied in this paper are $R_W=10.0$ when $\hbar \omega_0=0.40$ meV and $R_W=7.07$ when 
$\hbar \omega_0=0.80$ meV.

\subsection{The reference many-body Hamiltonian}
\label{sec:mbh}

We consider $N$ electrons in a double quantum dot under a low magnetic field $(B)$ (including the
case of a vanishing magnetic field). The corresponding many-body Hamiltonian, 
\begin{equation}
H_{\rm MB} = \sum_{i=1}^N H_{\rm TCO}(i) + 
\sum_{i=1}^{N} \sum_{j>i}^{N} V({\bf r}_i,{\bf r}_j),
\label{mbhd}
\end{equation}
is the sum of a single-particle part $H_{\rm TCO}(i)$ and the two-particle interaction 
$V({\bf r}_i,{\bf r}_j)$. 

Naturally, for the case of electrons, the two-body interaction is given by the Coulomb repulsion,
\begin{equation}
V({\bf r}_i,{\bf r}_j)=\frac{e^2}{\kappa |{\bf r}_i-{\bf r}_j|},
\label{tbie}
\end{equation}
where $\kappa$ is the dielectric constant of the semiconductor material ($\kappa=11.4$ for Si). 

The single-particle Hamiltonian is given by
\begin{equation}
H_{\rm TCO}=T+ V_{\rm TCO}(x,y) + g^* \mu_B B\sigma,
\label{hsp}
\end{equation}
where we dropped the particle index $i$. The last term in Eq.\ (\ref{hsp}) is the Zeeman interaction, 
with $g^*$ being the effective Land\'{e} factor ($g^*=2$ for Si), $\mu_B$ the Bohr magneton, $B$ the 
perpendicular magnetic field, and $\sigma=\pm 1/2$ the spin projection of an individual electron. 

The kinetic contribution in Eq.\ (\ref{hsp}) is given by 
\begin{equation}
T=\frac{[{\bf p} - (e/c) {\bf A}({\bf r})]^2}{2m^*}, 
\label{hkin}
\end{equation}
with $m^*$ being the in-plane effective mass (0.191$m_e$ for Si) and   
the vector potential ${\bf A} ({\bf r}) = 0.5(-By\hat{\imath}+Bx\hat{\jmath})$
being taken according to the symmetric gauge, where ${\bf r} =x \hat{\imath} + y \hat{\jmath}$. 

\begin{figure}[t]
\centering\includegraphics[width=7.5cm]{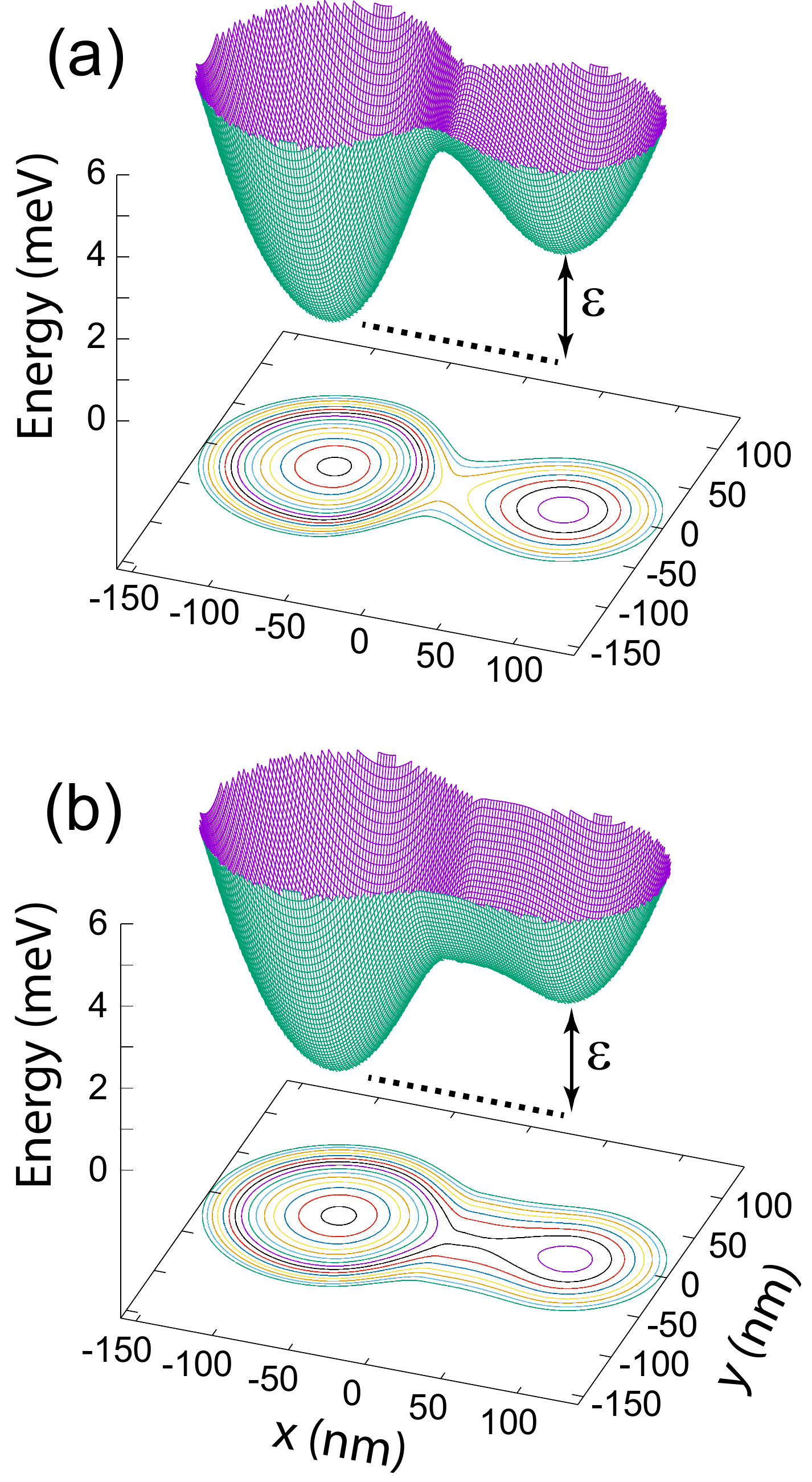}
\caption{
The TCO external confining potential, $V_{\rm TCO}(x,y)$, including the smooth neck. (a) A high interdot 
barrier corresponding to a barrier control parameter of $\epsilon_1^{b,{\rm inp}}=0.65$. (b) A lower 
interdot barrier corresponding to a barrier control parameter of $\epsilon_1^{b,{\rm inp}}=0.45$. The 
remaining parameters are: $\hbar\omega_{x1}=\hbar\omega_{x2}=\hbar\omega_y=\hbar \omega_0=0.80$ meV, 
interdot separation $d=150$ nm (with $-x_1=x_2=75$ nm), effective mass $m^*=0.191m_e$ (appropriate for 
Si), and detuning parameter $\ve=2.71$ meV.}
\label{tco3d}
\end{figure}

The external confining potential is denoted as $V_{\rm TCO} (x,y)$. 
For the two-dimensional DQDs considered in this paper, the confining potential is determined by the 
following two-center-oscillator expression:
\cite{yann99,yann02.1,yann02.2,yann09,yann22,yann22.2}
\begin{equation}
V_{\rm TCO} = \frac{1}{2} m^* \omega^2_y y^2
    + \frac{1}{2} m^* \omega^2_{x k} x^{\prime 2}_k + V_{\rm neck}(x) +h_k,
\label{vtco}
\end{equation}
where $x_k^\prime=x-x_k$ with $k=1$ for $x<0$ (left) and $k=2$ for 
$x>0$ (right), and the $h_k$'s control the relative depth of the two wells, with the detuning
defined as $\ve=h_2-h_1$. $y$ denotes the coordinate perpendicular to the
interdot axis ($x$). A notable property of $V_{\rm TCO}$ is the fact that it allows for the formation
of a smooth interwell barrier between the individual wells whose height $V_b$ can be varied independently
from the interdot distance $d=x_2-x_1$; see Fig.\ \ref{tco3d} for an illustration.
The most general shapes described by $V_{\rm TCO}$ are two semiellipses connected by the smooth neck, 
$V_{\rm neck}(x)$. $x_1 < 0$ and $x_2 > 0$ are the centers of these semiellipses. In this paper, we
take $\omega_{x1}=\omega_{x2}=\omega_y=\omega_0$ in all instances.

For the smooth neck, we use 
\begin{align}
V_{\rm neck}(x) = \frac{1}{2} m^* \omega^2_{x k} 
\Big[ {\cal C}_k x^{\prime 3}_k + {\cal D}_k x^{\prime 4}_k \Big] \theta(|x|-|x_k|),
\label{vneck}
\end{align}
where $\theta(u)=0$ for $u>0$ and $\theta(u)=1$ for $u<0$.
The four constants ${\cal C}_k$ and ${\cal D}_k$ can be expressed via two
parameters, as follows: ${\cal C}_k= (2-4\epsilon_k^b)/x_k$ and
${\cal D}_k=(1-3\epsilon_k^b)/x_k^2$, where the barrier-control parameters
$\epsilon_k^b=(V_{b}-h_k)/V_{0k}$ are related to the height of the targeted 
interdot barrier ($V_{b}$, measured from the origin of the energy scale), 
and $V_{0k}=m^* \omega_{x k}^2 x_k^2/2$. We note that measured from the bottom of 
the left ($k=1$) or right ($k=2$) well the interdot barrier is $V_{b}-h_k$.

We note that in all calculations in this paper we used nonnegative values of detuning ($\ve \geq 0$),
namely the left well was kept in all instances lower than the right one. In addition, for convenience, we
set $h_1=0$. In this case, it was advantageous to use a modified barrier-control parameter 
$\epsilon_1^{b,{\rm inp}}$ as an input parameter. Specifically, $\epsilon_1^b$ and $\epsilon_1^{b,{\rm inp}}$ 
are related as $\epsilon_1^b = \epsilon_1^{b,{\rm inp}}(V_{02}+\ve)/V_{01}$. 

Neglecting the term $V_{\rm neck}$ for the smooth neck, the eigenstates of $H_{\rm TCO}$ at $B=0$ are used 
to construct the space orbitals $\varphi_j (x,y)$ of the single-particle basis employed in the CI method; 
see Eq.\ (\ref{chi1234}) below. How to solve for the eigenvalues and eigenstates of the ensuing auxiliary 
Hamiltonian,
\begin{equation}
H_{\rm aux}=\frac{{\bf p}^2}{2m^*} + \frac{1}{2} m^* \omega_y^2 y^2
      + \frac{1}{2} m^* \omega_{xk}^2 x_k^{\prime 2}+h_k,
\label{haux}
\end{equation}
is described in Appendix \ref{sec:aux}.

Finally, the Hamiltonian term implementing the spin-isospin coupling is described in Sec.\ \ref{sec:sic}, 
after the introduction in the next section of the $\sigma_q$ and $\tau_q$, $q=x,y,z$, Pauli matrices that 
correspond to the regular spin and to the valley isospin, respectively. 

\subsection{The valleytronic FCI approach}
\label{sec:ci}

As aforementioned, we use the method of configuration interaction for determining the solution of the 
many-body problem specified by the Hamiltonians $H_{\rm MB}+H_{\rm VS}$, $H_{\rm MB}+H_{\rm VS}+H_{\rm VOC}$, 
or $H_{\rm MB}+H_{\rm VS}+H_{\rm SIC}$.

In the CI method, one writes the many-body wave function 
$\Phi^{\rm CI}_N ({\bf r}_1, {\bf r}_2, \ldots , {\bf r}_N)$ as a linear
superposition of Slater determinants 
$\Psi^N({\bf r}_1, {\bf r}_2, \ldots , {\bf r}_N)$ that span the many-body
Hilbert space and are constructed out of the single-particle 
{\it spin-isospin-orbitals\/} \footnote{
This is an apparent generalization of the term spin-orbital used in chemistry and molecular physics
\cite{Note3}.} 
\begin{align}
\begin{split}
& \chi_j (\br) = \varphi_j (x,y) \alpha\zeta, \mbox{~~~~~~if~~~}  ~1\leq j \leq K,\\
& \chi_j (\br) = \varphi_{j-K} (x,y) \beta\zeta, \mbox{~~~if~~~}  ~K < j \leq 2K, \\
& \chi_j (\br) = \varphi_{j-2K} (x,y) \alpha\eta,  \mbox{~~~if~~~} 2K < j \leq 3K, \\ 
& \chi_j (\br) = \varphi_{j-3K} (x,y) \beta\eta, \mbox{~~~if~~~} 3K < j \leq 4K, 
\end{split}
\label{chi1234} 
\end{align}
where $\alpha (\beta)$ denote up (down) spins, $\zeta (\eta)$ denote up (down) isospins [i.e., electrons 
in the first (second) valley], and the spatial orbitals $\varphi_j(x,y)$  are given by the $K$ lowest-energy 
solutions of the auxiliary single-particle Hamiltonian in Eq.\ (\ref{haux}). For clarity and convenience, 
these solutions are sorted in ascending energy.

We note that, in analogy with the case of the Pauli spin matrices 
$\sigma_x$, $\sigma_y$, and $\sigma_z$, three additional Pauli matrices $\tau_x$, $\tau_y$, and $\tau_z$, 
associated with the valley isospin, can be defined, yielding the relations $\tau_x \zeta = \eta$, 
$\tau_x \eta = \zeta$, $\tau_y \zeta =i \eta$, $\tau_y \eta = -i \zeta$, $\tau_z \zeta = \zeta$, and 
$\tau_z \eta = -\eta$. 

Making contact with the effective mass theory (continuum model) 
\cite{kohn55} for semiconductor heterostructures, we identify the $\varphi_j(x,y)$'s as the envelope 
functions of this theory as applied to gated finite-size semiconductor and carbon nanostructures 
\cite{ [{For the application of the effective mass theory to the band structure of single carbon-nanotube 
QDs, see Ref.\ \cite{maks12}(b) and (a) }][]sech10,*[{for the application of the effective mass theory to 
the band structure of single graphene QDs, see Ref.\ \cite{maks12}(a) and (b) }] []bece14,
*[{for the application of the effective mass theory to the band structure of Si QDs,
see (c) }][{; (d) }]eto03,*frie07,*[{and (e) }][]frie10}.

The isospin functions $\zeta$ and $\eta$ are orthornormal, in analogy with the regular spin functions
$\alpha$ and $\beta$. Unlike the exact orthornormality of $\alpha$ and $\beta$, however, the orthornormality 
of $\zeta$ and $\eta$ is an approximate property, which nonetheless is highly accurate when the confining 
(gate) potentials vary slowly over the distance defined by the lattice constant $a$ of the material. This 
follows from the fact \footnote{
See Ref.\ \cite{erca21.2}; (a), (b), and (c) in Ref.\ \cite{sech10}; (a) and (b) in Ref.\ \cite{zare13}.}
that the Bloch functions that multiply the envelope functions in the effective-mass approach 
are varying rapidly in space with a period determined by the material's lattice constant 
$a$, whereas the envelope functions vary slowly over the much larger extent defined by the size of the 
nanostructure.

Specifically, the many-body wave function is written as
\begin{equation}
\Phi^{\rm CI}_{N,q} ({\bf r}_1, \ldots , {\bf r}_N) = 
\sum_I C_I^q \Psi^N_I({\bf r}_1, \ldots , {\bf r}_N),
\label{mbwf}
\end{equation}
where 
\begin{equation}
\Psi^N_I = \frac{1}{\sqrt{N!}}
\left\vert
\begin{array}{ccc}
\chi_{j_1}({\bf r}_1) & \dots & \chi_{j_N}({\bf r}_1) \\
\vdots & \ddots & \vdots \\
\chi_{j_1}({\bf r}_N) & \dots & \chi_{j_N}({\bf r}_N) \\
\end{array}
\right\vert,
\label{detexd}
\end{equation}
and the master index $I$ counts the number of arrangements $\{j_1,j_2,\ldots,j_N\}$ under 
the restriction that $1 \leq j_1 < j_2 <\ldots < j_N \leq 4K$. $I$ specifies the dimension of
the many-body Hilbert space spanned by the basis of Slater determinants.
Of course, $q=1,2,\ldots$ counts the excitation spectrum, with $q=1$ corresponding to the ground state.

The many-body Schr\"{o}dinger equation 
\begin{equation}
{\cal H} \Phi^{\rm CI}_{N,q} = E^{\rm CI}_{N,q} \Phi^{\rm CI}_{N,q}
\label{mbsch}
\end{equation}
transforms into a matrix diagonalization problem, which yields the coefficients $C_I^q$ and the 
eigenenergies $E^{\rm CI}_{N,q}$. Because the resulting matrix is sparse, we implement its numerical 
diagonalization employing the well known ARPACK solver \cite{arpack} which uses implicitly restarted 
Arnoldi methods. Convergence of the many-body solutions is guaranteed by using a large enough value 
for the dimension $K$ of the single-particle basis; we used here $K \sim 50$. The attribute ``full'' 
is usually used for such well converged CI solutions, which naturally contain all possible
$n$-particle $-$ $n$-hole basis Slater determinants, $n$ being an integer.

The matrix elements $\langle \Psi^N_{I} | H_{\rm MB} + H_{\rm VS} + H_{\rm SIC} | \Psi^N_{J} \rangle$, or
the simpler ones $\langle \Psi^N_{I} | H_{\rm MB} + H_{\rm VS} | \Psi^N_{J} \rangle$,
between the basis Slater determinants [see Eq.\ (\ref{detexd})] are calculated using the Slater–Condon rules
\cite{yann09,szabo,slat29,cond30}; for the spin- and/or isospin-dependent Hamiltonian terms $H_{\rm VS}$ and
$H_{\rm SIC}$, see Sec.\ \ref{sec:sic} below. 

Naturally, an important ingredient in this respect are the matrix elements of the two-body interaction,
\begin{equation}
\int_{-\infty}^{\infty} \int_{-\infty}^{\infty} d{\bf r}_1 d{\bf r}_2
\varphi^*_i({\bf r}_1) \varphi^*_j({\bf r}_2) V({\bf r}_1,{\bf r}_2)
\varphi_k({\bf r}_1) \varphi_l({\bf r}_2),
\label{clme}
\end{equation}
in the basis formed out of the single-particle spatial orbitals 
$\varphi_i({\bf r})$, $i=1,2,\ldots,K$ [see Eq.\ (\ref{chi1234})]. In our approach, 
these matrix elements are determined numerically and stored separately.

Taken individually, the Slater determinants $\Psi^N_I$ [see Eq.\ (\ref{detexd})] preserve the third 
projections $S_z$ and $V_z$,  but not necessarily the square $\hat{\bf S}^2$ and $\hat{\bf V}^2$ of the total 
spin and total isospin. However, because $\hat{\bf S}^2$ and $\hat{\bf V}^2$ commute with the many-body 
Hamiltonians $H_{\rm MB}$ and $H_{\rm MB}+H_{\rm VS}$, the associated exact many-body solutions are 
eigenstates of both $\hat{\bf S}^2$ and $\hat{\bf V}^2$ with eigenvalues $\cs(\cs+1)$ and $\cv(\cv+1)$, 
respectively. 

With the VFCI solution at hand [Eq.\ (\ref{mbwf}), which numerically approximates the exact many-body one], 
one calculates the expectation values 
\begin{align}
\langle \Phi^{\rm CI}_N| \hat{\bf S}^2 |\Phi^{\rm CI}_N \rangle=
\sum_I \sum_J C_I^* C_J \langle \Psi_I^N | \hat{\bf S}^2 | \Psi_J^N \rangle,
\label{s2_exp_val}
\end{align}
and similarly for $\hat{\bf V}^2$; for simplicity, in Eq.\ (\ref{s2_exp_val}), we dropped the index $q$.
The ARPACK diagonalization provides the numerical $C_I$ and $C_J$ coefficients, and the matrix elements
of $\hat{\bf S}^2$ and $\hat{\bf V}^2$ between the basis Slater determinants are determined by using the
relations 
\begin{equation}
\hat{{\bf S}}^2 \Psi^N_I = 
\left [(N_\alpha - N_\beta)^2/4 + N/2 + \sum_{i<j} \varpi_{ij} \right ] 
\Psi^N_I,
\label{s2}
\end{equation}
and
\begin{equation}
\hat{{\bf V}}^2 \Psi^N_I = 
\left [(N_\zeta - N_\eta)^2/4 + N/2 + \sum_{i<j} \varpi_{ij}^{\rm iso} \right ] 
\Psi^N_I,
\label{v2}
\end{equation}
where the operator $\varpi_{ij}$ ($\varpi_{ij}^{\rm iso}$) interchanges the spins (isospins) 
of fermions $i$ and $j$ provided that 
these spins (isospins) are different; $N_\alpha$ ($N_\zeta$) and $N_\beta$ ($N_\eta$) denote the number of 
spin-up (isospin-up) and spin-down (isospin-down) fermions, respectively.
Formula (\ref{v2}) for the square of the isospin is introduced here in complete analogy with
the familiar expression \cite{pauncz} for the square of the regular spin. 

Furthermore, the VFCI expectation values for the total-spin projection are calculated using the
formula:
\begin{align}
\langle \Phi^{\rm CI}_N| S_z |\Phi^{\rm CI}_N \rangle=
\sum_I C_I^* C_I \langle \Psi_I^N | S_z | \Psi_I^N \rangle,
\label{sz_exp_val}
\end{align}
and similarly for the total-isospin projection $V_z$.

We note that the VFCI solutions of the reference many-body Hamiltonian (\ref{mbhd}), as well those of
the $H_{\rm MB}+H_{\rm VS}$ Hamiltonian, preserve automatically the spin and isospin quantum numbers as long 
as they are not members of an energy degeneracy. To enforce that the VFCI solutions of these Hamiltonians 
preserve the spin and isospin quantum numbers in all instances, including the case of degeneracies, we add to 
$H_{\rm MB}$, or to $H_{\rm MB}+H_{\rm VS}$, a very small perturbing term $H_{\rm SIC}$, which 
lifts the energy degeneracies to an imperceptible amount, but it produces the desired effect. An example of 
the success of this approach is presented in Table \ref{vfci_si} in Appendix \ref{sec:vfci_si}, where the
deviations of the expectation values of the $\hat{{\bf S}}^2$ and $\hat{{\bf V}}^2$ from the expected
$\cs(\cs+1)$ and $\cv(\cv+1)$ integer values, i.e., 0 or 2 for two electrons in both cases, appear at most
in the fifth decimal point for all the 16 states listed. Similarly, the deviations of the expectation values 
of $S_z$ and $V_z$ from the expected $\pm 1$ or 0 integer values for two electrons appear again at most
in the fifth decimal point for all the 16 states listed.  

\subsection{The spin-isospin coupling and the valley splitting} 
\label{sec:sic}

Motivated by the large body of experimental evidence 
\cite{dzur13,taha14,dzur17,nich21} that a spin-valley coupling is operational in Si qubits, we implement 
in the VFCI an appropriate spin- and isospin-dependent coupling, referred to in this paper as spin-isospin 
coupling, by adding the following (one-body) term in the many-body Hamiltonian:
\begin{align}
H_{\rm SIC}=H_{\rm VOC} + H_{\rm SVOC},
\label{hsic}
\end{align}
where
\begin{align}
H_{\rm VOC} = \sum_{i=1}^N \Delta e^{i\phi_0} \widehat{O}(x,y) \tau_x(i),
\label{hvoc}
\end{align}
and
\begin{align}
H_{\rm SVOC} = \sum_{i=1}^N \Delta e^{i\phi_0} \widehat{O}(x,y)  \sigma_x(i)\tau_x(i).
\label{hsvoc}
\end{align}
with $\Delta$ being the strength and $\phi_0$ being the phase of the coupling parameter. When calculating the
$\langle \chi_{j_1}({\bf r})|H_{\rm SIC}| \chi_{j_2}({\bf r}) \rangle$ matrix elements, we approximate
the integrals over the space variables as 
$\langle \varphi_{j_1}(x,y) |\widehat{O}(x,y)| \varphi_{j_2}(x,y) \rangle \approx
\delta_{j_1-j_2,\pm 1}$, where $1 < j_1 \mbox{~or~} j_2 \leq K$ \footnote{
The Kronecker delta explicitly couples space orbitals with different indices and thus it implements a 
simplification of the effect from the space derivatives present in $\widehat{O}(x,y)$ according to the 
original expression for the spin-orbit coupling suggested in Yu. A. Bychkov and E. I. Rashba, JETP Lett. 
{\bf 39}, 78 (1984)}. 

We note that the first term, $H_{\rm VOC}$, in Eq.\ (\ref{hsic}) implements a pure intervalley coupling 
(referred to often as valley-orbit coupling) by keeping the spin indices unaltered. The second term, 
$H_{\rm SVOC}$, in Eq.\ (\ref{hsic}) flips both the valley and regular-spin indices, thus corresponding to 
a combined VOC and spin-orbit coupling; it is referred to as spin-valley-orbit coupling, or simply
as spin-valley coupling. A pure $H_{\rm VOC}$ coupling is implented in the VFCI by 
restricting the Hilbert space to the sector that preserves the total spin projection, $S_z$. To implement in
addition the $H_{\rm SVOC}$ coupling, one needs to remove all restrictions on $S_z$ and $V_z$ when building the
basis of Slater determinants. This requires substantially larger Hilbert spaces, e.g., for the case of $K=54$ 
employed in the calculations of Fig.\ \ref{figmf}, the dimension of the Hilbert space (the master index $I$)
increases from 5778 (using an $S_z=0$ restriction only) to 23220 basis Slater determinants (with no
$S_z$ and $V_z$ restrictions).

Finally, the valley gap (valley splitting) is described by the following (one-body) Hamiltonian term:
\begin{align}
H_{\rm VS} = \sum_{i=1}^N \frac{E_V}{2} \tau_z(i).
\label{hvg}
\end{align}

\begin{figure*}[t]
\centering\includegraphics[width=16cm]{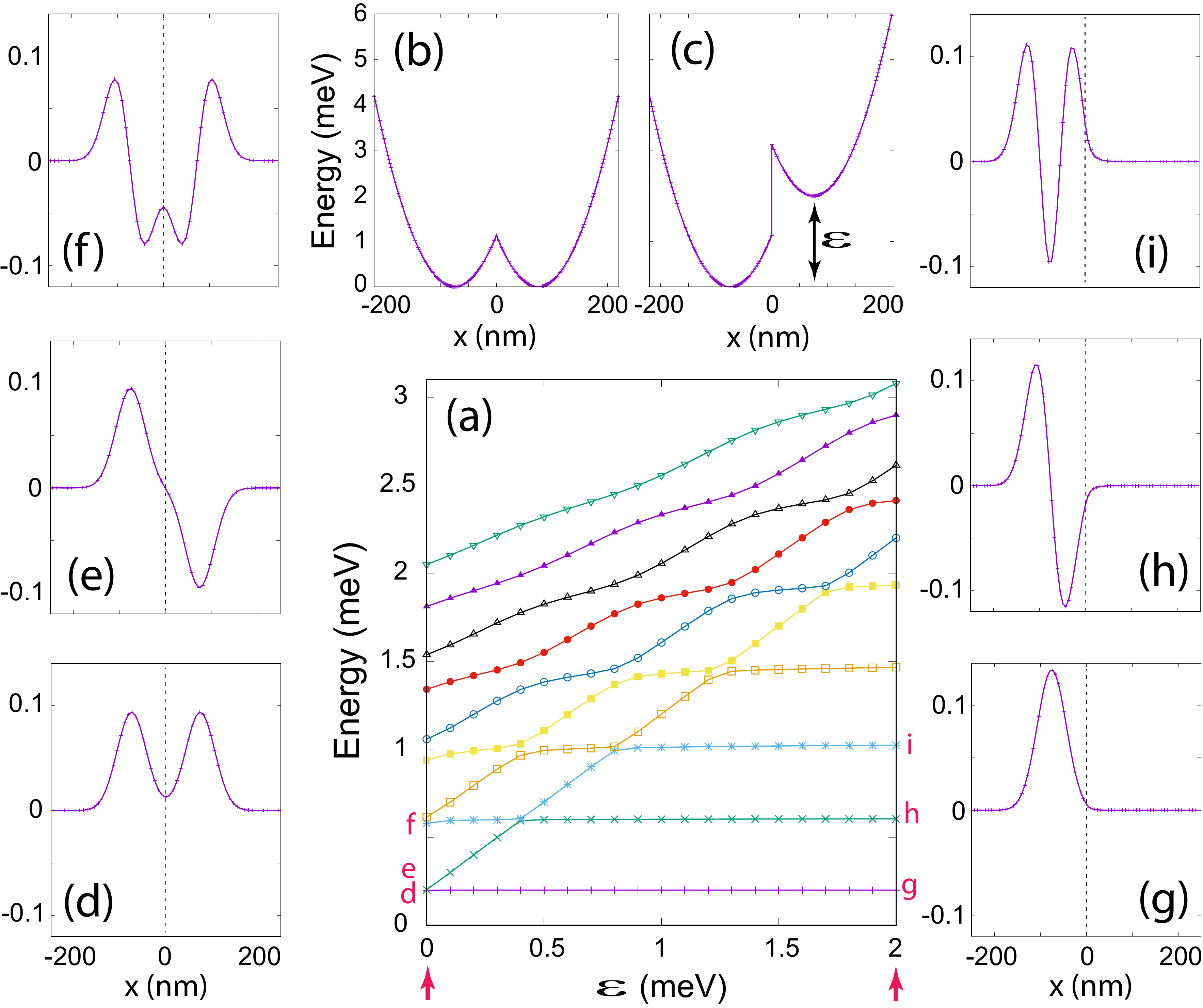}
\caption{
Illustration of the single-particle basis. (a) Energies, as a function of the detuning $\ve$, of the first 
ten states of the one-dimensional Hamiltonian $H_{\rm aux}^{(x)}$ [see Eq.\ (\ref{haux2})] that 
participate in the single-particle basis associated with the 2D Hamiltonian $H_{\rm aux}$. Frames (b) and 
(c) display the corresponding potential confinements, $m^* \omega_{xk}^2 x_k^{\prime 2}/2+h_k$, at two 
values of the detuning (indicated by the two upwards-pointing arrows), i.e, at $\ve=0$ and $\ve=2.0$ meV, 
respectively; here $h_1=0$ and $h_2=\ve$.
The associated single-particle wave functions $X_{\mu(j)}(x)$ for $j=1,\ldots,3$ [see Eq.\ 
(\ref{spphi})] are also displayed at these two points; see panels (d-f) for $\ve=0$ and panels
(g-i) for $\ve=2.0$ meV.
We mention that the smooth-neck Hamiltonian term, $V_{\rm neck}(x)$, is not included in $H_{\rm aux}$.
The parameters entering in $H_{\rm aux}^{(x)}$ were chosen as: confinement $\hbar \omega_{x1}
=\hbar \omega_{x2}=\hbar \omega_0=0.40$ meV, interwell distance $d=150$ nm (with $-x_1=x_2=75$ nm), and 
effective mass $m^*=0.191m_e$ (appropriate for Si). 
The wave functions in the panels (d-f) and (g-i) are in units of $1/\sqrt{nm}$.
}
\label{bas}
\end{figure*}

\section{Summary}

In this paper, a valleytronic FCI has been introduced that integrates in its formalism the SU(4) $\supset$ 
SU(2) $\times$ SU(2) group-theoretical organizational principles underlying the variety of multiplicities 
in the electronic spectra of a Si DQD qubit. A first application was presented concerning a detailed and 
complete analysis of the spectra of a 2e-DQD qubit. 

In the case of the two operational low-energy valleys of a Si qubit, this was achieved by exploiting the 
fact that the valley degree of freedom can, to a very good approximation, be treated as an isospin 
\cite{hein70} in complete analogy with the regular spin $-$ as was to be intuitively expected from well-known
quantum systems in other fields consisting of four species of fermions, such as atomic 
nuclei \cite{wign37} and metal ions of many transition metal oxides \cite{shi98}.

Using the effective mass treatment of the low-energy valleys of Si nanodevices in conjunction with a 
highly adaptable TCO emulation of the artificial gate confinement in a Si DQD qubit, we have introduced an 
appropriate single-particle set of space orbitals, which, when augmented through multiplication with the 
spin and isospin up and down functions, are used as the input of spin-isospin-orbitals in the construction 
of the many-body CI basis of Slater determinants (Sec.\ \ref{sec:meth}).

We demonstrated that our VFCI is able to offer a unified analysis for the spectra of a Si 2e-DQD that 
encompasses all three cases considered. The first two cases concerned the full spectra, including the
most important avoided crossings, as a function of detuning in the transition range from the (1,1) to the 
(2,0) charge configuration when: (i) the formation of a strong Wigner molecule (see Fig.\ \ref{sp046550}) 
suppresses the energy of the first-excited state, $E_{\rm ST}^{\mbox \circleps}$, 
within the same low-energy valley below the valley gap, $E_V$ 
(Sec.\ \ref{sec:same}), and (ii) in conjunction with the formation of a weaker WM, the valley gap $E_V$
determines the energy of the first-excited state (Sec.\ \ref{sec:diff}). The third case concerned the 
evolution of the spectrum in the (1,1) configuration as a function of the magnetic field, while keeping the 
detuning parameter constant (Sec.\ \ref{sec:mfs}).

When the many-body Hamiltonian accounts for the valley splitting, but does not include any valley-orbit or
spin-valley coupling, the 16 hallmark (Sec.\ \ref{sec:count}) low-energy states of the SU(4) $\supset$ 
SU(2) $\times$ SU(2) chain are organized in multiplets according to a 4 \circlep $-$8 \circlez $-$4 \circlem 
~scheme for all three cases mentioned above and as long as the system remains in the (1,1) configuration,
which induces spin-singlet$-$spin-triplet degeneracies due to the large interdot distance; for convenience we 
reiterate here the definitions of the  symbols \circlep, \circlez, and \circlem, which indicate states with
both electrons in the low-energy valley, with the electrons in different valleys, and with both electrons in
the high-energy valley, respectively (see Sec.\ \ref{sec:same_spec}).

When the system transitions to the (2,0) charge configuration, this scheme is modified because the 
spin-singlet$-$spin-triplet degeneracies are lifted; however, the hallmark family of 16 states persists and is
easily traceable in the spectra (see Figs.\ \ref{sp046550} and \ref{sp080065}), as it fans out from the
4 \circlep $-$8 \circlez $-$4 \circlem ~scheme in response to the increasing values of the detuning, $\ve$.
For the cases (i) and (ii), the enhancement of the avoided-crossing gaps in response to a reduced interdot
barrier has been demonstrated explicitly  [see Figs.\ \ref{sp046550}(b) and \ref{sp080065}(b)]. Furthermore,
the transformation of additional simple crossings to prominent avoided ones upon consideration of a 
valley-orbit coupling has been discussed in Sec.\ \ref{sec:diff} and illustrated in Fig.\ \ref{sp080065}(c).
    
Of particular interest are the magnetic-field-dependent VFCI spectra (third case considered, see Sec.\ 
\ref{sec:mfs}), which illustrate the influence of the full SIC coupling, including a spin-valley coupling 
\cite{dzur13,dzur17,taha14,hu14,guo20} which flips both the valley and spin indices. In particular, the VFCI 
magnetic-field spectra in the (1,1) configuration do confirm the appearance of avoided crossings [see Fig.\ 
\ref{figmf}(b)] at the point where the Zeeman energy, $E_Z$, equals the valley splitting, $E_V$.

As elaborated in the main text, the VFCI results presented here are in agreement with the many trends 
revealed in experimental measurements on actual DQD artificial devices that aim at establishing the 
proof-of-principle feasibility of solid-state qubits and logical gates; specifically, among others, the VFCI 
results were shown to emulate trends reported in Refs.\ \cite{corr21,dods22,pett22,nich21,taha14}.    

{\it In conclusion:\/}
With respect to the broader picture, the present paper takes a definitive step towards remedying the current 
incomplete understanding of the complexity of the spectra of Si solid-state qubits. Indeed, it has succeeded 
in integrating under the same framework of an efficient microscopic approach (namely the valleytronic FCI, 
see Sec.\ \ref{sec:ci}) the following pertinent aspects: 1) the valley degree of freedom as an isospin, in 
complete analogy with the regular spin, 2) the SU(4) $\supset$ SU(2) $\times$ SU(2) group-theoretical 
organization of the spectra, containing the salient features of avoided crossings, 3) the effect of strong 
$e-e$ correlations and of the ensuing formation of WMs in the experimentally relevant context of realistic 
double-well confining potentials, which strongly suppresses the spin-singlet$-$spin-triplet gaps within the 
same valley, and 4) the influence of valley-orbit and spin-valley Hamiltonian terms, in particular under an 
applied magnetic field. This valleytronic FCI, demonstrated for the case of two electrons confined in a 
tunable double quantum dot, offers also a most effective tool for analyzing the spectra of Si qubits with 
more than two wells and/or more than two electrons; it can also be straightforwardly extended 
\footnote{
C. Yannouleas and U. Landman, unpublished.} 
to the case of bilayer graphene QDs \cite{enss19,stam21}.

\section{acknowledgments}

This work has been supported by a grant from the Air Force Office of Scientific Research (AFOSR) 
under Award No. FA9550-21-1-0198. Calculations were carried out at the GATECH Center for 
Computational Materials Science.

\appendix

\section{Solving the auxiliary Hamiltonian eigenvalue problem}
\label{sec:aux}

For a given interwell separation $d$, the spatial orbitals $\varphi_i({\bf r})$, $i=1,\ldots,K$ that form 
the single-particle basis [see Eq.\ (\ref{chi1234})] are obtained by a semi-analytic diagonalization of the 
auxiliary single-particle Hamiltonian specified in Eq.\ (\ref{haux}).

Specifically, the eigenvalue problem associated with the auxiliary Hamiltonian [Eq.\ (\ref{haux})] 
is separable in the $x$ and $y$ variables, i.e., one has
\begin{align}
H_{\rm aux}=H_{\rm aux}^{(x)} + H_{\rm aux}^{(y)},
\label{haux2}
\end{align}
and as a result the single-particle wave functions are written as 
\begin{equation}
\varphi_j (x,y)= X_\mu (x) Y_n (y),
\label{spphi}
\end{equation}
with $j \equiv \{\mu,n\}$, $j=1,2,\ldots,K$. As mentioned earlier, $K$ specifies the size of the 
single-particle basis.

\begin{table}[t]
\caption{\label{vfci_si}
The VFCI calculated expectation values $\langle \ldots \rangle$ for the 
total spin and total isospin associated with the VFCI states 
in Fig.\ \ref{figmf}(a) at a magnetic-field value of $B=0.05$ T. The 16 displayed states are labeled in this
Table in ascending energy-eigenvalue order, with the ground state being labeled as no. 1. 
Deviations from the expected group-theoretical values, i.e., $\langle \hat{{\bf S}}^2 \rangle_{\rm exact}=
\cs(\cs+1)$ and $\langle \hat{{\bf V}}^2 \rangle_{\rm exact}=\cv(\cv+1)$ (0 or 2), and 
$\langle S_z \rangle_{\rm exact}=0$, $\pm1$, $\langle V_z \rangle_{\rm exact}=0$, $\pm1$, appear at most 
at the fifth decimal point.  
The symbols \circlep, \circlez, and \circlem ~have the same meaning as in Fig.\ \ref{sp046550}.}
\begin{ruledtabular}
\begin{tabular}{rcccccc}
~   & Energy/$h$ (GHz)~ & $\langle \hat{{\bf S}}^2 \rangle$   & $\langle \hat{{\bf V}}^2 \rangle$  & 
$\langle S_z \rangle$   & $\langle V_z \rangle$  & ~~   \\ \hline
   1 &  201.722160 &  2.00000 &  2.00000 & -1.0000  &  1.0000  & \circlep  \\  
   2 &  203.121755 &  0.00000 &  2.00000 & -0.0000  &  1.0000  & \circlep  \\
   3 &  203.121774 &  2.00000 &  2.00000 & -0.0000  &  1.0000  & \circlep  \\
   4 &  204.521388 &  2.00000 &  2.00000 &  1.0000  &  1.0000  & \circlep  \\
   5 &  216.229963 &  2.00000 &  0.00001 & -1.0000  &  0.0000  & \circlez  \\
   6 &  216.229983 &  2.00000 &  1.99999 & -1.0000  &  0.0000  & \circlez  \\
   7 &  217.629577 &  0.00000 &  1.99999 & -0.0000  &  0.0000  & \circlez  \\
   8 &  217.629577 &  2.00000 &  0.00001 & -0.0000  &  0.0000  & \circlez  \\
   9 &  217.629596 &  0.00000 &  0.00001 & -0.0000  &  0.0000  & \circlez  \\
  10 &  217.629596 &  2.00000 &  1.99999 & -0.0000  &  0.0000  & \circlez  \\
  11 &  219.029191 &  2.00000 &  0.00001 &  1.0000  & -0.0000  & \circlez  \\
  12 &  219.029210 &  2.00000 &  1.99999 &  1.0000  & -0.0000  & \circlez  \\
  13 &  230.737805 &  2.00000 &  2.00000 & -1.0000  & -1.0000  & \circlem  \\
  14 &  232.137399 &  0.00000 &  2.00000 &  0.0000  & -1.0000  & \circlem  \\
  15 &  232.137419 &  2.00000 &  2.00000 &  0.0000  & -1.0000  & \circlem  \\
  16 &  233.537032 &  2.00000 &  2.00000 &  1.0000  & -1.0000  & \circlem
\end{tabular}
\end{ruledtabular}
\end{table}
 
The $Y_n (y)$ are the eigenfunctions of a one-dimensional oscillator in the $y$ direction, and the
$X_\mu (x \leq 0)$ and $X_\mu (x>0)$ can be expressed through the parabolic
cylinder functions $U[\gamma_k, (-1)^k \xi_k]$ \cite{para}, where
$\xi_k = x^\prime_k \sqrt{2m^* \omega_{xk}/\hbar}$, 
$\gamma_k=(-E_x+h_k)/(\hbar \omega_{xk})$, 
and $E_x=(\mu+0.5)\hbar \omega_{x1} + h_1$ denotes the $x$-eigenvalues.
The matching conditions at $x=0$ for the left ($k=1$) and right ($k=2$) domains yield the 
$x$-eigenvalues and the eigenfunctions $X_\mu (x)$. The $n$ indices are integer numbers. 
The $\mu$ indices are in general real numbers, but their number is finite.

An advantage of the single-particle orbital basis described in this section is the fact that
it adapts continuously to both the interwell separation $d$ and the detuning parameter $\ve$. As a result,
a very efficient convergence is achieved for any $d$ and $\ve$. The adaptability of our
single-particle orbital basis is illustrated in Fig.\ \ref{bas}. In particular, Fig.\ \ref{bas}(a) 
displays eigenvalues of the non-trivial auxiliary Hamiltonian $H_{\rm aux}^{(x)}$ [see Eq.\ (\ref{haux2})]
which implements the TCO confinement along the $x$ direction. One observes that for larger values of 
detuning, these eigenvalues become constant (as was to be expected), and they run parallel to the $\ve$ 
axis. 

Two cases of the TCO potential confinement are also displayed in Figs.\ \ref{bas}(b) and \ref{bas}(c),
the former corresponding to a symmetric double well ($\ve=0$) and the latter to the case when the right 
well is strongly higher by $\ve=2.0$ meV. The corresponding three lowest-energy eigenfunctions are also
displayed in the triad of Figs.\ \ref{bas}(d-f) and the triad of Figs.\ \ref{bas}(g-i) for these
two values of $\ve$, respectively. It is seen that the eigenfunctions in Figs.\ \ref{bas}(d-f) preserve
the parity around the origin and extend over both wells (as was to be expected for a symmetric double well),
whereas those in Figs.\ \ref{bas}(g-i) are restricted within the left well (as was to be expected again for 
a highly tilted double well) \footnote{
For another illustration of the adaptability of our single-particle orbital basis in the case of a 
symmetric double well ($\ve=0$) as the interwell distance $d$ is varied from zero (``unified atom'') to 
large values (``separated atoms''), see Fig.\ 9 in Ref.\ \cite{yann09}.}.  

We mention again that the smooth-neck Hamiltonian term, $V_{\rm neck}(x)$, is not included in 
$H_{\rm aux}$.

The contributions in the many-body Hamiltonian from the smooth-neck term, the magnetic-field-dependent 
terms, and the spin- and isospin-dependent terms are calculated as part of the many-body exact 
diagonalization by using the Slater-Condon rules for one-body operators between pairs of the Slater 
determinants $\Psi^N_I$ [see Eq.\ (\ref{detexd})].

\section{An example of VFCI results concerning the spin and isospin quantum numbers}
\label{sec:vfci_si}

In this Appendix, we give an example (see Table \ref{vfci_si}) of VFCI calculated expectation values that
correspond to complete sets of the four integer quantum numbers that are expected from group theoretical
considerations for the total spin and the total valley isospin. Deviations from the appropriate integer 
values, if any, appear at most at the fifth decimal digit.\\

\nocite{*}
\bibliographystyle{apsrev4-2}
\bibliography{mycontrols,DoubleDot_2e_valley_Si}

\begin{thebibliography}{91}%
\makeatletter
\providecommand \@ifxundefined [1]{%
 \@ifx{#1\undefined}
}%
\providecommand \@ifnum [1]{%
 \ifnum #1\expandafter \@firstoftwo
 \else \expandafter \@secondoftwo
 \fi
}%
\providecommand \@ifx [1]{%
 \ifx #1\expandafter \@firstoftwo
 \else \expandafter \@secondoftwo
 \fi
}%
\providecommand \natexlab [1]{#1}%
\providecommand \enquote  [1]{``#1''}%
\providecommand \bibnamefont  [1]{#1}%
\providecommand \bibfnamefont [1]{#1}%
\providecommand \citenamefont [1]{#1}%
\providecommand \href@noop [0]{\@secondoftwo}%
\providecommand \href [0]{\begingroup \@sanitize@url \@href}%
\providecommand \@href[1]{\@@startlink{#1}\@@href}%
\providecommand \@@href[1]{\endgroup#1\@@endlink}%
\providecommand \@sanitize@url [0]{\catcode `\\12\catcode `\$12\catcode
  `\&12\catcode `\#12\catcode `\^12\catcode `\_12\catcode `\%12\relax}%
\providecommand \@@startlink[1]{}%
\providecommand \@@endlink[0]{}%
\providecommand \url  [0]{\begingroup\@sanitize@url \@url }%
\providecommand \@url [1]{\endgroup\@href {#1}{\urlprefix }}%
\providecommand \urlprefix  [0]{URL }%
\providecommand \Eprint [0]{\href }%
\providecommand \doibase [0]{https://doi.org/}%
\providecommand \selectlanguage [0]{\@gobble}%
\providecommand \bibinfo  [0]{\@secondoftwo}%
\providecommand \bibfield  [0]{\@secondoftwo}%
\providecommand \translation [1]{[#1]}%
\providecommand \BibitemOpen [0]{}%
\providecommand \bibitemStop [0]{}%
\providecommand \bibitemNoStop [0]{.\EOS\space}%
\providecommand \EOS [0]{\spacefactor3000\relax}%
\providecommand \BibitemShut  [1]{\csname bibitem#1\endcsname}%
\let\auto@bib@innerbib\@empty
\bibitem [{\citenamefont {Zwanenburg}\ \emph {et~al.}(2013)\citenamefont
  {Zwanenburg}, \citenamefont {Dzurak}, \citenamefont {Morello}, \citenamefont
  {Simmons}, \citenamefont {Hollenberg}, \citenamefont {Klimeck}, \citenamefont
  {Rogge}, \citenamefont {Coppersmith},\ and\ \citenamefont
  {Eriksson}}]{copp13}%
  \BibitemOpen
  \bibfield  {author} {\bibinfo {author} {\bibfnamefont {F.~A.}\ \bibnamefont
  {Zwanenburg}}, \bibinfo {author} {\bibfnamefont {A.~S.}\ \bibnamefont
  {Dzurak}}, \bibinfo {author} {\bibfnamefont {A.}~\bibnamefont {Morello}},
  \bibinfo {author} {\bibfnamefont {M.~Y.}\ \bibnamefont {Simmons}}, \bibinfo
  {author} {\bibfnamefont {L.~C.~L.}\ \bibnamefont {Hollenberg}}, \bibinfo
  {author} {\bibfnamefont {G.}~\bibnamefont {Klimeck}}, \bibinfo {author}
  {\bibfnamefont {S.}~\bibnamefont {Rogge}}, \bibinfo {author} {\bibfnamefont
  {S.~N.}\ \bibnamefont {Coppersmith}},\ and\ \bibinfo {author} {\bibfnamefont
  {M.~A.}\ \bibnamefont {Eriksson}},\ }\bibfield  {title} {\bibinfo {title}
  {Silicon quantum electronics},\ }\href
  {https://doi.org/10.1103/RevModPhys.85.961} {\bibfield  {journal} {\bibinfo
  {journal} {Rev. Mod. Phys.}\ }\textbf {\bibinfo {volume} {85}},\ \bibinfo
  {pages} {961--1019} (\bibinfo {year} {2013})}\BibitemShut {NoStop}%
\bibitem [{\citenamefont {Vandersypen}\ and\ \citenamefont
  {Eriksson}(2019)}]{vand19}%
  \BibitemOpen
  \bibfield  {author} {\bibinfo {author} {\bibfnamefont {L.~M.~K.}\
  \bibnamefont {Vandersypen}}\ and\ \bibinfo {author} {\bibfnamefont {M.~A.}\
  \bibnamefont {Eriksson}},\ }\bibfield  {title} {\bibinfo {title} {Quantum
  computing with semiconductor spins},\ }\href
  {https://doi.org/10.1063/PT.3.4270} {\bibfield  {journal} {\bibinfo
  {journal} {Phys. Today}\ }\textbf {\bibinfo {volume} {72}},\ \bibinfo {pages}
  {38--45} (\bibinfo {year} {2019})}\BibitemShut {NoStop}%
\bibitem [{\citenamefont {Burkard}\ \emph {et~al.}()\citenamefont {Burkard},
  \citenamefont {Ladd}, \citenamefont {Nichol}, \citenamefont {Pan},\ and\
  \citenamefont {Petta}}]{burk21}%
  \BibitemOpen
  \bibfield  {author} {\bibinfo {author} {\bibfnamefont {G.}~\bibnamefont
  {Burkard}}, \bibinfo {author} {\bibfnamefont {T.~D.}\ \bibnamefont {Ladd}},
  \bibinfo {author} {\bibfnamefont {J.~M.}\ \bibnamefont {Nichol}}, \bibinfo
  {author} {\bibfnamefont {A.}~\bibnamefont {Pan}},\ and\ \bibinfo {author}
  {\bibfnamefont {J.~R.}\ \bibnamefont {Petta}},\ }\bibfield  {title} {\bibinfo
  {title} {{Semiconductor Spin Qubits}},\ }\href
  {https://doi.org/10.48550/arXiv.2112.08863} {\ }\Eprint
  {https://arxiv.org/abs/arXiv:2112.08863} {arXiv:2112.08863} \BibitemShut
  {NoStop}%
\bibitem [{\citenamefont {Chatterjee}\ \emph {et~al.}(2021)\citenamefont
  {Chatterjee}, \citenamefont {Stevenson}, \citenamefont {De~Franceschi},
  \citenamefont {Morello}, \citenamefont {de~Leon},\ and\ \citenamefont
  {Kuemmeth}}]{kuem21}%
  \BibitemOpen
  \bibfield  {author} {\bibinfo {author} {\bibfnamefont {A.}~\bibnamefont
  {Chatterjee}}, \bibinfo {author} {\bibfnamefont {P.}~\bibnamefont
  {Stevenson}}, \bibinfo {author} {\bibfnamefont {S.}~\bibnamefont
  {De~Franceschi}}, \bibinfo {author} {\bibfnamefont {A.}~\bibnamefont
  {Morello}}, \bibinfo {author} {\bibfnamefont {N.~P.}\ \bibnamefont
  {de~Leon}},\ and\ \bibinfo {author} {\bibfnamefont {F.}~\bibnamefont
  {Kuemmeth}},\ }\bibfield  {title} {\bibinfo {title} {Semiconductor qubits in
  practice},\ }\href {https://doi.org/10.1038/s42254-021-00283-9} {\bibfield
  {journal} {\bibinfo  {journal} {Nature Reviews Physics}\ }\textbf {\bibinfo
  {volume} {3}},\ \bibinfo {pages} {157--177} (\bibinfo {year}
  {2021})}\BibitemShut {NoStop}%
\bibitem [{\citenamefont {Petta}\ \emph {et~al.}(2005)\citenamefont {Petta},
  \citenamefont {Johnson}, \citenamefont {Taylor}, \citenamefont {Laird},
  \citenamefont {Yacoby}, \citenamefont {Lukin}, \citenamefont {Marcus},
  \citenamefont {Hanson},\ and\ \citenamefont {Gossard}}]{pett05}%
  \BibitemOpen
  \bibfield  {author} {\bibinfo {author} {\bibfnamefont {J.~R.}\ \bibnamefont
  {Petta}}, \bibinfo {author} {\bibfnamefont {A.~C.}\ \bibnamefont {Johnson}},
  \bibinfo {author} {\bibfnamefont {J.~M.}\ \bibnamefont {Taylor}}, \bibinfo
  {author} {\bibfnamefont {E.~A.}\ \bibnamefont {Laird}}, \bibinfo {author}
  {\bibfnamefont {A.}~\bibnamefont {Yacoby}}, \bibinfo {author} {\bibfnamefont
  {M.~D.}\ \bibnamefont {Lukin}}, \bibinfo {author} {\bibfnamefont {C.~M.}\
  \bibnamefont {Marcus}}, \bibinfo {author} {\bibfnamefont {M.~P.}\
  \bibnamefont {Hanson}},\ and\ \bibinfo {author} {\bibfnamefont {A.~C.}\
  \bibnamefont {Gossard}},\ }\bibfield  {title} {\bibinfo {title} {Coherent
  manipulation of coupled electron spins in semiconductor quantum dots},\
  }\href {https://doi.org/10.1126/science.1116955} {\bibfield  {journal}
  {\bibinfo  {journal} {Science}\ }\textbf {\bibinfo {volume} {309}},\ \bibinfo
  {pages} {2180--2184} (\bibinfo {year} {2005})}\BibitemShut {NoStop}%
\bibitem [{\citenamefont {Ellenberger}\ \emph {et~al.}(2006)\citenamefont
  {Ellenberger}, \citenamefont {Ihn}, \citenamefont {Yannouleas}, \citenamefont
  {Landman}, \citenamefont {Ensslin}, \citenamefont {Driscoll},\ and\
  \citenamefont {Gossard}}]{yann06}%
  \BibitemOpen
  \bibfield  {author} {\bibinfo {author} {\bibfnamefont {C.}~\bibnamefont
  {Ellenberger}}, \bibinfo {author} {\bibfnamefont {T.}~\bibnamefont {Ihn}},
  \bibinfo {author} {\bibfnamefont {C.}~\bibnamefont {Yannouleas}}, \bibinfo
  {author} {\bibfnamefont {U.}~\bibnamefont {Landman}}, \bibinfo {author}
  {\bibfnamefont {K.}~\bibnamefont {Ensslin}}, \bibinfo {author} {\bibfnamefont
  {D.}~\bibnamefont {Driscoll}},\ and\ \bibinfo {author} {\bibfnamefont
  {A.~C.}\ \bibnamefont {Gossard}},\ }\bibfield  {title} {\bibinfo {title}
  {{Excitation Spectrum of Two Correlated Electrons in a Lateral Quantum Dot
  with Negligible Zeeman Splitting}},\ }\href
  {https://doi.org/10.1103/PhysRevLett.96.126806} {\bibfield  {journal}
  {\bibinfo  {journal} {Phys. Rev. Lett.}\ }\textbf {\bibinfo {volume} {96}},\
  \bibinfo {pages} {126806} (\bibinfo {year} {2006})}\BibitemShut {NoStop}%
\bibitem [{\citenamefont {Jang}\ \emph {et~al.}(2021)\citenamefont {Jang},
  \citenamefont {Cho}, \citenamefont {Jang}, \citenamefont {Kim}, \citenamefont
  {Park}, \citenamefont {Kim}, \citenamefont {Kang}, \citenamefont {Jung},
  \citenamefont {Umansky},\ and\ \citenamefont {Kim}}]{kim21}%
  \BibitemOpen
  \bibfield  {author} {\bibinfo {author} {\bibfnamefont {W.}~\bibnamefont
  {Jang}}, \bibinfo {author} {\bibfnamefont {M.-K.}\ \bibnamefont {Cho}},
  \bibinfo {author} {\bibfnamefont {H.}~\bibnamefont {Jang}}, \bibinfo {author}
  {\bibfnamefont {J.}~\bibnamefont {Kim}}, \bibinfo {author} {\bibfnamefont
  {J.}~\bibnamefont {Park}}, \bibinfo {author} {\bibfnamefont {G.}~\bibnamefont
  {Kim}}, \bibinfo {author} {\bibfnamefont {B.}~\bibnamefont {Kang}}, \bibinfo
  {author} {\bibfnamefont {H.}~\bibnamefont {Jung}}, \bibinfo {author}
  {\bibfnamefont {V.}~\bibnamefont {Umansky}},\ and\ \bibinfo {author}
  {\bibfnamefont {D.}~\bibnamefont {Kim}},\ }\bibfield  {title} {\bibinfo
  {title} {{Single-Shot Readout of a Driven Hybrid Qubit in a GaAs Double
  Quantum Dot}},\ }\href {https://doi.org/10.1021/acs.nanolett.1c00783}
  {\bibfield  {journal} {\bibinfo  {journal} {Nano Letters}\ }\textbf {\bibinfo
  {volume} {21}},\ \bibinfo {pages} {4999--5005} (\bibinfo {year}
  {2021})}\BibitemShut {NoStop}%
\bibitem [{\citenamefont {Maune}\ \emph {et~al.}(2012)\citenamefont {Maune},
  \citenamefont {Borselli}, \citenamefont {Huang}, \citenamefont {Ladd},
  \citenamefont {Deelman}, \citenamefont {Holabird}, \citenamefont {Kiselev},
  \citenamefont {Alvarado-Rodriguez}, \citenamefont {Ross}, \citenamefont
  {Schmitz}, \citenamefont {Sokolich}, \citenamefont {Watson}, \citenamefont
  {Gyure},\ and\ \citenamefont {Hunter}}]{maun12}%
  \BibitemOpen
  \bibfield  {author} {\bibinfo {author} {\bibfnamefont {B.~M.}\ \bibnamefont
  {Maune}}, \bibinfo {author} {\bibfnamefont {M.~G.}\ \bibnamefont {Borselli}},
  \bibinfo {author} {\bibfnamefont {B.}~\bibnamefont {Huang}}, \bibinfo
  {author} {\bibfnamefont {T.~D.}\ \bibnamefont {Ladd}}, \bibinfo {author}
  {\bibfnamefont {P.~W.}\ \bibnamefont {Deelman}}, \bibinfo {author}
  {\bibfnamefont {K.~S.}\ \bibnamefont {Holabird}}, \bibinfo {author}
  {\bibfnamefont {A.~A.}\ \bibnamefont {Kiselev}}, \bibinfo {author}
  {\bibfnamefont {I.}~\bibnamefont {Alvarado-Rodriguez}}, \bibinfo {author}
  {\bibfnamefont {R.~S.}\ \bibnamefont {Ross}}, \bibinfo {author}
  {\bibfnamefont {A.~E.}\ \bibnamefont {Schmitz}}, \bibinfo {author}
  {\bibfnamefont {M.}~\bibnamefont {Sokolich}}, \bibinfo {author}
  {\bibfnamefont {C.~A.}\ \bibnamefont {Watson}}, \bibinfo {author}
  {\bibfnamefont {M.~F.}\ \bibnamefont {Gyure}},\ and\ \bibinfo {author}
  {\bibfnamefont {A.~T.}\ \bibnamefont {Hunter}},\ }\bibfield  {title}
  {\bibinfo {title} {Coherent singlet-triplet oscillations in a silicon-based
  double quantum dot},\ }\href {https://doi.org/10.1038/nature10707} {\bibfield
   {journal} {\bibinfo  {journal} {Nature}\ }\textbf {\bibinfo {volume}
  {481}},\ \bibinfo {pages} {344--347} (\bibinfo {year} {2012})}\BibitemShut
  {NoStop}%
\bibitem [{\citenamefont {Kawakami}\ \emph {et~al.}(2014)\citenamefont
  {Kawakami}, \citenamefont {Scarlino}, \citenamefont {Ward}, \citenamefont
  {Braakman}, \citenamefont {Savage}, \citenamefont {Lagally}, \citenamefont
  {Friesen}, \citenamefont {Coppersmith}, \citenamefont {Eriksson},\ and\
  \citenamefont {Vandersypen}}]{kawa14}%
  \BibitemOpen
  \bibfield  {author} {\bibinfo {author} {\bibfnamefont {E.}~\bibnamefont
  {Kawakami}}, \bibinfo {author} {\bibfnamefont {P.}~\bibnamefont {Scarlino}},
  \bibinfo {author} {\bibfnamefont {D.~R.}\ \bibnamefont {Ward}}, \bibinfo
  {author} {\bibfnamefont {F.~R.}\ \bibnamefont {Braakman}}, \bibinfo {author}
  {\bibfnamefont {D.~E.}\ \bibnamefont {Savage}}, \bibinfo {author}
  {\bibfnamefont {M.~G.}\ \bibnamefont {Lagally}}, \bibinfo {author}
  {\bibfnamefont {M.}~\bibnamefont {Friesen}}, \bibinfo {author} {\bibfnamefont
  {S.~N.}\ \bibnamefont {Coppersmith}}, \bibinfo {author} {\bibfnamefont
  {M.~A.}\ \bibnamefont {Eriksson}},\ and\ \bibinfo {author} {\bibfnamefont
  {L.~M.~K.}\ \bibnamefont {Vandersypen}},\ }\bibfield  {title} {\bibinfo
  {title} {Electrical control of a long-lived spin qubit in a {Si/SiGe} quantum
  dot},\ }\href {https://doi.org/10.1038/nnano.2014.153} {\bibfield  {journal}
  {\bibinfo  {journal} {Nature Nanotechnology}\ }\textbf {\bibinfo {volume}
  {9}},\ \bibinfo {pages} {666--670} (\bibinfo {year} {2014})}\BibitemShut
  {NoStop}%
\bibitem [{\citenamefont {McJunkin}\ \emph {et~al.}()\citenamefont {McJunkin},
  \citenamefont {Harpt}, \citenamefont {Feng}, \citenamefont {Losert},
  \citenamefont {Rahman}, \citenamefont {Dodson}, \citenamefont {Wolfe},
  \citenamefont {Savage}, \citenamefont {Lagally}, \citenamefont {Coppersmith},
  \citenamefont {Friesen}, \citenamefont {Joynt},\ and\ \citenamefont
  {Eriksson}}]{coppwigg21}%
  \BibitemOpen
  \bibfield  {author} {\bibinfo {author} {\bibfnamefont {T.}~\bibnamefont
  {McJunkin}}, \bibinfo {author} {\bibfnamefont {B.}~\bibnamefont {Harpt}},
  \bibinfo {author} {\bibfnamefont {Y.}~\bibnamefont {Feng}}, \bibinfo {author}
  {\bibfnamefont {M.}~\bibnamefont {Losert}}, \bibinfo {author} {\bibfnamefont
  {R.}~\bibnamefont {Rahman}}, \bibinfo {author} {\bibfnamefont {J.~P.}\
  \bibnamefont {Dodson}}, \bibinfo {author} {\bibfnamefont {M.~A.}\
  \bibnamefont {Wolfe}}, \bibinfo {author} {\bibfnamefont {D.~E.}\ \bibnamefont
  {Savage}}, \bibinfo {author} {\bibfnamefont {M.~G.}\ \bibnamefont {Lagally}},
  \bibinfo {author} {\bibfnamefont {S.~N.}\ \bibnamefont {Coppersmith}},
  \bibinfo {author} {\bibfnamefont {M.}~\bibnamefont {Friesen}}, \bibinfo
  {author} {\bibfnamefont {R.}~\bibnamefont {Joynt}},\ and\ \bibinfo {author}
  {\bibfnamefont {M.~A.}\ \bibnamefont {Eriksson}},\ }\bibfield  {title}
  {\bibinfo {title} {{SiGe} quantum wells with oscillating {Ge} concentrations
  for quantum dot qubits},\ }\href {https://doi.org/10.48550/arXiv.2112.09765}
  {\ }\Eprint {https://arxiv.org/abs/arXiv:2112.09765} {arXiv:2112.09765}
  \BibitemShut {NoStop}%
\bibitem [{\citenamefont {Burkard}\ and\ \citenamefont {Petta}(2016)}]{pett16}%
  \BibitemOpen
  \bibfield  {author} {\bibinfo {author} {\bibfnamefont {G.}~\bibnamefont
  {Burkard}}\ and\ \bibinfo {author} {\bibfnamefont {J.~R.}\ \bibnamefont
  {Petta}},\ }\bibfield  {title} {\bibinfo {title} {Dispersive readout of
  valley splittings in cavity-coupled silicon quantum dots},\ }\href
  {https://doi.org/10.1103/PhysRevB.94.195305} {\bibfield  {journal} {\bibinfo
  {journal} {Phys. Rev. B}\ }\textbf {\bibinfo {volume} {94}},\ \bibinfo
  {pages} {195305} (\bibinfo {year} {2016})}\BibitemShut {NoStop}%
\bibitem [{\citenamefont {Mi}\ \emph {et~al.}(2018)\citenamefont {Mi},
  \citenamefont {Kohler},\ and\ \citenamefont {Petta}}]{pett18}%
  \BibitemOpen
  \bibfield  {author} {\bibinfo {author} {\bibfnamefont {X.}~\bibnamefont
  {Mi}}, \bibinfo {author} {\bibfnamefont {S.}~\bibnamefont {Kohler}},\ and\
  \bibinfo {author} {\bibfnamefont {J.~R.}\ \bibnamefont {Petta}},\ }\bibfield
  {title} {\bibinfo {title} {{Landau-Zener interferometry of valley-orbit
  states in Si/SiGe double quantum dots}},\ }\href
  {https://doi.org/10.1103/PhysRevB.98.161404} {\bibfield  {journal} {\bibinfo
  {journal} {Phys. Rev. B}\ }\textbf {\bibinfo {volume} {98}},\ \bibinfo
  {pages} {161404} (\bibinfo {year} {2018})}\BibitemShut {NoStop}%
\bibitem [{\citenamefont {Cai}\ \emph {et~al.}()\citenamefont {Cai},
  \citenamefont {Connors},\ and\ \citenamefont {Nichol}}]{nich21}%
  \BibitemOpen
  \bibfield  {author} {\bibinfo {author} {\bibfnamefont {X.}~\bibnamefont
  {Cai}}, \bibinfo {author} {\bibfnamefont {E.~J.}\ \bibnamefont {Connors}},\
  and\ \bibinfo {author} {\bibfnamefont {J.~M.}\ \bibnamefont {Nichol}},\
  }\bibfield  {title} {\bibinfo {title} {Coherent spin-valley oscillations in
  silicon},\ }\href {https://doi.org/10.48550/arXiv.2111.14847} {\ }\Eprint
  {https://arxiv.org/abs/arXiv:2111.14847} {arXiv:2111.14847} \BibitemShut
  {NoStop}%
\bibitem [{\citenamefont {Ercan}\ \emph {et~al.}(2022)\citenamefont {Ercan},
  \citenamefont {Friesen},\ and\ \citenamefont {Coppersmith}}]{erca22}%
  \BibitemOpen
  \bibfield  {author} {\bibinfo {author} {\bibfnamefont {H.~E.}\ \bibnamefont
  {Ercan}}, \bibinfo {author} {\bibfnamefont {M.}~\bibnamefont {Friesen}},\
  and\ \bibinfo {author} {\bibfnamefont {S.~N.}\ \bibnamefont {Coppersmith}},\
  }\bibfield  {title} {\bibinfo {title} {{Charge-Noise Resilience of
  Two-Electron Quantum Dots in $\mathrm{Si}/\mathrm{SiGe}$ Heterostructures}},\
  }\href {https://doi.org/10.1103/PhysRevLett.128.247701} {\bibfield  {journal}
  {\bibinfo  {journal} {Phys. Rev. Lett.}\ }\textbf {\bibinfo {volume} {128}},\
  \bibinfo {pages} {247701} (\bibinfo {year} {2022})}\BibitemShut {NoStop}%
\bibitem [{\citenamefont {Denisov}\ \emph {et~al.}()\citenamefont {Denisov},
  \citenamefont {Oh}, \citenamefont {Fuchs}, \citenamefont {Mills},
  \citenamefont {Chen}, \citenamefont {Anderson}, \citenamefont {Gyure},
  \citenamefont {Barnard},\ and\ \citenamefont {Petta}}]{pett22}%
  \BibitemOpen
  \bibfield  {author} {\bibinfo {author} {\bibfnamefont {A.~O.}\ \bibnamefont
  {Denisov}}, \bibinfo {author} {\bibfnamefont {S.~W.}\ \bibnamefont {Oh}},
  \bibinfo {author} {\bibfnamefont {G.}~\bibnamefont {Fuchs}}, \bibinfo
  {author} {\bibfnamefont {A.~R.}\ \bibnamefont {Mills}}, \bibinfo {author}
  {\bibfnamefont {P.}~\bibnamefont {Chen}}, \bibinfo {author} {\bibfnamefont
  {C.~R.}\ \bibnamefont {Anderson}}, \bibinfo {author} {\bibfnamefont {M.~F.}\
  \bibnamefont {Gyure}}, \bibinfo {author} {\bibfnamefont {A.~W.}\ \bibnamefont
  {Barnard}},\ and\ \bibinfo {author} {\bibfnamefont {J.~R.}\ \bibnamefont
  {Petta}},\ }\bibfield  {title} {\bibinfo {title} {Microwave-frequency
  scanning gate microscopy of a {Si/SiGe} double quantum dot},\ }\href
  {https://doi.org/10.48550/arXiv.2203.05912} {\ }\Eprint
  {https://arxiv.org/abs/arXiv:2203.05912} {arXiv:2203.05912} \BibitemShut
  {NoStop}%
\bibitem [{\citenamefont {Yannouleas}\ and\ \citenamefont
  {Landman}(1999)}]{yann99}%
  \BibitemOpen
  \bibfield  {author} {\bibinfo {author} {\bibfnamefont {C.}~\bibnamefont
  {Yannouleas}}\ and\ \bibinfo {author} {\bibfnamefont {U.}~\bibnamefont
  {Landman}},\ }\bibfield  {title} {\bibinfo {title} {{Spontaneous Symmetry
  Breaking in Single and Molecular Quantum Dots}},\ }\href
  {https://doi.org/10.1103/PhysRevLett.82.5325} {\bibfield  {journal} {\bibinfo
   {journal} {Phys. Rev. Lett.}\ }\textbf {\bibinfo {volume} {82}},\ \bibinfo
  {pages} {5325--5328} (\bibinfo {year} {1999})}\BibitemShut {NoStop}%
\bibitem [{\citenamefont {Egger}\ \emph {et~al.}(1999)\citenamefont {Egger},
  \citenamefont {{H\"ausler}}, \citenamefont {Mak},\ and\ \citenamefont
  {Grabert}}]{grab99}%
  \BibitemOpen
  \bibfield  {author} {\bibinfo {author} {\bibfnamefont {R.}~\bibnamefont
  {Egger}}, \bibinfo {author} {\bibfnamefont {W.}~\bibnamefont {{H\"ausler}}},
  \bibinfo {author} {\bibfnamefont {C.~H.}\ \bibnamefont {Mak}},\ and\ \bibinfo
  {author} {\bibfnamefont {H.}~\bibnamefont {Grabert}},\ }\bibfield  {title}
  {\bibinfo {title} {{Crossover from Fermi Liquid to Wigner Molecule Behavior
  in Quantum Dots}},\ }\href {https://doi.org/10.1103/PhysRevLett.82.3320}
  {\bibfield  {journal} {\bibinfo  {journal} {Phys. Rev. Lett.}\ }\textbf
  {\bibinfo {volume} {82}},\ \bibinfo {pages} {3320--3323} (\bibinfo {year}
  {1999})}\BibitemShut {NoStop}%
\bibitem [{\citenamefont {Yannouleas}\ and\ \citenamefont
  {Landman}(2000{\natexlab{a}})}]{yann00}%
  \BibitemOpen
  \bibfield  {author} {\bibinfo {author} {\bibfnamefont {C.}~\bibnamefont
  {Yannouleas}}\ and\ \bibinfo {author} {\bibfnamefont {U.}~\bibnamefont
  {Landman}},\ }\bibfield  {title} {\bibinfo {title} {{Collective and
  Independent-Particle Motion in Two-Electron Artificial Atoms}},\ }\href
  {https://doi.org/10.1103/PhysRevLett.85.1726} {\bibfield  {journal} {\bibinfo
   {journal} {Phys. Rev. Lett.}\ }\textbf {\bibinfo {volume} {85}},\ \bibinfo
  {pages} {1726--1729} (\bibinfo {year} {2000}{\natexlab{a}})}\BibitemShut
  {NoStop}%
\bibitem [{\citenamefont {Yannouleas}\ and\ \citenamefont
  {Landman}(2000{\natexlab{b}})}]{yann00.2}%
  \BibitemOpen
  \bibfield  {author} {\bibinfo {author} {\bibfnamefont {C.}~\bibnamefont
  {Yannouleas}}\ and\ \bibinfo {author} {\bibfnamefont {U.}~\bibnamefont
  {Landman}},\ }\bibfield  {title} {\bibinfo {title} {Formation and control of
  electron molecules in artificial atoms: Impurity and magnetic-field
  effects},\ }\href {https://doi.org/10.1103/PhysRevB.61.15895} {\bibfield
  {journal} {\bibinfo  {journal} {Phys. Rev. B}\ }\textbf {\bibinfo {volume}
  {61}},\ \bibinfo {pages} {15895--15904} (\bibinfo {year}
  {2000}{\natexlab{b}})}\BibitemShut {NoStop}%
\bibitem [{\citenamefont {Filinov}\ \emph {et~al.}(2001)\citenamefont
  {Filinov}, \citenamefont {Bonitz},\ and\ \citenamefont {Lozovik}}]{fili01}%
  \BibitemOpen
  \bibfield  {author} {\bibinfo {author} {\bibfnamefont {A.~V.}\ \bibnamefont
  {Filinov}}, \bibinfo {author} {\bibfnamefont {M.}~\bibnamefont {Bonitz}},\
  and\ \bibinfo {author} {\bibfnamefont {Y.~E.}\ \bibnamefont {Lozovik}},\
  }\bibfield  {title} {\bibinfo {title} {{Wigner Crystallization in Mesoscopic
  2D Electron Systems}},\ }\href {https://doi.org/10.1103/PhysRevLett.86.3851}
  {\bibfield  {journal} {\bibinfo  {journal} {Phys. Rev. Lett.}\ }\textbf
  {\bibinfo {volume} {86}},\ \bibinfo {pages} {3851--3854} (\bibinfo {year}
  {2001})}\BibitemShut {NoStop}%
\bibitem [{\citenamefont {Yannouleas}\ and\ \citenamefont
  {Landman}(2002{\natexlab{a}})}]{yann02.1}%
  \BibitemOpen
  \bibfield  {author} {\bibinfo {author} {\bibfnamefont {C.}~\bibnamefont
  {Yannouleas}}\ and\ \bibinfo {author} {\bibfnamefont {U.}~\bibnamefont
  {Landman}},\ }\bibfield  {title} {\bibinfo {title} {Magnetic-field
  manipulation of chemical bonding in artificial molecules},\ }\href
  {https://doi.org/https://doi.org/10.1002/qua.980} {\bibfield  {journal}
  {\bibinfo  {journal} {International Journal of Quantum Chemistry}\ }\textbf
  {\bibinfo {volume} {90}},\ \bibinfo {pages} {699--708} (\bibinfo {year}
  {2002}{\natexlab{a}})}\BibitemShut {NoStop}%
\bibitem [{\citenamefont {Yannouleas}\ and\ \citenamefont
  {Landman}(2002{\natexlab{b}})}]{yann02.2}%
  \BibitemOpen
  \bibfield  {author} {\bibinfo {author} {\bibfnamefont {C.}~\bibnamefont
  {Yannouleas}}\ and\ \bibinfo {author} {\bibfnamefont {U.}~\bibnamefont
  {Landman}},\ }\bibfield  {title} {\bibinfo {title} {Strongly correlated
  wavefunctions for artificial atoms and molecules},\ }\href
  {https://doi.org/10.1088/0953-8984/14/34/101} {\bibfield  {journal} {\bibinfo
   {journal} {Journal of Physics: Condensed Matter}\ }\textbf {\bibinfo
  {volume} {14}},\ \bibinfo {pages} {L591--L598} (\bibinfo {year}
  {2002}{\natexlab{b}})}\BibitemShut {NoStop}%
\bibitem [{\citenamefont {Tavernier}\ \emph {et~al.}(2003)\citenamefont
  {Tavernier}, \citenamefont {Anisimovas}, \citenamefont {Peeters},
  \citenamefont {Szafran}, \citenamefont {Adamowski},\ and\ \citenamefont
  {Bednarek}}]{szaf03}%
  \BibitemOpen
  \bibfield  {author} {\bibinfo {author} {\bibfnamefont {M.~B.}\ \bibnamefont
  {Tavernier}}, \bibinfo {author} {\bibfnamefont {E.}~\bibnamefont
  {Anisimovas}}, \bibinfo {author} {\bibfnamefont {F.~M.}\ \bibnamefont
  {Peeters}}, \bibinfo {author} {\bibfnamefont {B.}~\bibnamefont {Szafran}},
  \bibinfo {author} {\bibfnamefont {J.}~\bibnamefont {Adamowski}},\ and\
  \bibinfo {author} {\bibfnamefont {S.}~\bibnamefont {Bednarek}},\ }\bibfield
  {title} {\bibinfo {title} {Four-electron quantum dot in a magnetic field},\
  }\href {https://doi.org/10.1103/PhysRevB.68.205305} {\bibfield  {journal}
  {\bibinfo  {journal} {Phys. Rev. B}\ }\textbf {\bibinfo {volume} {68}},\
  \bibinfo {pages} {205305} (\bibinfo {year} {2003})}\BibitemShut {NoStop}%
\bibitem [{\citenamefont {Rontani}\ \emph {et~al.}(2006)\citenamefont
  {Rontani}, \citenamefont {Cavazzoni}, \citenamefont {Bellucci},\ and\
  \citenamefont {Goldoni}}]{ront06}%
  \BibitemOpen
  \bibfield  {author} {\bibinfo {author} {\bibfnamefont {M.}~\bibnamefont
  {Rontani}}, \bibinfo {author} {\bibfnamefont {C.}~\bibnamefont {Cavazzoni}},
  \bibinfo {author} {\bibfnamefont {D.}~\bibnamefont {Bellucci}},\ and\
  \bibinfo {author} {\bibfnamefont {G.}~\bibnamefont {Goldoni}},\ }\bibfield
  {title} {\bibinfo {title} {Full configuration interaction approach to the
  few-electron problem in artificial atoms},\ }\href
  {https://doi.org/10.1063/1.2179418} {\bibfield  {journal} {\bibinfo
  {journal} {The Journal of Chemical Physics}\ }\textbf {\bibinfo {volume}
  {124}},\ \bibinfo {pages} {124102} (\bibinfo {year} {2006})}\BibitemShut
  {NoStop}%
\bibitem [{\citenamefont {Li}\ \emph {et~al.}(2007)\citenamefont {Li},
  \citenamefont {Yannouleas},\ and\ \citenamefont {Landman}}]{yann07.2}%
  \BibitemOpen
  \bibfield  {author} {\bibinfo {author} {\bibfnamefont {Y.}~\bibnamefont
  {Li}}, \bibinfo {author} {\bibfnamefont {C.}~\bibnamefont {Yannouleas}},\
  and\ \bibinfo {author} {\bibfnamefont {U.}~\bibnamefont {Landman}},\
  }\bibfield  {title} {\bibinfo {title} {Three-electron anisotropic quantum
  dots in variable magnetic fields: Exact results for excitation spectra, spin
  structures, and entanglement},\ }\href
  {https://doi.org/10.1103/PhysRevB.76.245310} {\bibfield  {journal} {\bibinfo
  {journal} {Phys. Rev. B}\ }\textbf {\bibinfo {volume} {76}},\ \bibinfo
  {pages} {245310} (\bibinfo {year} {2007})}\BibitemShut {NoStop}%
\bibitem [{\citenamefont {Yannouleas}\ and\ \citenamefont
  {Landman}(2007)}]{yann07}%
  \BibitemOpen
  \bibfield  {author} {\bibinfo {author} {\bibfnamefont {C.}~\bibnamefont
  {Yannouleas}}\ and\ \bibinfo {author} {\bibfnamefont {U.}~\bibnamefont
  {Landman}},\ }\bibfield  {title} {\bibinfo {title} {Symmetry breaking and
  quantum correlations in finite systems: {Studies} of quantum dots and
  ultracold {Bose} gases and related nuclear and chemical methods},\ }\href
  {https://doi.org/10.1088/0034-4885/70/12/r02} {\bibfield  {journal} {\bibinfo
   {journal} {Reports on Progress in Physics}\ }\textbf {\bibinfo {volume}
  {70}},\ \bibinfo {pages} {2067--2148} (\bibinfo {year} {2007})}\BibitemShut
  {NoStop}%
\bibitem [{\citenamefont {Li}\ \emph {et~al.}(2009)\citenamefont {Li},
  \citenamefont {Yannouleas},\ and\ \citenamefont {Landman}}]{yann09}%
  \BibitemOpen
  \bibfield  {author} {\bibinfo {author} {\bibfnamefont {Y.}~\bibnamefont
  {Li}}, \bibinfo {author} {\bibfnamefont {C.}~\bibnamefont {Yannouleas}},\
  and\ \bibinfo {author} {\bibfnamefont {U.}~\bibnamefont {Landman}},\
  }\bibfield  {title} {\bibinfo {title} {Artificial quantum-dot helium
  molecules: Electronic spectra, spin structures, and {Heisenberg} clusters},\
  }\href {https://doi.org/10.1103/PhysRevB.80.045326} {\bibfield  {journal}
  {\bibinfo  {journal} {Phys. Rev. B}\ }\textbf {\bibinfo {volume} {80}},\
  \bibinfo {pages} {045326} (\bibinfo {year} {2009})}\BibitemShut {NoStop}%
\bibitem [{\citenamefont {Corrigan}\ \emph {et~al.}(2021)\citenamefont
  {Corrigan}, \citenamefont {Dodson}, \citenamefont {Ercan}, \citenamefont
  {Abadillo-Uriel}, \citenamefont {Thorgrimsson}, \citenamefont {Knapp},
  \citenamefont {Holman}, \citenamefont {McJunkin}, \citenamefont {Neyens},
  \citenamefont {MacQuarrie}, \citenamefont {Foote}, \citenamefont {Edge},
  \citenamefont {Friesen}, \citenamefont {Coppersmith},\ and\ \citenamefont
  {Eriksson}}]{corr21}%
  \BibitemOpen
  \bibfield  {author} {\bibinfo {author} {\bibfnamefont {J.}~\bibnamefont
  {Corrigan}}, \bibinfo {author} {\bibfnamefont {J.~P.}\ \bibnamefont
  {Dodson}}, \bibinfo {author} {\bibfnamefont {H.~E.}\ \bibnamefont {Ercan}},
  \bibinfo {author} {\bibfnamefont {J.~C.}\ \bibnamefont {Abadillo-Uriel}},
  \bibinfo {author} {\bibfnamefont {B.}~\bibnamefont {Thorgrimsson}}, \bibinfo
  {author} {\bibfnamefont {T.~J.}\ \bibnamefont {Knapp}}, \bibinfo {author}
  {\bibfnamefont {N.}~\bibnamefont {Holman}}, \bibinfo {author} {\bibfnamefont
  {T.}~\bibnamefont {McJunkin}}, \bibinfo {author} {\bibfnamefont {S.~F.}\
  \bibnamefont {Neyens}}, \bibinfo {author} {\bibfnamefont {E.~R.}\
  \bibnamefont {MacQuarrie}}, \bibinfo {author} {\bibfnamefont {R.~H.}\
  \bibnamefont {Foote}}, \bibinfo {author} {\bibfnamefont {L.~F.}\ \bibnamefont
  {Edge}}, \bibinfo {author} {\bibfnamefont {M.}~\bibnamefont {Friesen}},
  \bibinfo {author} {\bibfnamefont {S.~N.}\ \bibnamefont {Coppersmith}},\ and\
  \bibinfo {author} {\bibfnamefont {M.~A.}\ \bibnamefont {Eriksson}},\
  }\bibfield  {title} {\bibinfo {title} {{Coherent Control and Spectroscopy of
  a Semiconductor Quantum Dot Wigner Molecule}},\ }\href
  {https://doi.org/10.1103/PhysRevLett.127.127701} {\bibfield  {journal}
  {\bibinfo  {journal} {Phys. Rev. Lett.}\ }\textbf {\bibinfo {volume} {127}},\
  \bibinfo {pages} {127701} (\bibinfo {year} {2021})}\BibitemShut {NoStop}%
\bibitem [{\citenamefont {Ercan}\ \emph {et~al.}(2021)\citenamefont {Ercan},
  \citenamefont {Coppersmith},\ and\ \citenamefont {Friesen}}]{erca21.2}%
  \BibitemOpen
  \bibfield  {author} {\bibinfo {author} {\bibfnamefont {H.~E.}\ \bibnamefont
  {Ercan}}, \bibinfo {author} {\bibfnamefont {S.~N.}\ \bibnamefont
  {Coppersmith}},\ and\ \bibinfo {author} {\bibfnamefont {M.}~\bibnamefont
  {Friesen}},\ }\bibfield  {title} {\bibinfo {title} {Strong electron-electron
  interactions in {Si/SiGe} quantum dots},\ }\href
  {https://doi.org/10.1103/PhysRevB.104.235302} {\bibfield  {journal} {\bibinfo
   {journal} {Phys. Rev. B}\ }\textbf {\bibinfo {volume} {104}},\ \bibinfo
  {pages} {235302} (\bibinfo {year} {2021})}\BibitemShut {NoStop}%
\bibitem [{\citenamefont {Abadillo-Uriel}\ \emph {et~al.}(2021)\citenamefont
  {Abadillo-Uriel}, \citenamefont {Martinez}, \citenamefont {Filippone},\ and\
  \citenamefont {Niquet}}]{urie21}%
  \BibitemOpen
  \bibfield  {author} {\bibinfo {author} {\bibfnamefont {J.~C.}\ \bibnamefont
  {Abadillo-Uriel}}, \bibinfo {author} {\bibfnamefont {B.}~\bibnamefont
  {Martinez}}, \bibinfo {author} {\bibfnamefont {M.}~\bibnamefont
  {Filippone}},\ and\ \bibinfo {author} {\bibfnamefont {Y.-M.}\ \bibnamefont
  {Niquet}},\ }\bibfield  {title} {\bibinfo {title} {Two-body {Wigner}
  molecularization in asymmetric quantum dot spin qubits},\ }\href
  {https://doi.org/10.1103/PhysRevB.104.195305} {\bibfield  {journal} {\bibinfo
   {journal} {Phys. Rev. B}\ }\textbf {\bibinfo {volume} {104}},\ \bibinfo
  {pages} {195305} (\bibinfo {year} {2021})}\BibitemShut {NoStop}%
\bibitem [{\citenamefont {Corrigan}\ \emph {et~al.}()\citenamefont {Corrigan}
  \emph {et~al.}}]{corr21.2}%
  \BibitemOpen
  \bibfield  {author} {\bibinfo {author} {\bibfnamefont {J.}~\bibnamefont
  {Corrigan}} \emph {et~al.},\ }\href {https://arxiv.org/abs/2009.13572} {\
  }\Eprint {https://arxiv.org/abs/arXiv:2009.13572v1} {arXiv:2009.13572v1}
  \BibitemShut {NoStop}%
\bibitem [{\citenamefont {Yannouleas}\ and\ \citenamefont
  {Landman}(2022{\natexlab{a}})}]{yann22}%
  \BibitemOpen
  \bibfield  {author} {\bibinfo {author} {\bibfnamefont {C.}~\bibnamefont
  {Yannouleas}}\ and\ \bibinfo {author} {\bibfnamefont {U.}~\bibnamefont
  {Landman}},\ }\bibfield  {title} {\bibinfo {title} {{Wigner} molecules and
  hybrid qubits},\ }\href {https://doi.org/10.1088/1361-648x/ac5c28} {\bibfield
   {journal} {\bibinfo  {journal} {J. Phys.: Condens. Matter (Letter)}\
  }\textbf {\bibinfo {volume} {34}},\ \bibinfo {pages} {21LT01} (\bibinfo
  {year} {2022}{\natexlab{a}})}\BibitemShut {NoStop}%
\bibitem [{\citenamefont {Yannouleas}\ and\ \citenamefont
  {Landman}(2022{\natexlab{b}})}]{yann22.2}%
  \BibitemOpen
  \bibfield  {author} {\bibinfo {author} {\bibfnamefont {C.}~\bibnamefont
  {Yannouleas}}\ and\ \bibinfo {author} {\bibfnamefont {U.}~\bibnamefont
  {Landman}},\ }\bibfield  {title} {\bibinfo {title} {Molecular formations and
  spectra due to electron correlations in three-electron hybrid double-well
  qubits},\ }\href {https://doi.org/10.1103/PhysRevB.105.205302} {\bibfield
  {journal} {\bibinfo  {journal} {Phys. Rev. B}\ }\textbf {\bibinfo {volume}
  {105}},\ \bibinfo {pages} {205302} (\bibinfo {year}
  {2022}{\natexlab{b}})}\BibitemShut {NoStop}%
\bibitem [{\citenamefont {Shavitt}(1998)}]{shav98}%
  \BibitemOpen
  \bibfield  {author} {\bibinfo {author} {\bibfnamefont {I.}~\bibnamefont
  {Shavitt}},\ }\bibfield  {title} {\bibinfo {title} {The history and evolution
  of configuration interaction},\ }\href
  {https://doi.org/10.1080/002689798168303} {\bibfield  {journal} {\bibinfo
  {journal} {Molecular Physics}\ }\textbf {\bibinfo {volume} {94}},\ \bibinfo
  {pages} {3--17} (\bibinfo {year} {1998})}\BibitemShut {NoStop}%
\bibitem [{\citenamefont {Yannouleas}\ and\ \citenamefont
  {Landman}(2003)}]{yann03}%
  \BibitemOpen
  \bibfield  {author} {\bibinfo {author} {\bibfnamefont {C.}~\bibnamefont
  {Yannouleas}}\ and\ \bibinfo {author} {\bibfnamefont {U.}~\bibnamefont
  {Landman}},\ }\bibfield  {title} {\bibinfo {title} {Two-dimensional quantum
  dots in high magnetic fields: {Rotating-electron-molecule} versus
  composite-fermion approach},\ }\href
  {https://doi.org/10.1103/PhysRevB.68.035326} {\bibfield  {journal} {\bibinfo
  {journal} {Phys. Rev. B}\ }\textbf {\bibinfo {volume} {68}},\ \bibinfo
  {pages} {035326} (\bibinfo {year} {2003})}\BibitemShut {NoStop}%
\bibitem [{\citenamefont {Joecker}\ \emph {et~al.}(2021)\citenamefont
  {Joecker}, \citenamefont {Baczewski}, \citenamefont {Gamble}, \citenamefont
  {Pla}, \citenamefont {Saraiva},\ and\ \citenamefont {Morello}}]{more21}%
  \BibitemOpen
  \bibfield  {author} {\bibinfo {author} {\bibfnamefont {B.}~\bibnamefont
  {Joecker}}, \bibinfo {author} {\bibfnamefont {A.~D.}\ \bibnamefont
  {Baczewski}}, \bibinfo {author} {\bibfnamefont {J.~K.}\ \bibnamefont
  {Gamble}}, \bibinfo {author} {\bibfnamefont {J.~J.}\ \bibnamefont {Pla}},
  \bibinfo {author} {\bibfnamefont {A.}~\bibnamefont {Saraiva}},\ and\ \bibinfo
  {author} {\bibfnamefont {A.}~\bibnamefont {Morello}},\ }\bibfield  {title}
  {\bibinfo {title} {Full configuration interaction simulations of
  exchange-coupled donors in silicon using multi-valley effective mass
  theory},\ }\href {https://doi.org/10.1088/1367-2630/ac0abf} {\bibfield
  {journal} {\bibinfo  {journal} {New Journal of Physics}\ }\textbf {\bibinfo
  {volume} {23}},\ \bibinfo {pages} {073007} (\bibinfo {year}
  {2021})}\BibitemShut {NoStop}%
\bibitem [{\citenamefont {Rycerz}\ \emph {et~al.}(2007)\citenamefont {Rycerz},
  \citenamefont {Tworzyd{\l}o},\ and\ \citenamefont {Beenakker}}]{been07}%
  \BibitemOpen
  \bibfield  {author} {\bibinfo {author} {\bibfnamefont {A.}~\bibnamefont
  {Rycerz}}, \bibinfo {author} {\bibfnamefont {J.}~\bibnamefont
  {Tworzyd{\l}o}},\ and\ \bibinfo {author} {\bibfnamefont {C.~W.~J.}\
  \bibnamefont {Beenakker}},\ }\bibfield  {title} {\bibinfo {title} {Valley
  filter and valley valve in graphene},\ }\href
  {https://doi.org/10.1038/nphys547} {\bibfield  {journal} {\bibinfo  {journal}
  {Nature Physics}\ }\textbf {\bibinfo {volume} {3}},\ \bibinfo {pages}
  {172--175} (\bibinfo {year} {2007})}\BibitemShut {NoStop}%
\bibitem [{\citenamefont {Beenakker}(2008)}]{been08}%
  \BibitemOpen
  \bibfield  {author} {\bibinfo {author} {\bibfnamefont {C.~W.~J.}\
  \bibnamefont {Beenakker}},\ }\bibfield  {title} {\bibinfo {title}
  {Colloquium: Andreev reflection and klein tunneling in graphene},\ }\href
  {https://doi.org/10.1103/RevModPhys.80.1337} {\bibfield  {journal} {\bibinfo
  {journal} {Rev. Mod. Phys.}\ }\textbf {\bibinfo {volume} {80}},\ \bibinfo
  {pages} {1337--1354} (\bibinfo {year} {2008})}\BibitemShut {NoStop}%
\bibitem [{\citenamefont {Shevchenko}\ \emph {et~al.}(2018)\citenamefont
  {Shevchenko}, \citenamefont {Ryzhov},\ and\ \citenamefont {Nori}}]{nori18}%
  \BibitemOpen
  \bibfield  {author} {\bibinfo {author} {\bibfnamefont {S.~N.}\ \bibnamefont
  {Shevchenko}}, \bibinfo {author} {\bibfnamefont {A.~I.}\ \bibnamefont
  {Ryzhov}},\ and\ \bibinfo {author} {\bibfnamefont {F.}~\bibnamefont {Nori}},\
  }\bibfield  {title} {\bibinfo {title} {Low-frequency spectroscopy for quantum
  multilevel systems},\ }\href {https://doi.org/10.1103/PhysRevB.98.195434}
  {\bibfield  {journal} {\bibinfo  {journal} {Phys. Rev. B}\ }\textbf {\bibinfo
  {volume} {98}},\ \bibinfo {pages} {195434} (\bibinfo {year}
  {2018})}\BibitemShut {NoStop}%
\bibitem [{\citenamefont {Scuri}\ \emph {et~al.}(2020)\citenamefont {Scuri},
  \citenamefont {Andersen}, \citenamefont {Zhou}, \citenamefont {Wild},
  \citenamefont {Sung}, \citenamefont {Gelly}, \citenamefont {B\'erub\'e},
  \citenamefont {Heo}, \citenamefont {Shao}, \citenamefont {Joe}, \citenamefont
  {Mier~Valdivia}, \citenamefont {Taniguchi}, \citenamefont {Watanabe},
  \citenamefont {Lon\ifmmode~\check{c}\else \v{c}\fi{}ar}, \citenamefont {Kim},
  \citenamefont {Lukin},\ and\ \citenamefont {Park}}]{scur20}%
  \BibitemOpen
  \bibfield  {author} {\bibinfo {author} {\bibfnamefont {G.}~\bibnamefont
  {Scuri}}, \bibinfo {author} {\bibfnamefont {T.~I.}\ \bibnamefont {Andersen}},
  \bibinfo {author} {\bibfnamefont {Y.}~\bibnamefont {Zhou}}, \bibinfo {author}
  {\bibfnamefont {D.~S.}\ \bibnamefont {Wild}}, \bibinfo {author}
  {\bibfnamefont {J.}~\bibnamefont {Sung}}, \bibinfo {author} {\bibfnamefont
  {R.~J.}\ \bibnamefont {Gelly}}, \bibinfo {author} {\bibfnamefont
  {D.}~\bibnamefont {B\'erub\'e}}, \bibinfo {author} {\bibfnamefont
  {H.}~\bibnamefont {Heo}}, \bibinfo {author} {\bibfnamefont {L.}~\bibnamefont
  {Shao}}, \bibinfo {author} {\bibfnamefont {A.~Y.}\ \bibnamefont {Joe}},
  \bibinfo {author} {\bibfnamefont {A.~M.}\ \bibnamefont {Mier~Valdivia}},
  \bibinfo {author} {\bibfnamefont {T.}~\bibnamefont {Taniguchi}}, \bibinfo
  {author} {\bibfnamefont {K.}~\bibnamefont {Watanabe}}, \bibinfo {author}
  {\bibfnamefont {M.}~\bibnamefont {Lon\ifmmode~\check{c}\else \v{c}\fi{}ar}},
  \bibinfo {author} {\bibfnamefont {P.}~\bibnamefont {Kim}}, \bibinfo {author}
  {\bibfnamefont {M.~D.}\ \bibnamefont {Lukin}},\ and\ \bibinfo {author}
  {\bibfnamefont {H.}~\bibnamefont {Park}},\ }\bibfield  {title} {\bibinfo
  {title} {{Electrically Tunable Valley Dynamics in Twisted
  ${\mathrm{WSe}}_{2}/{\mathrm{WSe}}_{2}$ Bilayers}},\ }\href
  {https://doi.org/10.1103/PhysRevLett.124.217403} {\bibfield  {journal}
  {\bibinfo  {journal} {Phys. Rev. Lett.}\ }\textbf {\bibinfo {volume} {124}},\
  \bibinfo {pages} {217403} (\bibinfo {year} {2020})}\BibitemShut {NoStop}%
\bibitem [{\citenamefont {Mrudul}\ \emph {et~al.}(2021)\citenamefont {Mrudul},
  \citenamefont {\'{A}lvaro Jim\'{e}nez-Gal\'{a}n}, \citenamefont {Ivanov},\
  and\ \citenamefont {Dixit}}]{mrud21}%
  \BibitemOpen
  \bibfield  {author} {\bibinfo {author} {\bibfnamefont {M.~S.}\ \bibnamefont
  {Mrudul}}, \bibinfo {author} {\bibnamefont {\'{A}lvaro
  Jim\'{e}nez-Gal\'{a}n}}, \bibinfo {author} {\bibfnamefont {M.}~\bibnamefont
  {Ivanov}},\ and\ \bibinfo {author} {\bibfnamefont {G.}~\bibnamefont
  {Dixit}},\ }\bibfield  {title} {\bibinfo {title} {Light-induced valleytronics
  in pristine graphene},\ }\href {https://doi.org/10.1364/OPTICA.418152}
  {\bibfield  {journal} {\bibinfo  {journal} {Optica}\ }\textbf {\bibinfo
  {volume} {8}},\ \bibinfo {pages} {422--427} (\bibinfo {year}
  {2021})}\BibitemShut {NoStop}%
\bibitem [{\citenamefont {Heine}(1970)}]{hein70}%
  \BibitemOpen
  \bibfield  {author} {\bibinfo {author} {\bibfnamefont {V.}~\bibnamefont
  {Heine}},\ }\href@noop {} {\emph {\bibinfo {title} {Group Theory in Quantum
  Mechanics: An Introduction to Its Present Usage}}}\ (\bibinfo  {publisher}
  {Pergamon Press},\ \bibinfo {address} {London},\ \bibinfo {year}
  {1970})\BibitemShut {NoStop}%
\bibitem [{\citenamefont {Luttinger}\ and\ \citenamefont
  {Kohn}(1955)}]{kohn55}%
  \BibitemOpen
  \bibfield  {author} {\bibinfo {author} {\bibfnamefont {J.~M.}\ \bibnamefont
  {Luttinger}}\ and\ \bibinfo {author} {\bibfnamefont {W.}~\bibnamefont
  {Kohn}},\ }\bibfield  {title} {\bibinfo {title} {Motion of electrons and
  holes in perturbed periodic fields},\ }\href
  {https://doi.org/10.1103/PhysRev.97.869} {\bibfield  {journal} {\bibinfo
  {journal} {Phys. Rev.}\ }\textbf {\bibinfo {volume} {97}},\ \bibinfo {pages}
  {869--883} (\bibinfo {year} {1955})}\BibitemShut {NoStop}%
\bibitem [{\citenamefont {Secchi}\ and\ \citenamefont
  {Rontani}(2010)}]{sech10}%
  \BibitemOpen
  \bibfield  {author} {\bibinfo {author} {\bibfnamefont {A.}~\bibnamefont
  {Secchi}}\ and\ \bibinfo {author} {\bibfnamefont {M.}~\bibnamefont
  {Rontani}},\ }\bibfield  {title} {\bibinfo {title} {Wigner molecules in
  carbon-nanotube quantum dots},\ }\href
  {https://doi.org/10.1103/PhysRevB.82.035417} {\bibfield  {journal} {\bibinfo
  {journal} {Phys. Rev. B}\ }\textbf {\bibinfo {volume} {82}},\ \bibinfo
  {pages} {035417} (\bibinfo {year} {2010})}\BibitemShut {NoStop}%
\bibitem [{\citenamefont {Guerrero-Becerra}\ and\ \citenamefont
  {Rontani}(2014)}]{bece14}%
  \BibitemOpen
  \bibfield  {author} {\bibinfo {author} {\bibfnamefont {K.~A.}\ \bibnamefont
  {Guerrero-Becerra}}\ and\ \bibinfo {author} {\bibfnamefont {M.}~\bibnamefont
  {Rontani}},\ }\bibfield  {title} {\bibinfo {title} {Wigner localization in a
  graphene quantum dot with a mass gap},\ }\href
  {https://doi.org/10.1103/PhysRevB.90.125446} {\bibfield  {journal} {\bibinfo
  {journal} {Phys. Rev. B}\ }\textbf {\bibinfo {volume} {90}},\ \bibinfo
  {pages} {125446} (\bibinfo {year} {2014})}\BibitemShut {NoStop}%
\bibitem [{\citenamefont {Hada}\ and\ \citenamefont {Eto}(2003)}]{eto03}%
  \BibitemOpen
  \bibfield  {author} {\bibinfo {author} {\bibfnamefont {Y.}~\bibnamefont
  {Hada}}\ and\ \bibinfo {author} {\bibfnamefont {M.}~\bibnamefont {Eto}},\
  }\bibfield  {title} {\bibinfo {title} {Electronic states in silicon quantum
  dots: Multivalley artificial atoms},\ }\href
  {https://doi.org/10.1103/PhysRevB.68.155322} {\bibfield  {journal} {\bibinfo
  {journal} {Phys. Rev. B}\ }\textbf {\bibinfo {volume} {68}},\ \bibinfo
  {pages} {155322} (\bibinfo {year} {2003})}\BibitemShut {NoStop}%
\bibitem [{\citenamefont {Friesen}\ \emph {et~al.}(2007)\citenamefont
  {Friesen}, \citenamefont {Chutia}, \citenamefont {Tahan},\ and\ \citenamefont
  {Coppersmith}}]{frie07}%
  \BibitemOpen
  \bibfield  {author} {\bibinfo {author} {\bibfnamefont {M.}~\bibnamefont
  {Friesen}}, \bibinfo {author} {\bibfnamefont {S.}~\bibnamefont {Chutia}},
  \bibinfo {author} {\bibfnamefont {C.}~\bibnamefont {Tahan}},\ and\ \bibinfo
  {author} {\bibfnamefont {S.~N.}\ \bibnamefont {Coppersmith}},\ }\bibfield
  {title} {\bibinfo {title} {Valley splitting theory of {Si/SiGe} quantum
  wells},\ }\href {https://doi.org/10.1103/PhysRevB.75.115318} {\bibfield
  {journal} {\bibinfo  {journal} {Phys. Rev. B}\ }\textbf {\bibinfo {volume}
  {75}},\ \bibinfo {pages} {115318} (\bibinfo {year} {2007})}\BibitemShut
  {NoStop}%
\bibitem [{\citenamefont {Friesen}\ and\ \citenamefont
  {Coppersmith}(2010)}]{frie10}%
  \BibitemOpen
  \bibfield  {author} {\bibinfo {author} {\bibfnamefont {M.}~\bibnamefont
  {Friesen}}\ and\ \bibinfo {author} {\bibfnamefont {S.~N.}\ \bibnamefont
  {Coppersmith}},\ }\bibfield  {title} {\bibinfo {title} {Theory of
  valley-orbit coupling in a {Si/SiGe} quantum dot},\ }\href
  {https://doi.org/10.1103/PhysRevB.81.115324} {\bibfield  {journal} {\bibinfo
  {journal} {Phys. Rev. B}\ }\textbf {\bibinfo {volume} {81}},\ \bibinfo
  {pages} {115324} (\bibinfo {year} {2010})}\BibitemShut {NoStop}%
\bibitem [{\citenamefont {Wigner}(1937)}]{wign37}%
  \BibitemOpen
  \bibfield  {author} {\bibinfo {author} {\bibfnamefont {E.}~\bibnamefont
  {Wigner}},\ }\bibfield  {title} {\bibinfo {title} {On the consequences of the
  symmetry of the nuclear {Hamiltonian} on the spectroscopy of nuclei},\ }\href
  {https://doi.org/10.1103/PhysRev.51.106} {\bibfield  {journal} {\bibinfo
  {journal} {Phys. Rev.}\ }\textbf {\bibinfo {volume} {51}},\ \bibinfo {pages}
  {106--119} (\bibinfo {year} {1937})}\BibitemShut {NoStop}%
\bibitem [{\citenamefont {Li}\ \emph {et~al.}(1998)\citenamefont {Li},
  \citenamefont {Ma}, \citenamefont {Shi},\ and\ \citenamefont
  {Zhang}}]{shi98}%
  \BibitemOpen
  \bibfield  {author} {\bibinfo {author} {\bibfnamefont {Y.~Q.}\ \bibnamefont
  {Li}}, \bibinfo {author} {\bibfnamefont {M.}~\bibnamefont {Ma}}, \bibinfo
  {author} {\bibfnamefont {D.~N.}\ \bibnamefont {Shi}},\ and\ \bibinfo {author}
  {\bibfnamefont {F.~C.}\ \bibnamefont {Zhang}},\ }\bibfield  {title} {\bibinfo
  {title} {{SU(4) Theory for Spin Systems with Orbital Degeneracy}},\ }\href
  {https://doi.org/10.1103/PhysRevLett.81.3527} {\bibfield  {journal} {\bibinfo
   {journal} {Phys. Rev. Lett.}\ }\textbf {\bibinfo {volume} {81}},\ \bibinfo
  {pages} {3527--3530} (\bibinfo {year} {1998})}\BibitemShut {NoStop}%
\bibitem [{\citenamefont {Arovas}\ and\ \citenamefont
  {Auerbach}(1995)}]{auer95}%
  \BibitemOpen
  \bibfield  {author} {\bibinfo {author} {\bibfnamefont {D.~P.}\ \bibnamefont
  {Arovas}}\ and\ \bibinfo {author} {\bibfnamefont {A.}~\bibnamefont
  {Auerbach}},\ }\bibfield  {title} {\bibinfo {title}
  {Tetrakis(dimethylamino)ethylene-{${\mathrm{C}}_{60}$}: {Multicomponent}
  superexchange and {Mott} ferromagnetism},\ }\href
  {https://doi.org/10.1103/PhysRevB.52.10114} {\bibfield  {journal} {\bibinfo
  {journal} {Phys. Rev. B}\ }\textbf {\bibinfo {volume} {52}},\ \bibinfo
  {pages} {10114--10121} (\bibinfo {year} {1995})}\BibitemShut {NoStop}%
\bibitem [{\citenamefont {Capponi}\ \emph {et~al.}(2016)\citenamefont
  {Capponi}, \citenamefont {Lecheminant},\ and\ \citenamefont
  {Totsuka}}]{capp16}%
  \BibitemOpen
  \bibfield  {author} {\bibinfo {author} {\bibfnamefont {S.}~\bibnamefont
  {Capponi}}, \bibinfo {author} {\bibfnamefont {P.}~\bibnamefont
  {Lecheminant}},\ and\ \bibinfo {author} {\bibfnamefont {K.}~\bibnamefont
  {Totsuka}},\ }\bibfield  {title} {\bibinfo {title} {Phases of one-dimensional
  {SU($N$)} cold atomic {Fermi} gases—from molecular {Luttinger} liquids to
  topological phases},\ }\href
  {https://doi.org/https://doi.org/10.1016/j.aop.2016.01.011} {\bibfield
  {journal} {\bibinfo  {journal} {Annals of Physics}\ }\textbf {\bibinfo
  {volume} {367}},\ \bibinfo {pages} {50--95} (\bibinfo {year}
  {2016})}\BibitemShut {NoStop}%
\bibitem [{\citenamefont {Honerkamp}\ and\ \citenamefont
  {Hofstetter}(2004)}]{hone04}%
  \BibitemOpen
  \bibfield  {author} {\bibinfo {author} {\bibfnamefont {C.}~\bibnamefont
  {Honerkamp}}\ and\ \bibinfo {author} {\bibfnamefont {W.}~\bibnamefont
  {Hofstetter}},\ }\bibfield  {title} {\bibinfo {title} {{Ultracold Fermions
  and the SU($N$) Hubbard Model}},\ }\href
  {https://doi.org/10.1103/PhysRevLett.92.170403} {\bibfield  {journal}
  {\bibinfo  {journal} {Phys. Rev. Lett.}\ }\textbf {\bibinfo {volume} {92}},\
  \bibinfo {pages} {170403} (\bibinfo {year} {2004})}\BibitemShut {NoStop}%
\bibitem [{\citenamefont {Zarenia}\ \emph {et~al.}(2013)\citenamefont
  {Zarenia}, \citenamefont {Partoens}, \citenamefont {Chakraborty},\ and\
  \citenamefont {Peeters}}]{zare13}%
  \BibitemOpen
  \bibfield  {author} {\bibinfo {author} {\bibfnamefont {M.}~\bibnamefont
  {Zarenia}}, \bibinfo {author} {\bibfnamefont {B.}~\bibnamefont {Partoens}},
  \bibinfo {author} {\bibfnamefont {T.}~\bibnamefont {Chakraborty}},\ and\
  \bibinfo {author} {\bibfnamefont {F.~M.}\ \bibnamefont {Peeters}},\
  }\bibfield  {title} {\bibinfo {title} {Electron-electron interactions in
  bilayer graphene quantum dots},\ }\href
  {https://doi.org/10.1103/PhysRevB.88.245432} {\bibfield  {journal} {\bibinfo
  {journal} {Phys. Rev. B}\ }\textbf {\bibinfo {volume} {88}},\ \bibinfo
  {pages} {245432} (\bibinfo {year} {2013})}\BibitemShut {NoStop}%
\bibitem [{\citenamefont {Roy}\ and\ \citenamefont {Maksym}(2012)}]{maks12}%
  \BibitemOpen
  \bibfield  {author} {\bibinfo {author} {\bibfnamefont {M.}~\bibnamefont
  {Roy}}\ and\ \bibinfo {author} {\bibfnamefont {P.~A.}\ \bibnamefont
  {Maksym}},\ }\bibfield  {title} {\bibinfo {title} {Effective mass theory of
  interacting electron states in semiconducting carbon nanotube quantum dots},\
  }\href {https://doi.org/10.1103/PhysRevB.85.205432} {\bibfield  {journal}
  {\bibinfo  {journal} {Phys. Rev. B}\ }\textbf {\bibinfo {volume} {85}},\
  \bibinfo {pages} {205432} (\bibinfo {year} {2012})}\BibitemShut {NoStop}%
\bibitem [{Note1()}]{Note1}%
  \BibitemOpen
  \bibinfo {note} {Also CI implementations (see, e.g., Ref.\ \cite {erca21.2})
  for Si/SiGe single QDs that employ lattice tight-binding single-particle
  bases, as well as CI simulations of exchange-coupled donors in silicon using
  multi-valley effective mass theory \cite {more21}, are using the regular spin
  indices $({\protect \cal S}, S_z)$ only.}\BibitemShut {Stop}%
\bibitem [{Note2()}]{Note2}%
  \BibitemOpen
  \bibinfo {note} {The notation $(n_L,n_R)$ indicates charge configurations
  with $n_L$ electrons in the left dot, and $n_R$ electrons in the right
  dot.}\BibitemShut {Stop}%
\bibitem [{\citenamefont {Dodson}\ \emph {et~al.}(2022)\citenamefont {Dodson},
  \citenamefont {Ercan}, \citenamefont {Corrigan}, \citenamefont {Losert},
  \citenamefont {Holman}, \citenamefont {McJunkin}, \citenamefont {Edge},
  \citenamefont {Friesen}, \citenamefont {Coppersmith},\ and\ \citenamefont
  {Eriksson}}]{dods22}%
  \BibitemOpen
  \bibfield  {author} {\bibinfo {author} {\bibfnamefont {J.~P.}\ \bibnamefont
  {Dodson}}, \bibinfo {author} {\bibfnamefont {H.~E.}\ \bibnamefont {Ercan}},
  \bibinfo {author} {\bibfnamefont {J.}~\bibnamefont {Corrigan}}, \bibinfo
  {author} {\bibfnamefont {M.~P.}\ \bibnamefont {Losert}}, \bibinfo {author}
  {\bibfnamefont {N.}~\bibnamefont {Holman}}, \bibinfo {author} {\bibfnamefont
  {T.}~\bibnamefont {McJunkin}}, \bibinfo {author} {\bibfnamefont {L.~F.}\
  \bibnamefont {Edge}}, \bibinfo {author} {\bibfnamefont {M.}~\bibnamefont
  {Friesen}}, \bibinfo {author} {\bibfnamefont {S.~N.}\ \bibnamefont
  {Coppersmith}},\ and\ \bibinfo {author} {\bibfnamefont {M.~A.}\ \bibnamefont
  {Eriksson}},\ }\bibfield  {title} {\bibinfo {title} {{How Valley-Orbit States
  in Silicon Quantum Dots Probe Quantum Well Interfaces}},\ }\href
  {https://doi.org/10.1103/PhysRevLett.128.146802} {\bibfield  {journal}
  {\bibinfo  {journal} {Phys. Rev. Lett.}\ }\textbf {\bibinfo {volume} {128}},\
  \bibinfo {pages} {146802} (\bibinfo {year} {2022})}\BibitemShut {NoStop}%
\bibitem [{\citenamefont {Hao}\ \emph {et~al.}(2014)\citenamefont {Hao},
  \citenamefont {Ruskov}, \citenamefont {Xiao}, \citenamefont {Tahan},\ and\
  \citenamefont {Jiang}}]{taha14}%
  \BibitemOpen
  \bibfield  {author} {\bibinfo {author} {\bibfnamefont {X.}~\bibnamefont
  {Hao}}, \bibinfo {author} {\bibfnamefont {R.}~\bibnamefont {Ruskov}},
  \bibinfo {author} {\bibfnamefont {M.}~\bibnamefont {Xiao}}, \bibinfo {author}
  {\bibfnamefont {C.}~\bibnamefont {Tahan}},\ and\ \bibinfo {author}
  {\bibfnamefont {H.}~\bibnamefont {Jiang}},\ }\bibfield  {title} {\bibinfo
  {title} {Electron spin resonance and spin-valley physics in a silicon double
  quantum dot},\ }\href {https://doi.org/10.1038/ncomms4860} {\bibfield
  {journal} {\bibinfo  {journal} {Nature Communications}\ }\textbf {\bibinfo
  {volume} {5}},\ \bibinfo {pages} {3860} (\bibinfo {year} {2014})}\BibitemShut
  {NoStop}%
\bibitem [{\citenamefont {Kurzmann}\ \emph {et~al.}(2019)\citenamefont
  {Kurzmann}, \citenamefont {Eich}, \citenamefont {Overweg}, \citenamefont
  {Mangold}, \citenamefont {Herman}, \citenamefont {Rickhaus}, \citenamefont
  {Pisoni}, \citenamefont {Lee}, \citenamefont {Garreis}, \citenamefont {Tong},
  \citenamefont {Watanabe}, \citenamefont {Taniguchi}, \citenamefont
  {Ensslin},\ and\ \citenamefont {Ihn}}]{enss19}%
  \BibitemOpen
  \bibfield  {author} {\bibinfo {author} {\bibfnamefont {A.}~\bibnamefont
  {Kurzmann}}, \bibinfo {author} {\bibfnamefont {M.}~\bibnamefont {Eich}},
  \bibinfo {author} {\bibfnamefont {H.}~\bibnamefont {Overweg}}, \bibinfo
  {author} {\bibfnamefont {M.}~\bibnamefont {Mangold}}, \bibinfo {author}
  {\bibfnamefont {F.}~\bibnamefont {Herman}}, \bibinfo {author} {\bibfnamefont
  {P.}~\bibnamefont {Rickhaus}}, \bibinfo {author} {\bibfnamefont
  {R.}~\bibnamefont {Pisoni}}, \bibinfo {author} {\bibfnamefont
  {Y.}~\bibnamefont {Lee}}, \bibinfo {author} {\bibfnamefont {R.}~\bibnamefont
  {Garreis}}, \bibinfo {author} {\bibfnamefont {C.}~\bibnamefont {Tong}},
  \bibinfo {author} {\bibfnamefont {K.}~\bibnamefont {Watanabe}}, \bibinfo
  {author} {\bibfnamefont {T.}~\bibnamefont {Taniguchi}}, \bibinfo {author}
  {\bibfnamefont {K.}~\bibnamefont {Ensslin}},\ and\ \bibinfo {author}
  {\bibfnamefont {T.}~\bibnamefont {Ihn}},\ }\bibfield  {title} {\bibinfo
  {title} {{Excited States in Bilayer Graphene Quantum Dots}},\ }\href
  {https://doi.org/10.1103/PhysRevLett.123.026803} {\bibfield  {journal}
  {\bibinfo  {journal} {Phys. Rev. Lett.}\ }\textbf {\bibinfo {volume} {123}},\
  \bibinfo {pages} {026803} (\bibinfo {year} {2019})}\BibitemShut {NoStop}%
\bibitem [{\citenamefont {M\"oller}\ \emph {et~al.}(2021)\citenamefont
  {M\"oller}, \citenamefont {Banszerus}, \citenamefont {Knothe}, \citenamefont
  {Steiner}, \citenamefont {Icking}, \citenamefont {Trellenkamp}, \citenamefont
  {Lentz}, \citenamefont {Watanabe}, \citenamefont {Taniguchi}, \citenamefont
  {Glazman}, \citenamefont {Fal'ko}, \citenamefont {Volk},\ and\ \citenamefont
  {Stampfer}}]{stam21}%
  \BibitemOpen
  \bibfield  {author} {\bibinfo {author} {\bibfnamefont {S.}~\bibnamefont
  {M\"oller}}, \bibinfo {author} {\bibfnamefont {L.}~\bibnamefont {Banszerus}},
  \bibinfo {author} {\bibfnamefont {A.}~\bibnamefont {Knothe}}, \bibinfo
  {author} {\bibfnamefont {C.}~\bibnamefont {Steiner}}, \bibinfo {author}
  {\bibfnamefont {E.}~\bibnamefont {Icking}}, \bibinfo {author} {\bibfnamefont
  {S.}~\bibnamefont {Trellenkamp}}, \bibinfo {author} {\bibfnamefont
  {F.}~\bibnamefont {Lentz}}, \bibinfo {author} {\bibfnamefont
  {K.}~\bibnamefont {Watanabe}}, \bibinfo {author} {\bibfnamefont
  {T.}~\bibnamefont {Taniguchi}}, \bibinfo {author} {\bibfnamefont {L.~I.}\
  \bibnamefont {Glazman}}, \bibinfo {author} {\bibfnamefont {V.~I.}\
  \bibnamefont {Fal'ko}}, \bibinfo {author} {\bibfnamefont {C.}~\bibnamefont
  {Volk}},\ and\ \bibinfo {author} {\bibfnamefont {C.}~\bibnamefont
  {Stampfer}},\ }\bibfield  {title} {\bibinfo {title} {{Probing Two-Electron
  Multiplets in Bilayer Graphene Quantum Dots}},\ }\href
  {https://doi.org/10.1103/PhysRevLett.127.256802} {\bibfield  {journal}
  {\bibinfo  {journal} {Phys. Rev. Lett.}\ }\textbf {\bibinfo {volume} {127}},\
  \bibinfo {pages} {256802} (\bibinfo {year} {2021})}\BibitemShut {NoStop}%
\bibitem [{\citenamefont {Cohen}\ and\ \citenamefont
  {Chelikowsky}(1988)}]{cohe88}%
  \BibitemOpen
  \bibfield  {author} {\bibinfo {author} {\bibfnamefont {M.~L.}\ \bibnamefont
  {Cohen}}\ and\ \bibinfo {author} {\bibfnamefont {J.~R.}\ \bibnamefont
  {Chelikowsky}},\ }\href@noop {} {\emph {\bibinfo {title} {{Electronic
  Structure and Optical Properties of Semiconductors}}}}\ (\bibinfo
  {publisher} {Springer-Verlag},\ \bibinfo {address} {Berlin},\ \bibinfo {year}
  {1988})\BibitemShut {NoStop}%
\bibitem [{\citenamefont {Yu}\ and\ \citenamefont {Cardona}(2001)}]{card01}%
  \BibitemOpen
  \bibfield  {author} {\bibinfo {author} {\bibfnamefont {P.~Y.}\ \bibnamefont
  {Yu}}\ and\ \bibinfo {author} {\bibfnamefont {M.}~\bibnamefont {Cardona}},\
  }\href@noop {} {\emph {\bibinfo {title} {{Fundamentals of Semiconductors}}}}\
  (\bibinfo  {publisher} {Springer-Verlag},\ \bibinfo {address} {Berlin},\
  \bibinfo {year} {2001})\ \bibinfo {note} {3rd ed}\BibitemShut {NoStop}%
\bibitem [{\citenamefont {Phillips}(1962)}]{phil62}%
  \BibitemOpen
  \bibfield  {author} {\bibinfo {author} {\bibfnamefont {J.~C.}\ \bibnamefont
  {Phillips}},\ }\bibfield  {title} {\bibinfo {title} {Band structure of
  {Silicon}, {Germanium}, and related semiconductors},\ }\href
  {https://doi.org/10.1103/PhysRev.125.1931} {\bibfield  {journal} {\bibinfo
  {journal} {Phys. Rev.}\ }\textbf {\bibinfo {volume} {125}},\ \bibinfo {pages}
  {1931--1936} (\bibinfo {year} {1962})}\BibitemShut {NoStop}%
\bibitem [{\citenamefont {Schäffler}(1997)}]{schf97}%
  \BibitemOpen
  \bibfield  {author} {\bibinfo {author} {\bibfnamefont {F.}~\bibnamefont
  {Schäffler}},\ }\bibfield  {title} {\bibinfo {title} {High-mobility {Si} and
  {Ge} structures},\ }\href {https://doi.org/10.1088/0268-1242/12/12/001}
  {\bibfield  {journal} {\bibinfo  {journal} {Semiconductor Science and
  Technology}\ }\textbf {\bibinfo {volume} {12}},\ \bibinfo {pages}
  {1515--1549} (\bibinfo {year} {1997})}\BibitemShut {NoStop}%
\bibitem [{\citenamefont {Ando}\ \emph {et~al.}(1982)\citenamefont {Ando},
  \citenamefont {Fowler},\ and\ \citenamefont {Stern}}]{ando82}%
  \BibitemOpen
  \bibfield  {author} {\bibinfo {author} {\bibfnamefont {T.}~\bibnamefont
  {Ando}}, \bibinfo {author} {\bibfnamefont {A.~B.}\ \bibnamefont {Fowler}},\
  and\ \bibinfo {author} {\bibfnamefont {F.}~\bibnamefont {Stern}},\ }\bibfield
   {title} {\bibinfo {title} {Electronic properties of two-dimensional
  systems},\ }\href {https://doi.org/10.1103/RevModPhys.54.437} {\bibfield
  {journal} {\bibinfo  {journal} {Rev. Mod. Phys.}\ }\textbf {\bibinfo {volume}
  {54}},\ \bibinfo {pages} {437--672} (\bibinfo {year} {1982})}\BibitemShut
  {NoStop}%
\bibitem [{Note3()}]{Note3}%
  \BibitemOpen
  \bibinfo {note} {These confinement-induced single-particle energy states are
  also referred to as ``orbitals'', see, e.g., the expressions ``atomic
  orbitals'', ``space orbitals'', and ``spin-orbitals'' in chemistry \cite
  {szabo}}\BibitemShut {NoStop}%
\bibitem [{Note4()}]{Note4}%
  \BibitemOpen
  \bibinfo {note} {The dots in all instances in this paper are placed at equal
  distances from the origin.}\BibitemShut {Stop}%
\bibitem [{Note5()}]{Note5}%
  \BibitemOpen
  \bibinfo {note} {A spin singlet has eigenvalues ${\protect \cal S}({\protect
  \cal S}+1)=0$ and $S_z=0$, whereas a spin triplet has eigenvalues ${\protect
  \cal S}({\protect \cal S}+1)=2$ and $S_z=0$ or $S_z=\pm 1$. Correspondingly,
  an isospin singlet has eigenvalues ${\protect \cal V}({\protect \cal V}+1)=0$
  and $V_z=0$, whereas an isospin triplet has eigenvalues ${\protect \cal
  V}({\protect \cal V}+1)=2$ and $V_z=0$ or $V_z=\pm 1$.}\BibitemShut {Stop}%
\bibitem [{\citenamefont {Yannouleas}\ and\ \citenamefont
  {Landman}(2001)}]{yann01}%
  \BibitemOpen
  \bibfield  {author} {\bibinfo {author} {\bibfnamefont {C.}~\bibnamefont
  {Yannouleas}}\ and\ \bibinfo {author} {\bibfnamefont {U.}~\bibnamefont
  {Landman}},\ }\bibfield  {title} {\bibinfo {title} {Coupling and dissociation
  in artificial molecules},\ }\href {https://doi.org/10.1007/s100530170133}
  {\bibfield  {journal} {\bibinfo  {journal} {The European Physical Journal D -
  Atomic, Molecular, Optical and Plasma Physics}\ }\textbf {\bibinfo {volume}
  {16}},\ \bibinfo {pages} {373--380} (\bibinfo {year} {2001})}\BibitemShut
  {NoStop}%
\bibitem [{\citenamefont {Yannouleas}\ and\ \citenamefont
  {Landman}(2006)}]{yann06.2}%
  \BibitemOpen
  \bibfield  {author} {\bibinfo {author} {\bibfnamefont {C.}~\bibnamefont
  {Yannouleas}}\ and\ \bibinfo {author} {\bibfnamefont {U.}~\bibnamefont
  {Landman}},\ }\bibfield  {title} {\bibinfo {title} {Electron and boson
  clusters in confined geometries: Symmetry breaking in quantum dots and
  harmonic traps},\ }\href {https://doi.org/10.1073/pnas.0509041103} {\bibfield
   {journal} {\bibinfo  {journal} {Proceedings of the National Academy of
  Sciences}\ }\textbf {\bibinfo {volume} {103}},\ \bibinfo {pages}
  {10600--10605} (\bibinfo {year} {2006})},\ \Eprint
  {https://arxiv.org/abs/https://www.pnas.org/doi/pdf/10.1073/pnas.0509041103}
  {https://www.pnas.org/doi/pdf/10.1073/pnas.0509041103} \BibitemShut {NoStop}%
\bibitem [{Note6()}]{Note6}%
  \BibitemOpen
  \bibinfo {note} {We have indeed checked the accuracy of this statement for a
  single elliptic Si QD using the VFCI code.}\BibitemShut {Stop}%
\bibitem [{\citenamefont {{S\'ark\'any}}\ \emph {et~al.}(2017)\citenamefont
  {{S\'ark\'any}}, \citenamefont {Szirmai}, \citenamefont {Moca}, \citenamefont
  {Glazman},\ and\ \citenamefont {{Zar\'and}}}]{sark17}%
  \BibitemOpen
  \bibfield  {author} {\bibinfo {author} {\bibfnamefont {L.}~\bibnamefont
  {{S\'ark\'any}}}, \bibinfo {author} {\bibfnamefont {E.}~\bibnamefont
  {Szirmai}}, \bibinfo {author} {\bibfnamefont {C.~P.}\ \bibnamefont {Moca}},
  \bibinfo {author} {\bibfnamefont {L.}~\bibnamefont {Glazman}},\ and\ \bibinfo
  {author} {\bibfnamefont {G.}~\bibnamefont {{Zar\'and}}},\ }\bibfield  {title}
  {\bibinfo {title} {Wigner crystal phases in confined carbon nanotubes},\
  }\href {https://doi.org/10.1103/PhysRevB.95.115433} {\bibfield  {journal}
  {\bibinfo  {journal} {Phys. Rev. B}\ }\textbf {\bibinfo {volume} {95}},\
  \bibinfo {pages} {115433} (\bibinfo {year} {2017})}\BibitemShut {NoStop}%
\bibitem [{Note7()}]{Note7}%
  \BibitemOpen
  \bibinfo {note} {An additional multiplet of 10 states with an antisymmetric
  space part appears also in a single elliptic Si QD at even higher energy.
  Indeed, for a circular dot, there are two degenerate antisymmetric space wave
  functions (associated with the two angular momenta $L=\pm 1$) and thus the
  second higher-energy multiplet consists of 20 states. Including the 6-member
  lower-energy multiplet, this results in 26 SU(4) states. For an elliptic dot,
  the circular symmetry of the space wave functions is lifted and the multiplet
  of 20 states splits in two 10-state multiplets. Although overlooked, and even
  misinterpreted in earlier CI calculations \cite {corr21}, this underlying
  SU(4) $\supset $ SU(2) $\times $ SU(2) organization of the spectrum of a
  single elliptic Si quantum dot is operational in other variants of CI
  calculations as well, as one can attest by a careful examination of the
  associated CI results. Indeed in Fig.\ 3 of Ref.\ \cite {corr21.2}, panel (b)
  contains 16 states and panel (c) contains 10 states, for a total of 26
  states.}\BibitemShut {Stop}%
\bibitem [{\citenamefont {Rohling}\ and\ \citenamefont
  {Burkard}(2012)}]{burk12}%
  \BibitemOpen
  \bibfield  {author} {\bibinfo {author} {\bibfnamefont {N.}~\bibnamefont
  {Rohling}}\ and\ \bibinfo {author} {\bibfnamefont {G.}~\bibnamefont
  {Burkard}},\ }\bibfield  {title} {\bibinfo {title} {Universal quantum
  computing with spin and valley states},\ }\href
  {https://doi.org/10.1088/1367-2630/14/8/083008} {\bibfield  {journal}
  {\bibinfo  {journal} {New Journal of Physics}\ }\textbf {\bibinfo {volume}
  {14}},\ \bibinfo {pages} {083008} (\bibinfo {year} {2012})}\BibitemShut
  {NoStop}%
\bibitem [{Note8()}]{Note8}%
  \BibitemOpen
  \bibinfo {note} {Ref.\ \cite {pett22} proposes a more general framework. The
  case of the transition from the (1,1) to the (2,0) configuration happens by
  setting $N_1=1$ and $N_2=0$ in Fig.\ 4(b) of Ref.\ \cite
  {pett22}.}\BibitemShut {Stop}%
\bibitem [{\citenamefont {Yang}\ \emph {et~al.}(2013)\citenamefont {Yang},
  \citenamefont {Rossi}, \citenamefont {Ruskov}, \citenamefont {Lai},
  \citenamefont {Mohiyaddin}, \citenamefont {Lee}, \citenamefont {Tahan},
  \citenamefont {Klimeck}, \citenamefont {Morello},\ and\ \citenamefont
  {Dzurak}}]{dzur13}%
  \BibitemOpen
  \bibfield  {author} {\bibinfo {author} {\bibfnamefont {C.~H.}\ \bibnamefont
  {Yang}}, \bibinfo {author} {\bibfnamefont {A.}~\bibnamefont {Rossi}},
  \bibinfo {author} {\bibfnamefont {R.}~\bibnamefont {Ruskov}}, \bibinfo
  {author} {\bibfnamefont {N.~S.}\ \bibnamefont {Lai}}, \bibinfo {author}
  {\bibfnamefont {F.~A.}\ \bibnamefont {Mohiyaddin}}, \bibinfo {author}
  {\bibfnamefont {S.}~\bibnamefont {Lee}}, \bibinfo {author} {\bibfnamefont
  {C.}~\bibnamefont {Tahan}}, \bibinfo {author} {\bibfnamefont
  {G.}~\bibnamefont {Klimeck}}, \bibinfo {author} {\bibfnamefont
  {A.}~\bibnamefont {Morello}},\ and\ \bibinfo {author} {\bibfnamefont {A.~S.}\
  \bibnamefont {Dzurak}},\ }\bibfield  {title} {\bibinfo {title} {Spin-valley
  lifetimes in a silicon quantum dot with tunable valley splitting},\ }\href
  {https://doi.org/10.1038/ncomms3069} {\bibfield  {journal} {\bibinfo
  {journal} {Nature Communications}\ }\textbf {\bibinfo {volume} {4}},\
  \bibinfo {pages} {2069} (\bibinfo {year} {2013})}\BibitemShut {NoStop}%
\bibitem [{\citenamefont {Hwang}\ \emph {et~al.}(2017)\citenamefont {Hwang},
  \citenamefont {Yang}, \citenamefont {Veldhorst}, \citenamefont {Hendrickx},
  \citenamefont {Fogarty}, \citenamefont {Huang}, \citenamefont {Hudson},
  \citenamefont {Morello},\ and\ \citenamefont {Dzurak}}]{dzur17}%
  \BibitemOpen
  \bibfield  {author} {\bibinfo {author} {\bibfnamefont {J.~C.~C.}\
  \bibnamefont {Hwang}}, \bibinfo {author} {\bibfnamefont {C.~H.}\ \bibnamefont
  {Yang}}, \bibinfo {author} {\bibfnamefont {M.}~\bibnamefont {Veldhorst}},
  \bibinfo {author} {\bibfnamefont {N.}~\bibnamefont {Hendrickx}}, \bibinfo
  {author} {\bibfnamefont {M.~A.}\ \bibnamefont {Fogarty}}, \bibinfo {author}
  {\bibfnamefont {W.}~\bibnamefont {Huang}}, \bibinfo {author} {\bibfnamefont
  {F.~E.}\ \bibnamefont {Hudson}}, \bibinfo {author} {\bibfnamefont
  {A.}~\bibnamefont {Morello}},\ and\ \bibinfo {author} {\bibfnamefont {A.~S.}\
  \bibnamefont {Dzurak}},\ }\bibfield  {title} {\bibinfo {title} {Impact of
  $g$-factors and valleys on spin qubits in a silicon double quantum dot},\
  }\href {https://doi.org/10.1103/PhysRevB.96.045302} {\bibfield  {journal}
  {\bibinfo  {journal} {Phys. Rev. B}\ }\textbf {\bibinfo {volume} {96}},\
  \bibinfo {pages} {045302} (\bibinfo {year} {2017})}\BibitemShut {NoStop}%
\bibitem [{Note9()}]{Note9}%
  \BibitemOpen
  \bibinfo {note} {This is an apparent generalization of the term spin-orbital
  used in chemistry and molecular physics \cite {Note3}.}\BibitemShut {Stop}%
\bibitem [{Note10()}]{Note10}%
  \BibitemOpen
  \bibinfo {note} {See Ref.\ \cite {erca21.2}; (a), (b), and (c) in Ref.\ \cite
  {sech10}; (a) and (b) in Ref.\ \cite {zare13}.}\BibitemShut {Stop}%
\bibitem [{\citenamefont {Lehoucq}\ \emph {et~al.}(1998)\citenamefont
  {Lehoucq}, \citenamefont {Sorensen},\ and\ \citenamefont {Yang}}]{arpack}%
  \BibitemOpen
  \bibfield  {author} {\bibinfo {author} {\bibfnamefont {R.~B.}\ \bibnamefont
  {Lehoucq}}, \bibinfo {author} {\bibfnamefont {D.~C.}\ \bibnamefont
  {Sorensen}},\ and\ \bibinfo {author} {\bibfnamefont {C.}~\bibnamefont
  {Yang}},\ }\href@noop {} {\emph {\bibinfo {title} {{ARPACK USERS’ GUIDE:
  Solution of Large-Scale Eigenvalue Problems with Implicitly Restarted ARNOLDI
  Methods}}}}\ (\bibinfo  {publisher} {SIAM},\ \bibinfo {address}
  {Philadelphia},\ \bibinfo {year} {1998})\BibitemShut {NoStop}%
\bibitem [{\citenamefont {Szabo}\ and\ \citenamefont {Ostlund}(1989)}]{szabo}%
  \BibitemOpen
  \bibfield  {author} {\bibinfo {author} {\bibfnamefont {A.}~\bibnamefont
  {Szabo}}\ and\ \bibinfo {author} {\bibfnamefont {N.~S.}\ \bibnamefont
  {Ostlund}},\ }\href@noop {} {\emph {\bibinfo {title} {{Modern Quantum
  Chemistry}}}}\ (\bibinfo  {publisher} {McGraw-Hill},\ \bibinfo {address} {New
  York},\ \bibinfo {year} {1989})\ \bibinfo {note} {{For the Slater-Condon
  rules, see Chap. 4}}\BibitemShut {NoStop}%
\bibitem [{\citenamefont {Slater}(1929)}]{slat29}%
  \BibitemOpen
  \bibfield  {author} {\bibinfo {author} {\bibfnamefont {J.~C.}\ \bibnamefont
  {Slater}},\ }\bibfield  {title} {\bibinfo {title} {The theory of complex
  spectra},\ }\href {https://doi.org/10.1103/PhysRev.34.1293} {\bibfield
  {journal} {\bibinfo  {journal} {Phys. Rev.}\ }\textbf {\bibinfo {volume}
  {34}},\ \bibinfo {pages} {1293--1322} (\bibinfo {year} {1929})}\BibitemShut
  {NoStop}%
\bibitem [{\citenamefont {Condon}(1930)}]{cond30}%
  \BibitemOpen
  \bibfield  {author} {\bibinfo {author} {\bibfnamefont {E.~U.}\ \bibnamefont
  {Condon}},\ }\bibfield  {title} {\bibinfo {title} {{The Theory of Complex
  Spectra}},\ }\href {https://doi.org/10.1103/PhysRev.36.1121} {\bibfield
  {journal} {\bibinfo  {journal} {Phys. Rev.}\ }\textbf {\bibinfo {volume}
  {36}},\ \bibinfo {pages} {1121--1133} (\bibinfo {year} {1930})}\BibitemShut
  {NoStop}%
\bibitem [{\citenamefont {Pauncz}(2000)}]{pauncz}%
  \BibitemOpen
  \bibfield  {author} {\bibinfo {author} {\bibfnamefont {R.}~\bibnamefont
  {Pauncz}},\ }\href@noop {} {\emph {\bibinfo {title} {{The Construction of
  Spin Eigenfunctions: An Exercise Book}}}}\ (\bibinfo  {publisher} {Kluwer
  Academic/Plenum Publishers},\ \bibinfo {address} {New York},\ \bibinfo {year}
  {2000})\BibitemShut {NoStop}%
\bibitem [{Note11()}]{Note11}%
  \BibitemOpen
  \bibinfo {note} {The Kronecker delta explicitly couples space orbitals with
  different indices and thus it implements a simplification of the effect from
  the space derivatives present in $\protect \widehat {O}(x,y)$ according to
  the original expression for the spin-orbit coupling suggested in Yu. A.
  Bychkov and E. I. Rashba, JETP Lett. {\protect \bf 39}, 78
  (1984)}\BibitemShut {NoStop}%
\bibitem [{\citenamefont {Huang}\ and\ \citenamefont {Hu}(2014)}]{hu14}%
  \BibitemOpen
  \bibfield  {author} {\bibinfo {author} {\bibfnamefont {P.}~\bibnamefont
  {Huang}}\ and\ \bibinfo {author} {\bibfnamefont {X.}~\bibnamefont {Hu}},\
  }\bibfield  {title} {\bibinfo {title} {Spin relaxation in a {Si} quantum dot
  due to spin-valley mixing},\ }\href
  {https://doi.org/10.1103/PhysRevB.90.235315} {\bibfield  {journal} {\bibinfo
  {journal} {Phys. Rev. B}\ }\textbf {\bibinfo {volume} {90}},\ \bibinfo
  {pages} {235315} (\bibinfo {year} {2014})}\BibitemShut {NoStop}%
\bibitem [{\citenamefont {Zhang}\ \emph {et~al.}(2020)\citenamefont {Zhang},
  \citenamefont {Hu}, \citenamefont {Li}, \citenamefont {Jing}, \citenamefont
  {Zhou}, \citenamefont {Ma}, \citenamefont {Ni}, \citenamefont {Luo},
  \citenamefont {Cao}, \citenamefont {Wang}, \citenamefont {Hu}, \citenamefont
  {Jiang}, \citenamefont {Guo},\ and\ \citenamefont {Guo}}]{guo20}%
  \BibitemOpen
  \bibfield  {author} {\bibinfo {author} {\bibfnamefont {X.}~\bibnamefont
  {Zhang}}, \bibinfo {author} {\bibfnamefont {R.-Z.}\ \bibnamefont {Hu}},
  \bibinfo {author} {\bibfnamefont {H.-O.}\ \bibnamefont {Li}}, \bibinfo
  {author} {\bibfnamefont {F.-M.}\ \bibnamefont {Jing}}, \bibinfo {author}
  {\bibfnamefont {Y.}~\bibnamefont {Zhou}}, \bibinfo {author} {\bibfnamefont
  {R.-L.}\ \bibnamefont {Ma}}, \bibinfo {author} {\bibfnamefont
  {M.}~\bibnamefont {Ni}}, \bibinfo {author} {\bibfnamefont {G.}~\bibnamefont
  {Luo}}, \bibinfo {author} {\bibfnamefont {G.}~\bibnamefont {Cao}}, \bibinfo
  {author} {\bibfnamefont {G.-L.}\ \bibnamefont {Wang}}, \bibinfo {author}
  {\bibfnamefont {X.}~\bibnamefont {Hu}}, \bibinfo {author} {\bibfnamefont
  {H.-W.}\ \bibnamefont {Jiang}}, \bibinfo {author} {\bibfnamefont {G.-C.}\
  \bibnamefont {Guo}},\ and\ \bibinfo {author} {\bibfnamefont {G.-P.}\
  \bibnamefont {Guo}},\ }\bibfield  {title} {\bibinfo {title} {{Giant
  Anisotropy of Spin Relaxation and Spin-Valley Mixing in a Silicon Quantum
  Dot}},\ }\href {https://doi.org/10.1103/PhysRevLett.124.257701} {\bibfield
  {journal} {\bibinfo  {journal} {Phys. Rev. Lett.}\ }\textbf {\bibinfo
  {volume} {124}},\ \bibinfo {pages} {257701} (\bibinfo {year}
  {2020})}\BibitemShut {NoStop}%
\bibitem [{Note12()}]{Note12}%
  \BibitemOpen
  \bibinfo {note} {C. Yannouleas and U. Landman, unpublished.}\BibitemShut
  {Stop}%
\bibitem [{\citenamefont {Abramowitz}\ and\ \citenamefont
  {Stegun}(1964)}]{para}%
  \BibitemOpen
  \bibfield  {author} {\bibinfo {author} {\bibfnamefont {M.}~\bibnamefont
  {Abramowitz}}\ and\ \bibinfo {author} {\bibfnamefont {I.~A.}\ \bibnamefont
  {Stegun}},\ }\href@noop {} {\emph {\bibinfo {title} {Handbook of Mathematical
  Functions with Formulas, Graphs, and Mathematical Tables}}}\ (\bibinfo
  {publisher} {Dover},\ \bibinfo {address} {New York},\ \bibinfo {year}
  {1964})\ \bibinfo {note} {{Ch.} 19}\BibitemShut {NoStop}%
\bibitem [{Note13()}]{Note13}%
  \BibitemOpen
  \bibinfo {note} {For another illustration of the adaptability of our
  single-particle orbital basis in the case of a symmetric double well
  ($\varepsilon =0$) as the interwell distance $d$ is varied from zero
  (``unified atom'') to large values (``separated atoms''), see Fig.\ 9 in
  Ref.\ \cite {yann09}.}\BibitemShut {Stop}%
\end{thebibliography}%

\end{document}